\pdfoutput=1
\documentclass[useAMS,usenatbib]{mn2e}

\usepackage{amssymb}
\usepackage{amsmath}
\usepackage{rotating}
\usepackage{lscape}
\usepackage{pifont}
\usepackage{placeins}
\usepackage{textcomp}
\newcommand{\cmark}{\ding{51}}
\newcommand{\xmark}{\ding{55}}

\title[Massive YSOs in the SMC]{\textit{K-}band integral field spectroscopy and optical spectroscopy of massive young stellar objects in the Small Magellanic Cloud
\thanks{
Based on data obtained with SINFONI at the European Southern Observatory's Very Large Telescope under programme 092.C-0723(A).
}
}
 \author[J.L. Ward, J. M. Oliveira, J.Th. van Loon, M. Sewi\l{}o]{J.L. Ward$^{1}$\thanks{E-mail:
j.l.ward@keele.ac.uk}, J.M. Oliveira$^{1}$, J.Th. van Loon$^{1}$ and M. Sewi\l{}o$^{2}$\\
$^{1}$Physics and Astrophysics, Lennard-Jones Laboratories, Keele University, Keele, ST5 5BG, UK\\
$^{2}$NASA Goddard Space Flight Center, 8800 Greenbelt Rd, Greenbelt, MD 20771, USA
}
\makeatother 

\begin{document}

\date{Accepted 2016 September 16. Received 2016 September 1; in original form 2016 April 28}

\pagerange{\pageref{firstpage}--\pageref{lastpage}} \pubyear{2002}

\maketitle

\label{firstpage}

\begin{abstract}
We present \textit{K}-band integral field spectroscopic observations towards 17 massive young stellar objects (YSOs) in the low metallicity Small Magellanic
Cloud (SMC) and two YSO candidates in the compact H\,{\sc ii} regions N81 and N88\,A (also in the SMC). These sources, originally identified using
\textit{Spitzer} photometry and/or spectroscopy, have been resolved into 29 \textit{K-}band continuum sources. By comparing Br$\gamma$ emission
luminosities with those presented for a Galactic sample of massive YSOs, we find tentative evidence for increased accretion rates in the SMC. 
Around half of our
targets exhibit emission line (Br$\gamma$, He\,{\sc i} and H$_2$) morphologies which extend significantly beyond the continuum source and we have mapped both
the emission morphologies and the
radial velocity fields. This analysis also reveals evidence for the existence of ionized low density regions in the centre outflows from massive YSOs.
Additionally we present an analysis of optical spectra towards a similar sample of massive YSOs in the SMC, revealing that the optical emission is photo-excited
and originates near the outer edges of molecular clouds, and is therefore consistent with a high mean-free path of UV photons in the interstellar medium (ISM) of the SMC. Finally, we
discuss the sample of YSOs in an evolutionary context incorporating the results of previous infrared and radio observations, as well as the near-infrared and
optical observations presented in this work. Our spectroscopic analysis in both the \textit{K-}band and the optical regimes, combined
with previously obtained infrared and radio data, exposes differences between properties of massive YSOs in our own Galaxy and the SMC,
including tracers of accretion, discs and YSO--ISM interactions.

\end{abstract}


\section{Introduction}

\begin{figure*}
 \begin{center}
\includegraphics{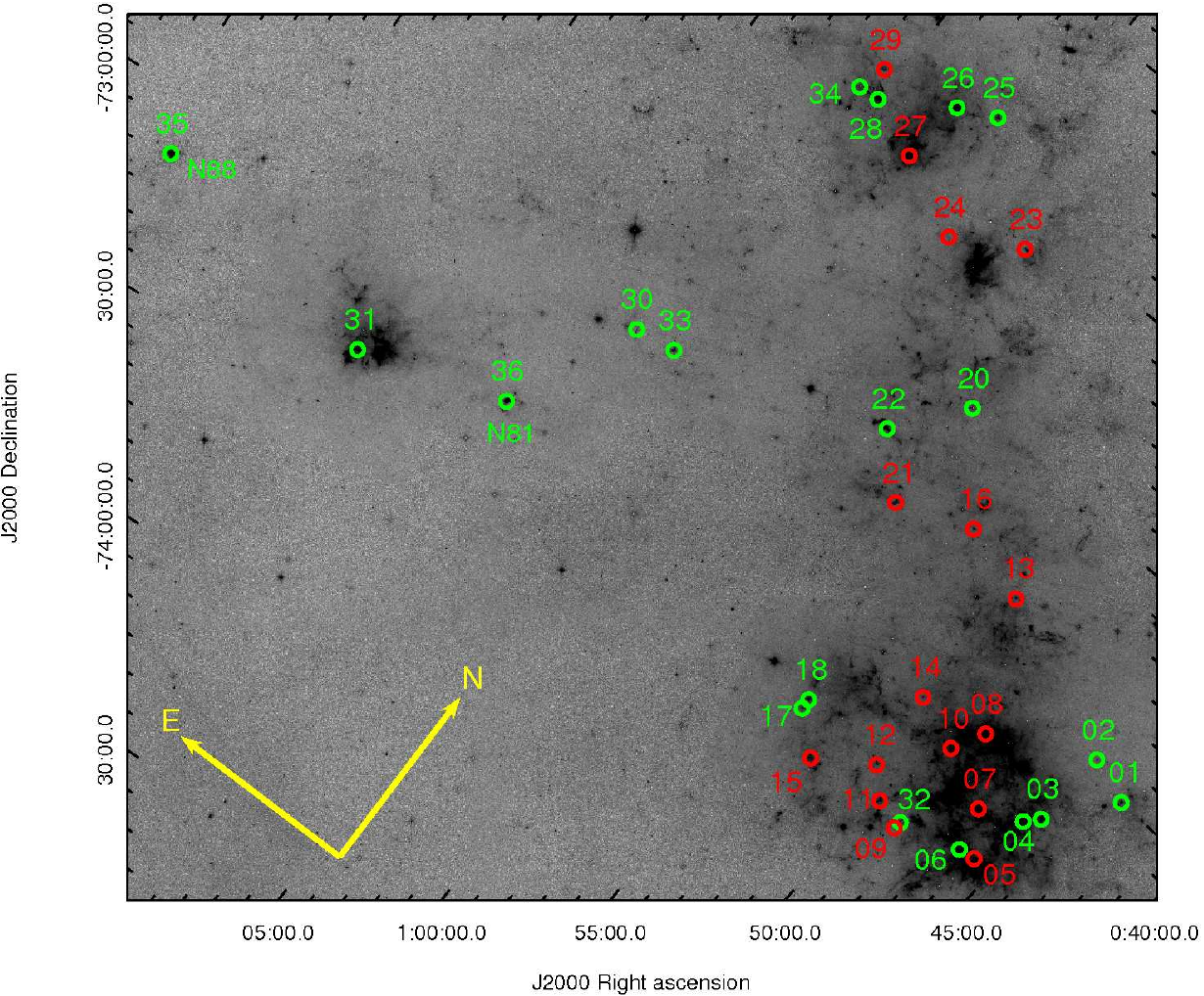}
\caption{SAGE-SMC 8.0\,$\mu$m mosaic of the Small Magellanic Cloud \citep{Gordon2011} showing the positions of all 33 
spectroscopic massive YSOs presented in \citet{Oliveira2013} with the addition of sources \#35 and 36.
The sources observed with SINFONI are in green with the remaining sources marked in red.
The H\,{\sc ii} regions N81 (\#36) and N88 (\#35) are also labelled.}
 \end{center}

\end{figure*}

The formation of massive stars relies on the balance between mass loss and accretion (see \citealt{Zinnecker2007} for a review). Both processes are heavily 
dependent on radiation pressure \citep{Krumholz2012},
the effects of which depend on the dust content of the circumstellar medium.
The metallicity of massive
star forming regions is an important parameter to consider; the heat dissipation necessary to develop dense pre-stellar cores
relies on radiation via the fine-structure lines of carbon and oxygen and the rotational transitions
of metallic molecules such as water and CO. Dust also plays a key role in massive star formation, driving the chemistry of molecular clouds,
forming self-shielding dusty discs, and allowing continued accretion to form stars in excess of 10\,M$_{\sun}$ (Kuiper et al. 2010).
It is therefore expected that massive star formation is
influenced by the initial conditions of the natal molecular cloud.
However the effects of metallicity on massive star formation remain poorly understood with the majority of studies being carried out in Galactic, 
approximately solar metallicity environments.
As a nearby ($\sim$60 kpc) gas-rich galaxy, the Small Magellanic Cloud (SMC) provides a valuable opportunity to study the formation of stars at lower metallicity ($Z_{\text{SMC}}\approx 0.2$\,Z$_{\sun}$; 
\citealt{Peimbert2000}) than in the Milky Way on both the scale of an entire galaxy and on the scale of individual stars.

The metallicity of a star forming region also has a profound effect on the structure and porosity of the interstellar medium (ISM). 
The lower dust abundance in low metallicity environments leads to a porous ISM and thus a larger mean free path length of UV photons \citep{Madden2006,Cormier2015,Dimaratos2015}
meaning that UV photons can permeate over longer distances and excite gas which lies farther from the source of the excitation. 
This in turn will have a significant effect on the feedback mechanisms we observe towards massive star forming regions as well
as the formation of the stars themselves.
Furthermore, the CO emission in the SMC is thought to arise only from high density clumps \citep{Rubio2004}
and the ISM in the similarly low metallicity galaxy NGC 1569 appears to be very clumpy \citep{Galliano2003}.

The \textquotedblleft\textit{Spitzer} Survey of the Small Magellanic Cloud\textquotedblright  (S$^{3}$MC, \citealt{Bolatto2007}) and the
 \textit{Spitzer} \textquotedblleft Surveying the Agents of Galaxy Evolution in the Tidally Stripped, Low Metallicity Small Magellanic Cloud\textquotedblright  
(SAGE--SMC, \citealt{Gordon2011}) survey have allowed the identification of a sample of 
YSO candidates in the SMC, \citet{Sewilo2013} identifying 742 high-reliability YSOs and 242 possible YSOs based on
colour--magnitude cuts, visual inspection of multi-wavelength images and spectral energy distribution (SED) fits to near-IR and mid-IR photometry.
A sample of 33 massive YSOs has been spectroscopically confirmed and studied by \citet{Oliveira2013}; a brief overview of this analysis is provided in Section 2.

In order to determine whether massive YSO properties vary between the Milky Way and the Magellanic 
Clouds, we will compare our results for massive YSOs in the SMC with those of the Red MSX\footnote{Midcourse Space Experiment \citep{Egan2003}} Source Survey 
(RMS Survey; \citealt{Lumsden2013}). The RMS survey provides the most comprehensive catalogue of massive YSOs and 
Ultra-Compact H\,{\sc ii} regions (UCH{\sc ii}s) to date. \textit{H}- and \textit{K}-band infrared spectroscopy of a large sample of these objects has been carried out 
by \citet{Cooper2013} which we will use as a Galactic dataset to compare our results to. 
A similar analysis for a small sample of Large Magellanic Cloud YSOs in LHA120-N113 was performed in \citet{Ward2016} and is also included
in our comparison.

In this work we present the \textit{K-}band integral field spectroscopic observations of 17 previously confirmed YSOs and two YSO candidates in the compact H\,{\sc ii} regions N81 and N88\,A in the SMC,
 obtained with SINFONI 
(Spectrograph for INtegral Field Observations in the Near Infrared; \citealt{Eisenhauer2003}) at the European Southern Observatory's (ESO)
Very Large Telescope (VLT). We also present an analysis of previously obtained  optical long slit spectroscopic observations of 
the H\,{\sc ii} regions surrounding our \textit{K}-band targets (originally presented in \citealt{Oliveira2013}), obtained with the Double-Beam spectrograph (DBS)
at the Australian National University Telescope. New spectroscopic data 
for N88\,A and N81 from the Robert Stobie Spectrograph (RSS) at the Southern African Large Telescope (SALT) are also presented.

Section 2 summarises the results of previous observations towards our sources and Section 3 describes the SINFONI observations presented in this
work and the data reduction procedure. The results of the SINFONI observations are presented in Section 4, with an analysis of the
optical spectra in Section 5. Discussion and conclusions follow in Sections 6 and 7.

\section{Previous observations and analysis}

The majority of the \textit{Spitzer} sources presented in this paper are a subset of the sample presented in \citet{Oliveira2013}. They used the 
S$^{3}$MC catalogue \citep{Bolatto2007} and colour and flux cuts to select YSO candidates
yielding 31 YSO candidates. Also added to the sample of \citet{Oliveira2013} were three
YSOs identified serendipitously by \citet{vanLoon2008} providing a total sample of 34 YSO candidates, all but one of which were spectroscopically confirmed as YSOs based
on properties of \textit{Spitzer} InfraRed Spectrograph (IRS) spectra.
\citet{Oliveira2013} also presented ancillary optical spectra which we perform additional analysis for in Section 5.
 
A simplified and truncated version of Table 3 from \citet{Oliveira2013} is given in Table 1 of this paper, showing only those \textit{Spitzer} YSOs 
for which SINFONI data were obtained. This table includes identified emission and absorption features based on \textit{Spitzer} and ground-based data including ice absorption
features as well as types based on the classification systems applied to LMC YSOs of \citet{Seale2009} and \citet{Woods2011}. 
The S, P and O types by \citet{Seale2009} indicate spectra dominated by silicate absorption, Polycyclic Aromatic Hydrocarbon 
(PAH) emission and silicate emission respectively, whilst an E following the primary classification 
indicates the presence of fine structure emission.
Note that the \citet{Seale2009} classification scheme does not consider the presence of ices, which are indicative of shielded colder regions
and are thus expected to be found in early stage YSOs.
The spectral types for YSOs from \citet{Woods2011}\footnote{The G1, G2, G3 and G4 notation used in this paper and in \citet{Oliveira2013}
is equivalent to the YSO-1, YSO-2, YSO-3 and YSO-4 classes in \citet{Woods2011}} are also based on the spectral features detected; ice absorption (G1), silicate
absorption (G2), PAH emission (G3) and silicate emission (G4). 
Also included in Table 1 are the bolometric luminosities 
of the \textit{Spitzer} sources obtained from SED fitting and whether or not radio emission is detected towards the source.
Optical types are also given in Table 1 based
on the identification of emission lines in the optical spectra presented in \citet{Oliveira2013};
type evolution I to V represents an increase in the number of 
high excitation energy transitions detected.

Point source catalogues \citep{Wong2011b, Wong2012} and high-resolution radio 20, 6 and 3 cm continuum images \citep{Wong2011a, Crawford2011} of 
the SMC have shown that 11 of the 33 YSOs in \citet{Oliveira2013} exhibit radio continuum emission. Radio free--free emission is commonly detected towards
UCH{\sc ii}s (e.g. \citealt{Hoare2007}) which are formed in the later stages of massive star formation.
 Of these 11 YSOs associated with radio emission, 7
are discussed in this work.

We have selected the subset of 17 spectroscopically confirmed YSOs from \citet{Oliveira2013} to form a sample of reasonable size
which includes a wide variety of \textit{Spitzer} classifications \citep{Seale2009,Woods2011} in order to create a relatively unbiased sample.
In addition to the 17 confirmed YSOs, we include in the sample two sources within the compact H\,{\sc ii} regions 
LHA 115-N81 and LHA 115-N88\,A (\citealt{Henize1956}; assigned the numbers 36 and 35, respectively in Table 1). Both of these regions have been previously studied and are
known to be associated with young massive main sequence stars and massive star formation. The locations of all of the sources discussed in \citet{Oliveira2013}
with the addition of sources \#35 and 36 are marked on an 8.0\,$\mu$m mosaic of the SMC in Fig. 1.

N81, located in the
Wing of the SMC was found by \citet{Koornneef1985} to be relatively free of dust and excited primarily by a 
central bright star of spectral type O6. \citet{Heydari-Malayeri1988} characterised N81 as a young, compact H\,{\sc ii} \textquotedblleft blob\textquotedblright with an 
extinction of $A_{V} =1.4$\,mag.
N81 has been studied in the Far-UV \citep{Heydari-Malayeri2002}, confirming the presence of very young O-type stars in the H\,{\sc ii} region 
and with \textit{JHK} imaging \citep{Heydari-Malayeri2003} resolving the central region into two sources 
(which we now resolve into five individual components; Section 4.1).

Since its identification by \citet{Testor1985} as a high-excitation compact H\,{\sc ii} region, N88\,A has become one of the best studied objects in the SMC and has been 
the subject of numerous imaging and spectroscopic studies 
\citep{Roche1987, Kurt1999, Heydari-Malayeri1999, Testor2003, Testor2005, Testor2010, Martin-Hernandez2008}.
N88\,A is a compact H\,{\sc ii} region ionized by four central stars with an H$_2$ emission shell approximately 3 arcsec in diameter. 
All four of these sources exhibit near-IR colours consistent with massive YSOs, but could also be consistent with reddened 
massive main sequence stars \citep{Testor2010}. 
N88\,A appears to be an exceptionally bright high-excitation blob in the SMC and would rival many of the H\,{\sc ii} regions in the LMC in this respect
 \citep{Charmandaris2008}.

\begin{table*}
 \begin{minipage}{175mm}
\caption{Previously known properties of YSOs in the SMC from \citet{Oliveira2013}, truncated to show only the YSOs observed with SINFONI.
The presence or absence of a feature is indicated by \cmark and \xmark, ? indicating doubt. 
Silicate emission is identified with $\wedge$. For H$_2$ emission, \cmark ? indicates objects
for which only two emission lines were detected in the \textit{Spitzer}-IRS range, rather than three to five. 
Also shown are the detections of either H$_2$O or CO$_2$ ice features at 3\,$\mu$m and 15.2\,$\mu$m, respectively.
The optical ionization classes are defined according to the emission lines present in the spectrum: Type I objects exhibit Balmer, Paschen and O\,{\sc i},
Type II objects show hydrogen, O\,{\sc i}, [N\,{\sc ii}], [O\,{\sc ii}] and [S\,{\sc ii}] emission, Type III objects show the same lines plus [O\,{\sc iii}], 
Type IV objects further add [S\,{\sc iii}] and
finally Type V objects show all these lines plus He\,{\sc i} emission. Some objects exhibit a stellar absorption spectrum and others only H$\alpha$ in emission (see \citealt{Oliveira2013}
for discussion). The following column indicates whether the object has been detected
at radio wavelengths; $^{\ast}$ signals an extended source. The sources are classified using features in their IRS spectrum, according to two classification
schemes previously applied to samples of LMC YSOs \citep{Seale2009, Woods2011}. The last column provides the luminosities determined from the SED fits.
See Section 2 for further information.}
\begin{center}
 \begin{tabular}{c c c c c c c c c c}
\hline
\# & PAH & Silicate & H$_2$ & H$_2$O/CO$_2$ ice & Optical & Radio & \multicolumn{2}{c}{YSO class.} & \textit{L} \\
   & emission &     & emission  & 3\,$\mu$m / 15.2\,$\mu$m & type & source & S09 & W11 & (10$^{3}$ L$_{\sun}$) \\
\hline
01 & \cmark & \xmark & \cmark & \xmark & IV/V & Y & PE & G3 & 16 \\
02 & {\tiny \checkmark} & \cmark & \cmark & \cmark & Only H$\alpha$ emission & Y & S & G1 & 19 \\
03 & {\tiny \checkmark} & \cmark & \cmark ? & \cmark & V & Y & S & G1 & 61 \\
04 & \cmark & \xmark & \cmark & \xmark & II & N & P & G3 & 2.3 \\
06 & \cmark & \cmark & \cmark & \cmark & Absorption lines & N & S & G1 & 5.8 \\
17 & {\tiny \checkmark} & \cmark & \cmark ? & \cmark & Only H$\alpha$ emission & N & S & G1 & 22 \\
18 & \xmark & \cmark & \xmark & \cmark & Only H$\alpha$ emission & N & S & G1 & 28 \\
20 & {\tiny \checkmark} & $\wedge$ & \cmark ? & \xmark & I & N & O & G4 & 1.5 \\
22 & \cmark & \xmark & \cmark & \cmark & IV/V & N & PE & G1 & 9.1 \\
25 & \cmark & \xmark & \cmark & \xmark & IV & Y & P & G3 & 17 \\
26 & \cmark & \xmark & \cmark & \xmark & V & Y & PE & G3 & 12 \\
28 & {\tiny \checkmark} & \cmark & \cmark ? & \xmark & IV/V & Y & S & G2 & 140 \\
30 & \cmark & \cmark ? & \cmark & \cmark & I & N & P & G1 & 7.9 \\
31 & \cmark & \xmark & \cmark ? & \xmark & No spectrum & Y$^{\ast}$ & PE & G3 & 6.7 \\
32 & {\tiny \checkmark} & \cmark & \cmark & \cmark & I/II & N & S & G1 & 3.5 \\
33 & {\tiny \checkmark} & \cmark & \cmark ? & \xmark & I/II & N & S & G2 & 26 \\
34 & {\tiny \checkmark} & \cmark & \cmark ? & \cmark & Only H$\alpha$ emission & N & S & G1 & 23 \\
\hline
 \end{tabular}
\end{center}
 \end{minipage}
\end{table*}

Masers are one of the key phenomena associated with massive star formation; however, relatively few masers have been discovered in the SMC.
Of particular interest are water masers, associated with outflows as observed in the earliest stage YSOs, and OH masers which are strongly 
associated with H\,{\sc ii} regions. Methanol masers also are one of the key tracers of massive star forming regions \citep{Ellingsen2006}.
To date six interstellar water masers have been detected towards the SMC \citep{Scalise1982, Oliveira2006, Breen2013}, only one of which lies close enough to 
one of the sources from \citet{Oliveira2013}, source \#03, to possibly be associated with it.
To date, no OH or methanol masers have been detected in the SMC \citep{vanLoon2012}.

\section{Observations and Data Reduction}

\subsection{SINFONI \textit{K--}band spectroscopy}

\textit{K}-band integral field spectroscopic observations were carried out for 19 \textit{Spitzer} YSO targets in the SMC using SINFONI
at the VLT under program 092.C-0723(A) (PI: Joana Oliveira). Each object was observed with a minimum of four 300 second integrations along with sky offset position observations in 
an ABBA pattern with jittering. Telluric B-type standard stars were also observed at regular intervals throughout each night in order to provide standard star
spectra for telluric correction and flux calibration. Dark, flat lamp, arc lamp and fibre calibration frames were observed during the daytime and 
linearity lamp frames were obtained from the ESO archive.
Observations were carried out with a 0.1$\times$0.05 arcsec spatial scale, providing a 3.2$\times$3.2 arcsec field-of-view (FOV) for the spectral range 1.95--2.45\,$\mu$m
and with a spectral resolving power of $\lambda / \Delta\lambda = 4000$, 
yielding a velocity resolution of $\sim$70\,km\,s$^{-1}$ (and therefore allowing the determination of Gaussian centroid positions to $<$4\,km\,s$^{-1}$).
Where possible the adaptive optics (AO) module was used. For sources \#17, 18 and 22 we were not able to use AO, yielding a seeing limited
spatial resolution for these sources (useful morphological information was still obtained).

The raw data were reduced using the standard SINFONI pipeline recipes in \textsc{gasgano}. 
Telluric and flux calibration were performed simultaneously for each cube using an \textsc{idl} script. For each pixel in the target cube, the target spectrum is
 divided through by the telluric spectrum removing the telluric spectral features. 
The target spectrum was then multiplied by a blackbody with a temperature appropriate for the spectral type of the standard star used. The blackbody spectra used in 
this calibration were generated using \textsc{pyraf}\footnote[3]{\textsc{pyraf} is a product of the Space Telescope Science Institute, which is operated by AURA for NASA.}. This process was looped to apply the same procedure to each spatial pixel (spaxel) in the cube.

SINFONI is a Cassegrain focus mounted instrument and as such it does suffer from a systematic
time-dependent wavelength shift during each night due to flexure. This is always small 
(less than 3 resolution elements)
so it only presents an issue when determining accurate absolute centroid velocity measurements. In order to 
account for this effect, a second wavelength calibration was performed on the final data cubes 
using the OH emission lines in the sky data cubes produced in the SINFONI pipeline.

Although sky line subtraction does form part of the standard SINFONI data reduction pipeline, 
due to the relatively long observing times in this study (8 $\times$ 5 minute exposures plus 
overheads), the 
variation in sky line intensities leads to sky line residuals remaining in the final data cubes.
The positions of these lines are shown in Ward et al. (2016; Fig. B2). Whilst aesthetically displeasing, the impact of
these residuals on the spectral analysis is actually small as none are coincident with any
emission lines of interest and the continuum measurements are calculated from models fitted to the 
continuum so noise and residuals are not an issue.

For source \#25 the data reduction sequence has yielded a data cube which exhibits negative continuum flux towards the red end of the spectrum. 
On inspection of
the sky cube this appears to be caused by an unusually strong, red continuum in the sky frames most likely caused by contamination from a
red continuum source nearby (see Fig. C1 in \citealt{Oliveira2013}). Whilst this has made continuum flux measurements of source \#25 unusable, it is unlikely to have affected 
the measurements of emission lines towards the source.

Following the data reduction we applied a process of re-sampling, Butterworth spatial filtering and instrumental fingerprint removal 
following the procedure outlined in \citet{Menezes2014} and \citet{Menezes2015}. 
First re-sampling was carried out with a 2$\times$2 rebin and a cubic spline interpolation. This introduces a high spatial-frequency component which
can be removed using Butterworth Spatial Filtering (BSF; \citealt{Gonzalez2002}) through the following steps.
\begin{itemize}
 \item Calculation of the Fourier transform of the image, $F(u,\nu)$.
 \item Multiplication of the Fourier transform by the corresponding Butterworth filter, $H(u,\nu)$.
 \item Calculation of the inverse Fourier transform of the product $F(u,\nu) \cdot H(u,\nu)$.
 \item Extraction of the real part of the calculated inverse Fourier transform.
\end{itemize}
\citet{Menezes2015} determined that a squared circular filter is most appropriate for \textit{K}-band SINFONI data at the 100 mas pixel scale;
\begin{equation}
 H(u,\nu) = \left\{1+\left[\sqrt{\left(\frac{u-u_{0}}{a}\right)^{2}+\left(\frac{\nu-\nu_{0}}{b}\right)^{2}}\right]^{2n} \right\} ^{-2}
\end{equation}
where n$=$2 and a$=$b$=$0.26 N$_y$ (where N$_y$ is the number of spaxels in the vertical direction).

The final stage of our truncated version of the data treatment procedure of \citet{Menezes2015} is instrumental fingerprint removal
using Principal Component Analysis (PCA) as summarised in the following steps.
\begin{itemize}
 \item All spectral lines are fitted and removed from the input data cube using Gaussian line profiles. The average pixel value for each slice, $Q_{\lambda}$ is then subtracted from every pixel
in that slice.
 \item The 3D datacube, $D_{0}$ is converted into a two dimensional array, I; $I[\beta,\lambda] = D_{0}[i,j,\lambda]$ where $\beta_{i,j}=\mu(i-1)+j$ 
and $\mu$ is equal to the number of elements in the first dimension.
 \item Application of PCA tomography to the 2D matrix as outlined in \citet{Steiner2009}. This yields the 2D tomogram matrix, $T$, and the 
eigenvector matrix, $E$.
 \item Selection of the eigenvectors related to the instrumental fingerprint. This is done by eye on the basis that the spectral signature will be 
that of the fingerprint and that the corresponding tomogram must have a spatial morphology including a large horizontal stripe at the bottom of the image
(see \citealt{Menezes2015} for details).
 \item Reconstruction of the datacube, excluding the eigenvector and tomogram related to the instrumental fingerprint; 
$I_{out} = \sum_{k} T_{\beta,K} \cdot [E_{\lambda,K}]^{T}$ for $K \neq K_{inst}$.
 \item The 2D array is converted back into a data cube and the properties subtracted in the first step are now added back in; 
$D_{out} = I_{i,j,\lambda}$ $+$ emission lines $+$ $ Q_{\lambda}$.
\end{itemize}
The total flux is conserved through this process within a fraction of a percent.
Whilst this procedure does not improve the spatial resolution nor change the observed morphologies and kinematics, it does provide
cleaner intensity and velocity maps.

1D spectra were extracted using \textsc{gasgano} for each of the detected continuum sources in the final data cubes.
Emission lines detected in these spectra were fitted with a Gaussian profile using the Starlink spectral analysis package {\sc splat}. 
The measured emission line fluxes are listed in Table C1.

\subsection{RSS optical spectroscopy}

In addition to the DBS optical spectra first presented in \citet{Oliveira2013}, optical spectra have been obtained for 
sources \#35 (N88\,A) and 36 (N81) using the Robert Stobie Spectrograph (RSS; \citealt{Kobulnicky2003}) 
at the Southern African Large Telescope (SALT; \citealt{Buckley2006}) under program 2014-1-UKSC-003 (PI: Jacob Ward) and these are also presented here 
(shown in Figs. B2, B3 and B4).
The optical spectra for N81 and N88\,A were obtained using both the pg900 grating centred at 6318.9 \AA{} to obtain a broad band low resolution spectrum for 
each source and the pg1800 grating centred on the positions of H$\alpha$ and H$\beta$ to obtain medium resolution spectra for these regions.
A slit width of 1 arcsec was used for all RSS spectroscopy presented here whilst the slit width used to obtain the DBS spectra was 2 arcsec.

The RSS spectra were reduced using the data reduction package {\sc figaro} for flat field correction, extraction and wavelength calibration.
Flux calibration was carried out using standard stars observed with RSS within one month of the observations.

\section{Results}

\subsection{Continuum emission and photometry}

For each spaxel in the final flux calibrated cubes, the continuum was fitted and summed for the spectral region spanning 2.028--2.290\,$\mu$m to produce
continuum flux maps without any contribution from line emission. 
The J2000 RA and Dec positions of each \textit{K}-band continuum source resolved in this work are given in Table 2.
The positions of each continuum source are marked in Fig. A1, the green circles 
show the regions from which 1D spectra were extracted from the cubes.
Values for the \textit{K}-band apparent magnitude (using the CIT photometric system where the \textit{K-}band zero point is at 620 Jy) of each of the sources were determined by fitting a polynomial to the continuum
of the 1D spectrum for each source and summing the total flux of the fitted continuum (Table 2).

\begin{table}
  \caption{Positions of each of the \textit{K}-band continuum sources as determined from the SINFONI continuum images, and
the \textit{K}-band apparent magnitude for each source with the visual extinction value determined from the ratio between
the 1-0Q(3) and 1-0S(1) H$_2$ emission lines.}
\scriptsize
\begin{tabular}{c c c c c}
\hline
& \multicolumn{2}{c}{J2000} & & \\
Source & RA (h:m:s) & Dec (\textdegree:$^{\prime}$:$^{\prime\prime}$) & \textit{K} (mag) & $A_V$ (mag) \\
\hline
01 & 00:43:12.885 & --72:59:58.18 & 16.1	$\pm$	0.01	&	2.6	$\pm$	10.9	\\
02\,A & 00:44:51.878 & --72:57:33.81 & 15.14	$\pm$	0.01	&	9.3	$\pm$	8.7	\\
02\,B &  00:44:52.094 & --72:57:34.06 & 15.75	$\pm$	0.01	&	4.3	$\pm$	24.4	\\
03 & 00:44:56.571 & --73:10:14.37 & 13.51	$\pm$	0.01	&	13.7	$\pm$	10.8	\\
04 & 00:45:21.283 & --73:12:18.39 & 14.52	$\pm$	0.01	&				\\
06 & 00:46:24.348 & --73:22:06.47 &  17.51	$\pm$	0.02	&	7.2	$\pm$	6.4	\\
17 & 00:54:02.269 & --73:21:19.02 & 15.87	$\pm$	0.01	&	47.5	$\pm$	10.0	\\
18 & 00:54:03.277 &--73:19:39.12 & 15.88	$\pm$	0.01	&	18.7	$\pm$	10.1	\\
20 & 00:56:06.411 & --72:28:27.62 & 15.96	$\pm$	0.01	&	24.1	$\pm$	13.2	\\
22\,A &	00:57:56.959 & --72:39:16.14 & 15.96	$\pm$	0.01	&	9.9	$\pm$	6.4	\\
22\,B &	00:57:57.182 & --72:39:15.54 & 17.6	$\pm$	0.01	&	22.5	$\pm$	16.6	\\
25& 01:01:31.678 & --71:50:38.48 & & $<$12.0	\\
26 & 01:02:48.441 & --71:53:17.15 & 16.68	$\pm$	0.03	&	14.2	$\pm$	16.6	\\
28\,A& 01:05:07.229 & --71:59:42.54 & 14.98	$\pm$	0.01	&	19.5	$\pm$	9.3	\\
28\,B& 01:05:07.315 & --71:59:41.85 & 17.63	$\pm$	0.01	&	25.2	$\pm$	12.5	\\
30& 01:06:59.656 & --72:50:42.82 & 14.22	$\pm$	0.01	&	13$\substack{+	55	 \\ -	25	}$ \\
31& 01:14:39.284 & --73:18:28.21 & 15.08	$\pm$	0.02	&	1.5	$\pm$	7.9	\\
32& 00:48:39.662 & --73:25:00.69 & 14.66	$\pm$	0.01	&	22.2	$\pm$	22.1	\\
33& 01:05:30.015 & --72:49:51.90 & 11.5	$\pm$	0.01	&	$<$5.0	\\
34& 01:05:49.349 & --71:59:48.59 & 14.12	$\pm$	0.01	&	16.0	$\pm$	8.0	\\
35\,A& 01:24:07.982 & --73:09:03.81 & 13.47	$\pm$	0.04	&	1.2	$\pm$	17.0	\\
35\,B& 01:24:07.901 & --73:09:03.66 & 14.65	$\pm$	0.06	&	$<$22.5	\\
35\,C& 01:24:07.867 & --73:09:04.36 & 14.85	$\pm$	0.08	&	$<$10.1	\\
35\,D& 01:24:07.982 & --73:09:03.06 & 14.91	$\pm$	0.06	&	1.8$\substack{+	12.0	 \\ -	15.8	}$ \\
36\,A& 01:09:12.912 & --73:11:38.53 & 14.93	$\pm$	0.01	&	$<$32.5	\\
36\,B& 01:09:12.889 & --73:11:38.28 & 15.48	$\pm$	0.01	&	13.3	$\pm$	12.8	\\ 
36\,C& 01:09:13.212 & --73:11:38.68 & 16.08	$\pm$	0.01	&	21.6	$\pm$	32.5	\\
36\,D& 01:09:13.131 & --73:11:38.68 & 16.54	$\pm$	0.01	&	35.3	$\pm$	23.3	\\
36\,E& 01:09:12.693 & --73:11:38.58 & 16.56	$\pm$	0.02	&	19.9	$\pm$	21.1	\\
\hline
\end{tabular}
\end{table}

Out of 19 observed sources, 14 consist of a single continuum source at the resolution obtained with this study (0.1--0.2 arcsec with
AO and $\sim$0.6 arcsec seeing limited),
 whilst the remaining five targets consist of multiple continuum sources. Sources \#02, 22 and 28 are resolved into two components whilst sources
\#35 and 36 are resolved into four and five \textit{K}-band continuum components, respectively (see Table 2 \& Fig. A1).

\subsection{Extinction}

In order to calculate values of extinction towards our sources, we utilise the H$_2$ lines in the extracted spectra. 
The 1-0S(1) / 1-0Q(3) flux ratio ($I_{S1}$/$I_{Q3}$ in equation 2) is used
due to its insensitivity to temperature and relatively large wavelength separation (e.g., \citealt{Davis2011}).
Visual extinction is given by:
\begin{equation}
 A_{V} = -114 \times \log(0.704[I_{S1}/I_{Q3}])
\end{equation}
Extinction corrections were calculated for each emission line individually 
using the Galactic mean $R_V$ dependent extinction law:
\begin{equation}
 [A(\lambda)/A(V)] = a(x) + b(x)/R_{V}
\end{equation}
where $x = 1 / \lambda$, $a(x) = 0.574x^{1.61}$ and $b(x) = -0.527x^{1.61}$ for the {\it K}-band \citep{Cardelli1989}.
An $R_{V}$ value of 3.1 has been adopted although $R_V$ has been found to range from 2.05$\pm$0.17 to 3.30$\pm$0.38 
in the SMC \citep{Gordon2003}.

The calculated extinction values using this method are given in the final column of Table 2. We find that the mean extinction value 
 towards the SMC sources (excluding limits and null values) presented here is $A_V =$ 15.8$\pm$3.3 mag, 
significantly lower than the average towards 139 Galactic massive YSOs obtained from \textit{HK} photometry (45.7$\pm$1.5 mag; \citealt{Cooper2013}) 
and that of N113 in the LMC (22.3$\pm$3.8 mag; \citealt{Ward2016}). 
Furthermore, the median values for extinction are $A_V =$ 43.8, 20.0 and 14.2 mag for the Galactic sample, the N113 sample and the SMC sample, respectively.
This is consistent with the lower dust to gas ratio in the SMC and suggests that extinction scales with metallicity. It should be noted that 
N113 may not be representative of the LMC (see \citealt{Ward2016}).

\subsection{\textit{K}-band emission features}

\subsubsection{H\,{\sc i} emission}

Br$\gamma$ emission (2.166\,$\mu$m) is most commonly associated with accretion in star formation studies. For intermediate mass YSOs the relation from \citet{Calvet2004}
can be used to estimate the accretion luminosity from Br$\gamma$ luminosity:
\begin{equation}
 \log(L_{\text{acc}}) = -0.7+0.9(\log(L_{\text{Br}\gamma})+4)
\end{equation}
It is possible that this relation breaks down for massive stars due to an additional emission component originating from the strong stellar 
winds associated with hot stars. 
Furthermore, Herbig Be stars do not appear to be modelled successfully through magnetospheric accretion,
suggesting a change of accretion mechanism at high masses (e.g., \citealt{Fairlamb2015}). 
For the purposes of comparing our sample with a Galactic sample, however, the above relation can be applied to gain an
equivalent accretion luminosity assuming that both samples cover the same range of evolutionary states and YSO masses. The models of \citet{Kudritzki2002} make the 
prediction that the effect of metallicity on the production of photons capable of ionizing hydrogen is almost negligible
due to the cancelling effect of metal line blocking and metal line blanketing,
 and thus the lower metallicity of the SMC should
not affect this relationship.
\begin{figure*}
\begin{minipage}{175mm}
\begin{center}
 \includegraphics[width=0.98\linewidth]{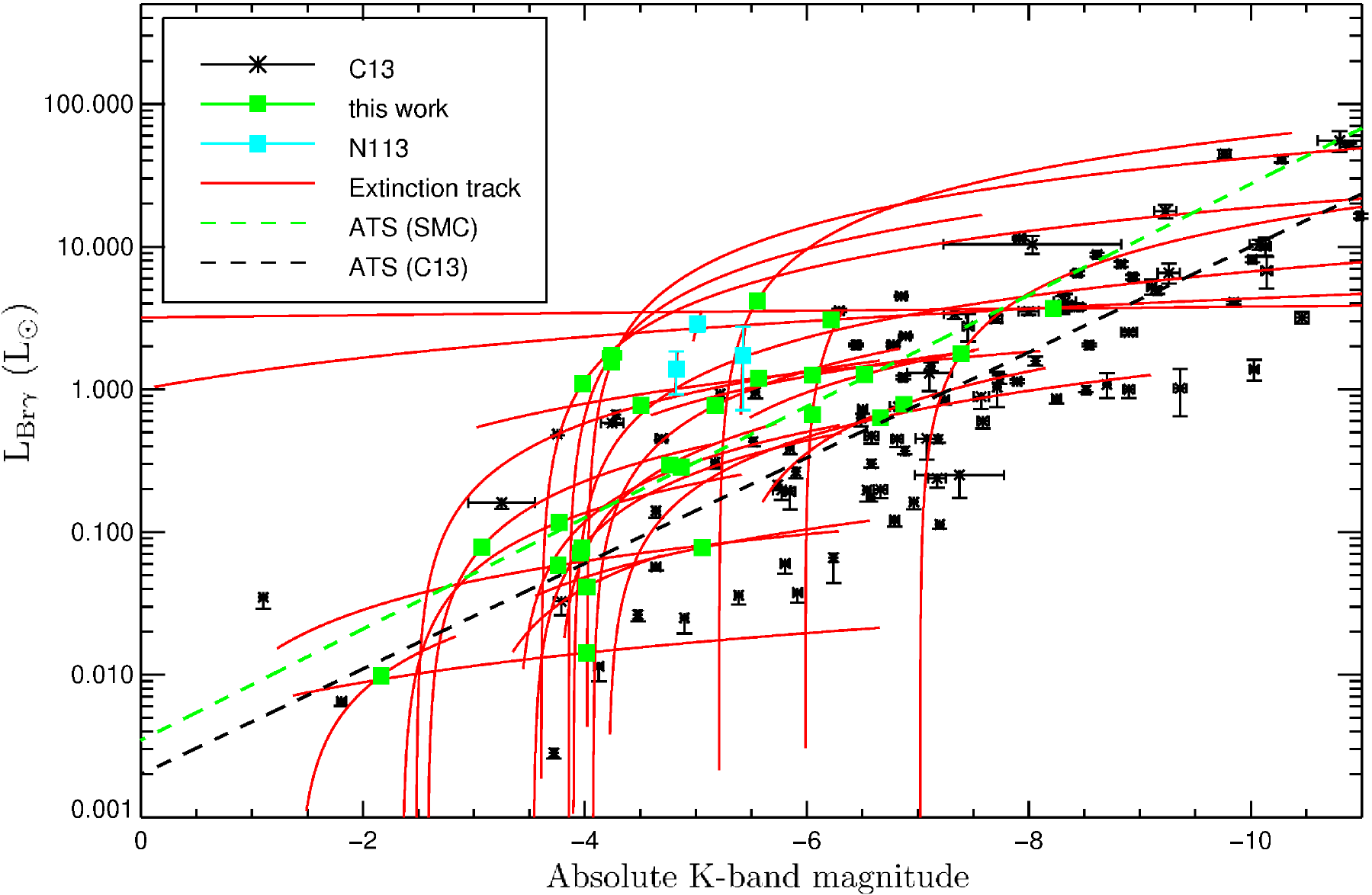}
\end{center}
\caption{Br$\gamma$ luminosity against absolute \textit{K}-band magnitude. Data from \citet{Cooper2013} and \citet{Ward2016} are also included for comparison, 
labelled as C13 and N113, respectively. The red extinction tracks are representative of the uncertainty due to the extinction correction (Table 2).
The green and black dashed lines indicate the ATS regressions fitted to the SMC and the \citet{Cooper2013} sources respectively.
}
\end{minipage}
\end{figure*}

Br$\gamma$ emission is detected towards 28 of the 29 SINFONI \textit{K-}band sources observed in this work, making it the most common spectral feature in the data set. Only the spectrum towards
source \#02\,B does not exhibit Br$\gamma$ emission.
Figure 2 shows the extinction corrected Br$\gamma$ luminosities plotted against absolute \textit{K--}band magnitude for all sources where Br$\gamma$ was detected.
Also included in this plot are the results of \citet{Cooper2013} and \citet{Ward2016} for YSOs in the Milky Way and N113 in the LMC, respectively,
along with extinction tracks which show the allowed positions of points in the diagram based on the uncertainty of the extinction correction (see Table 2).

Akritas-Theil-Sen (ATS; \citealt{Akritas1995}) regressions were fitted to the SMC sample and the Galactic sample.
The ATS regressions and Kendall's $\tau$ values were computed in the R-project statistics package using the \textsc{cenken} function 
from the \textsc{NADA} library \citep{Helsel2005}.
The ATS regression fit takes the form $y = 10^{ax + b}$; we compute the values of $a_{SMC} = -0.39$ and $b_{SMC} = -2.46$ for the SMC sample whilst the 
Galactic sample yielded values of $a_{MW} = -0.37$ and $b_{MW} = -2.69$.  
The associated Kendall's $\tau$  values are $\tau_{SMC} = -0.50$ and $\tau_{MW} = -0.70$ for the SMC and the \citet{Cooper2013}
samples respectively, indicating a stronger correlation for the Galactic sample than for the SMC sample.
The {\it p} values, which represent the likelihood of there being no correlation between $L_{\text{Br}\gamma}$ and \textit{K-}band magnitude, 
are 0.0004 and $<$1$\times$10$^{-8}$ for the SMC and Galactic samples respectively, indicative of a definite correlation for both sets of data.

Whilst consistent with the Galactic sample, our sample of SMC sources exhibit Br$\gamma$ emission luminosities that are high for their
absolute \textit{K}-band magnitudes compared with the \citet{Cooper2013} data.
Not accounting for any difference in the \textit{K}-band continuum emission or additional sources of excitation, 
assuming that Eqn. 4 holds true this suggests that YSOs in the SMC exhibit higher accretion rates than their counterparts in the Milky 
Way causing higher levels of Br$\gamma$ emission. This is discussed in detail in Section 6.1.

\begin{figure*}
 \begin{minipage}{175mm}
  \begin{center}
\includegraphics[width=0.32\linewidth]{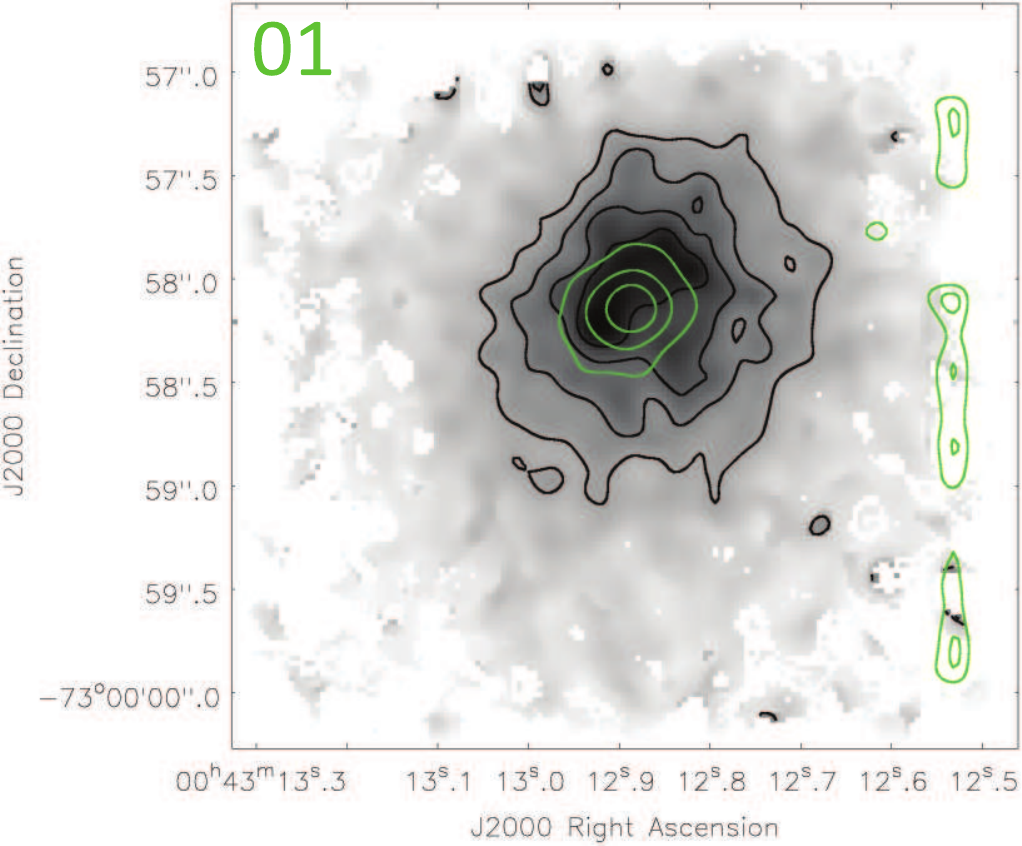}
\includegraphics[width=0.33\linewidth]{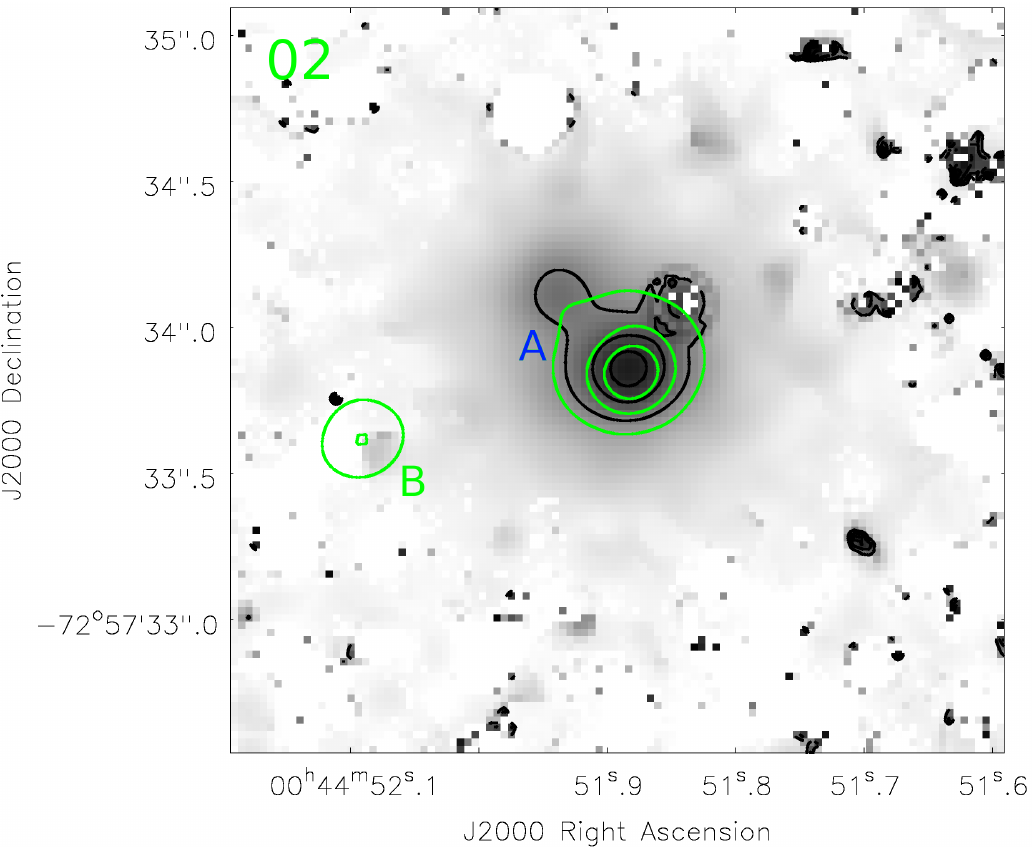}
\includegraphics[width=0.33\linewidth]{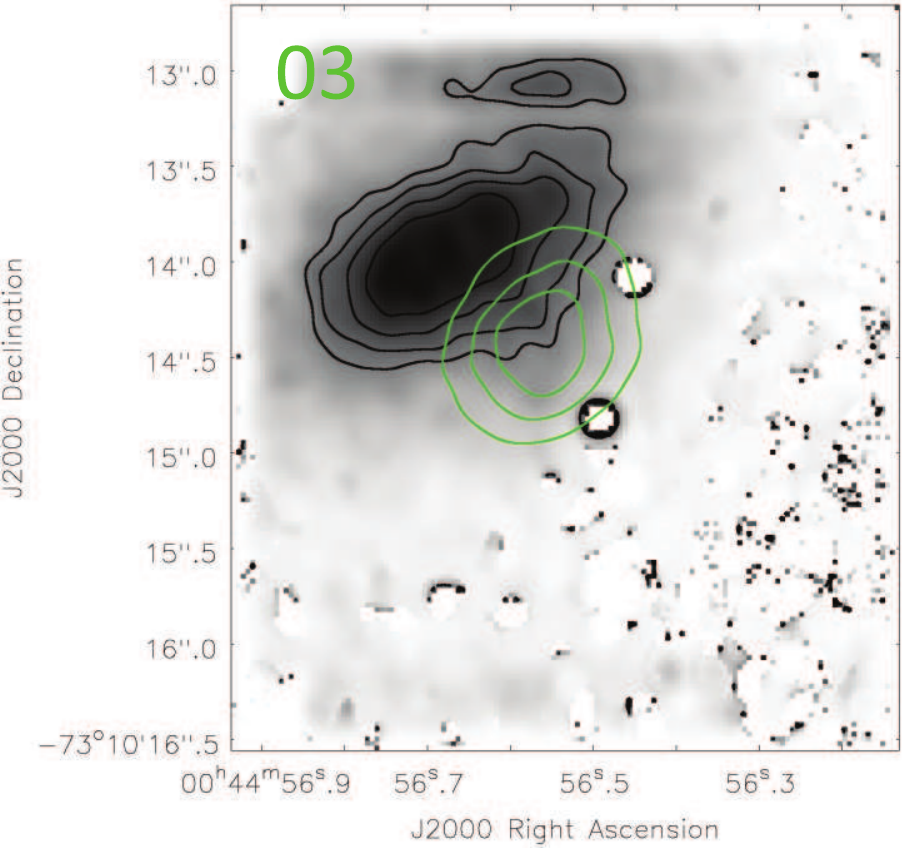}
\includegraphics[width=0.33\linewidth]{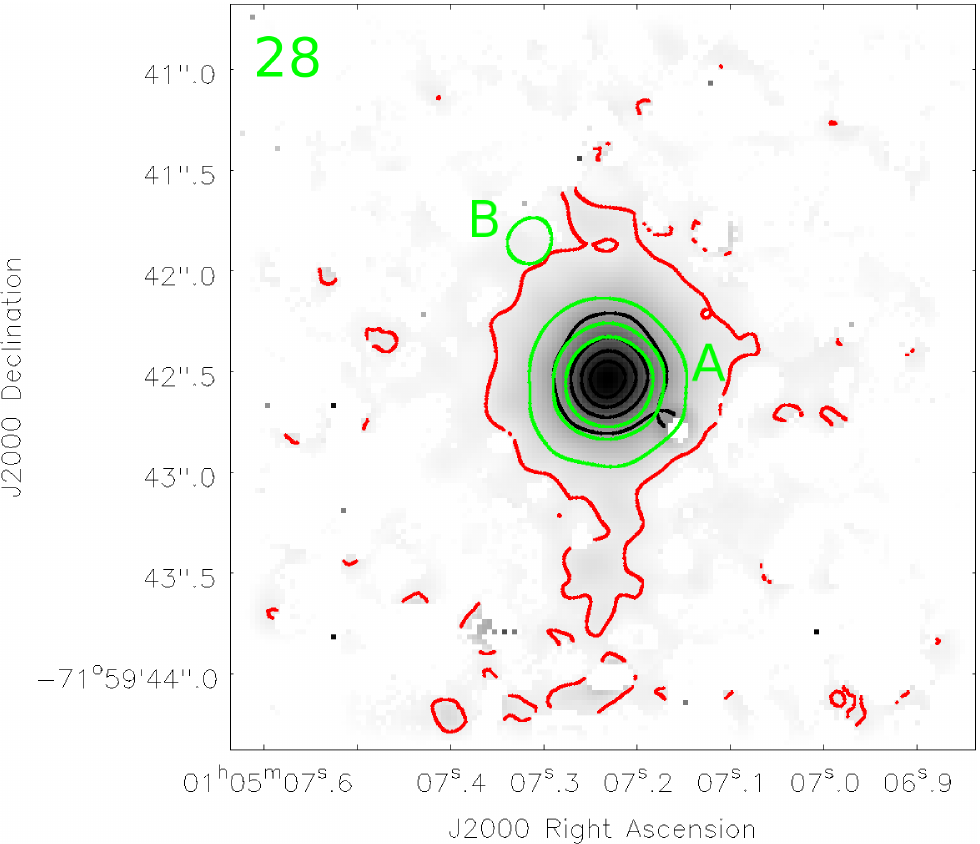}
\includegraphics[width=0.33\linewidth]{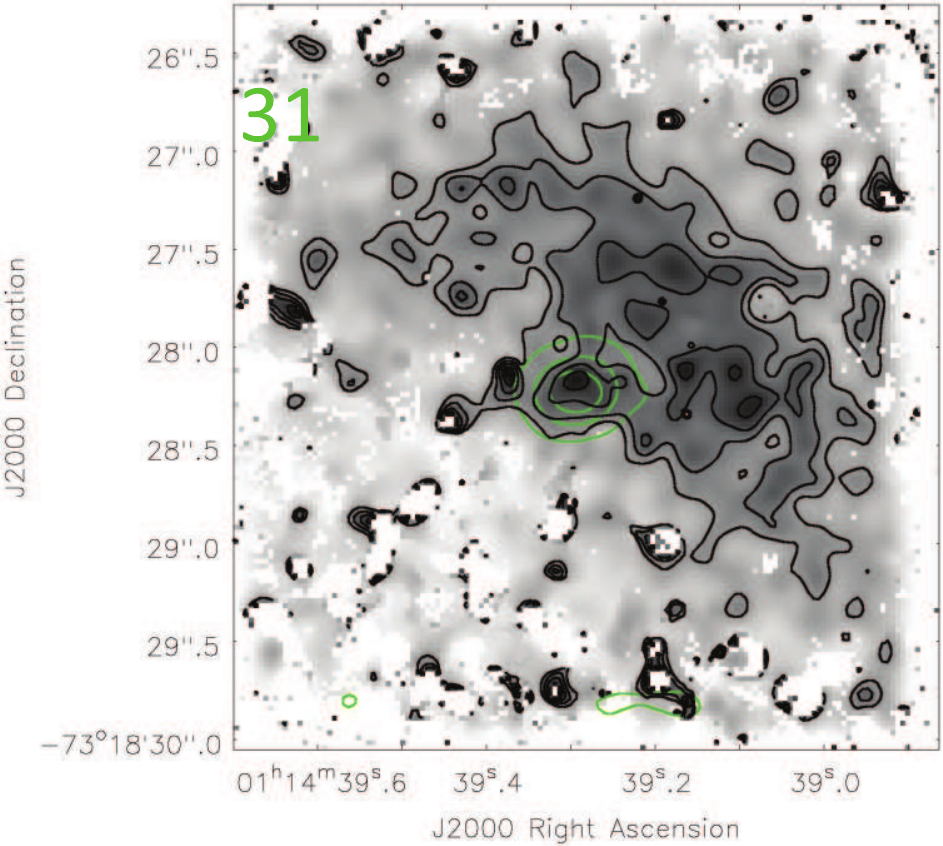}
\includegraphics[width=0.33\linewidth]{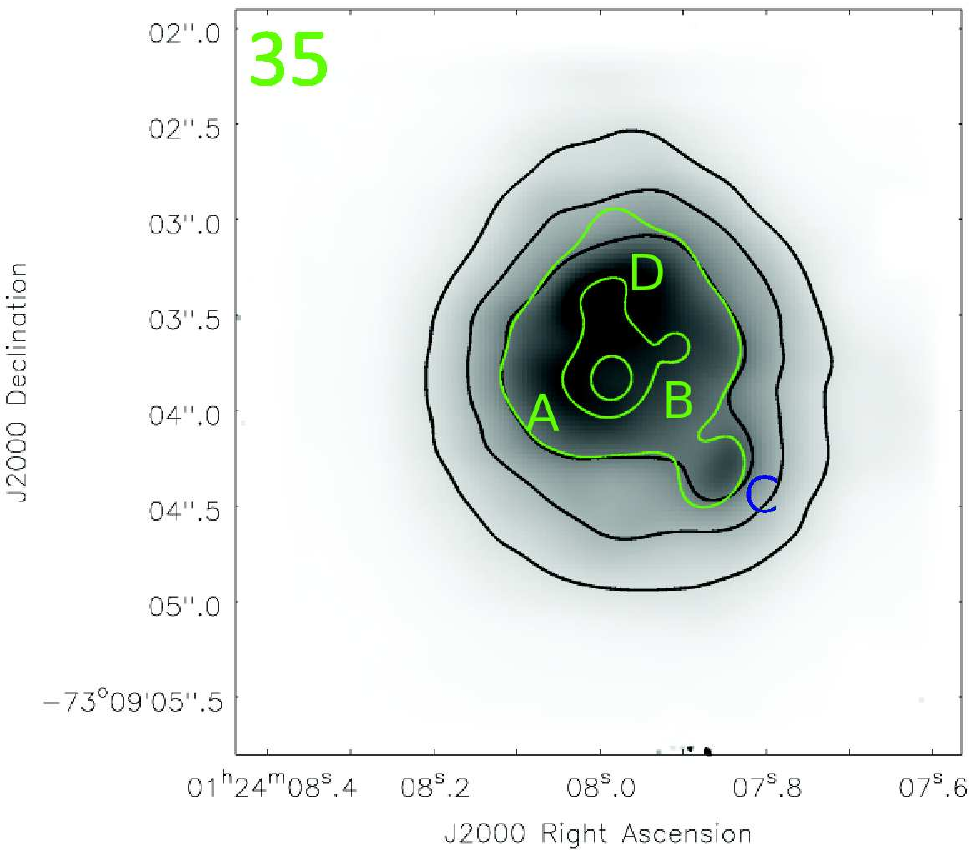}
\includegraphics[width=0.33\linewidth]{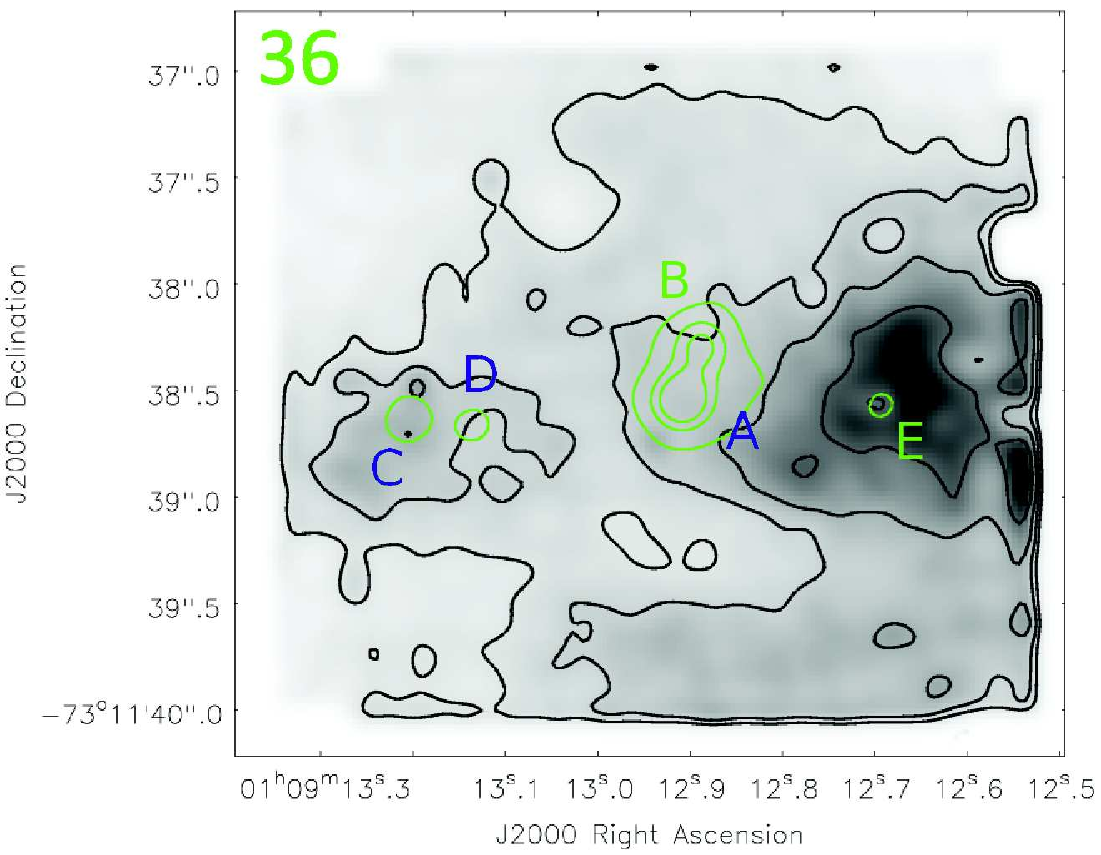}
   \caption{Extended Br$\gamma$ emission line morphologies for sources \#01, 02, 03, 28, 31, 35 (N88\,A) and 36 (N81). Black contours - [0.2,0.4,0.6,0.8] $\times$ maximum Br$\gamma$ 2.1661\,$\mu$m integrated flux, green contours - 
[0.25, 0.5, 0.75] $\times$ maximum continuum integrated flux.
The red contours in source \#28 indicate significantly weaker (4$\times$10$^{-22}$\,Wm$^{-2}$/spaxel), more extended Br$\gamma$ emission.}
  \end{center}

 \end{minipage}

\end{figure*}

In the majority of sources, the Br$\gamma$ emission is compact but in 7 of the 19 FOVs we see evidence of extended H\,{\sc i} emission 
as shown in Fig. 3.
In sources \#01, 28 and 35 (N88\,A) it extends roughly symmetrically from a central continuum source whilst in sources \#03 and 31 it is extended over a significant area and offset
to one side of a continuum source.
Source \#28 also exhibits weak elongated Br$\gamma$ emission extending to the south of the continuum source (Fig. 3, red contour).
 Source \#02 exhibits relatively compact and collimated extended Br$\gamma$ emission in one direction from the continuum source. 
Finally for source \#36 (N81), there appears to be relatively high levels of ambient Br$\gamma$ emission which is unsurprising given the 
nature of the N81 H\,{\sc ii} region (see Section 2). 
The extended Br$\gamma$ emission to the west most likely originates outside the FOV as the peak of the emission is at the edge of the observed region
 whilst that in the east appears to be associated with (or have a component associated with) source \#36\,C and possibly \#36\,D.

\begin{figure*}
 \begin{minipage}{175mm}
  \begin{center}
\includegraphics[width=0.48\linewidth]{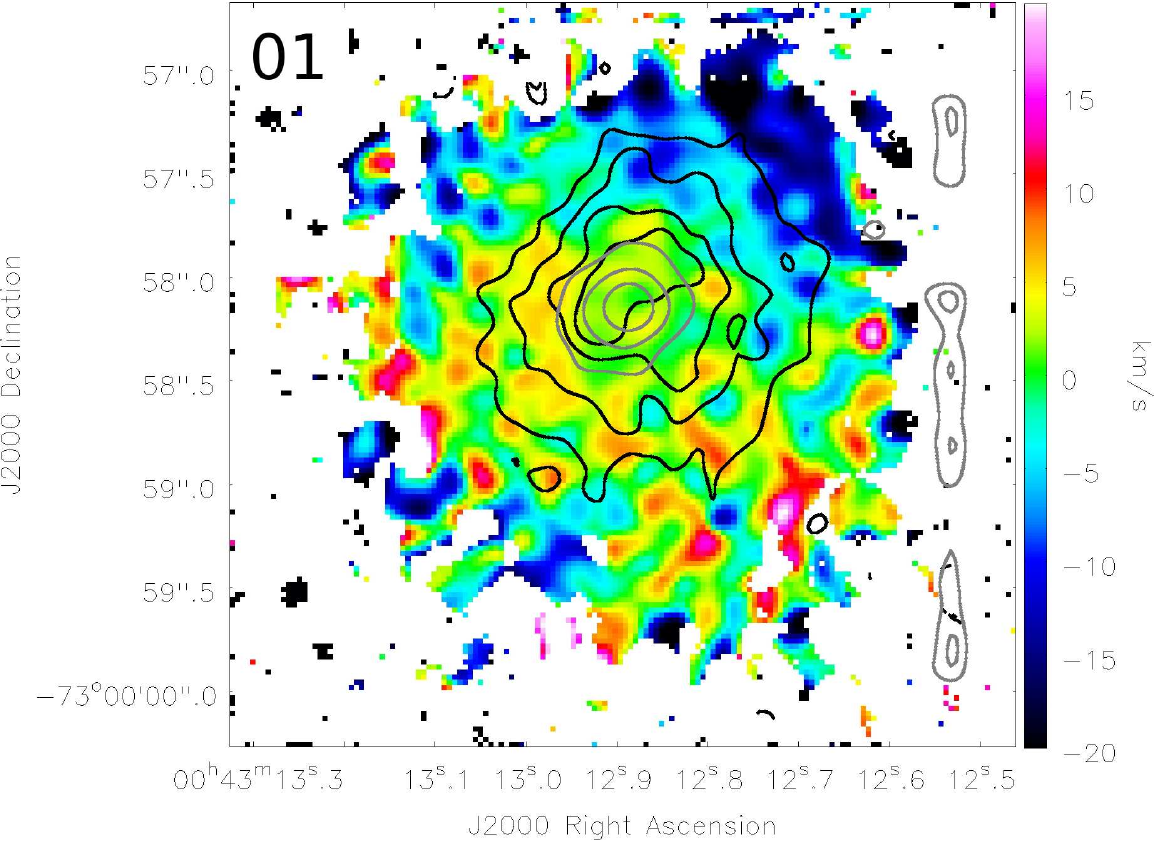}   
\includegraphics[width=0.48\linewidth]{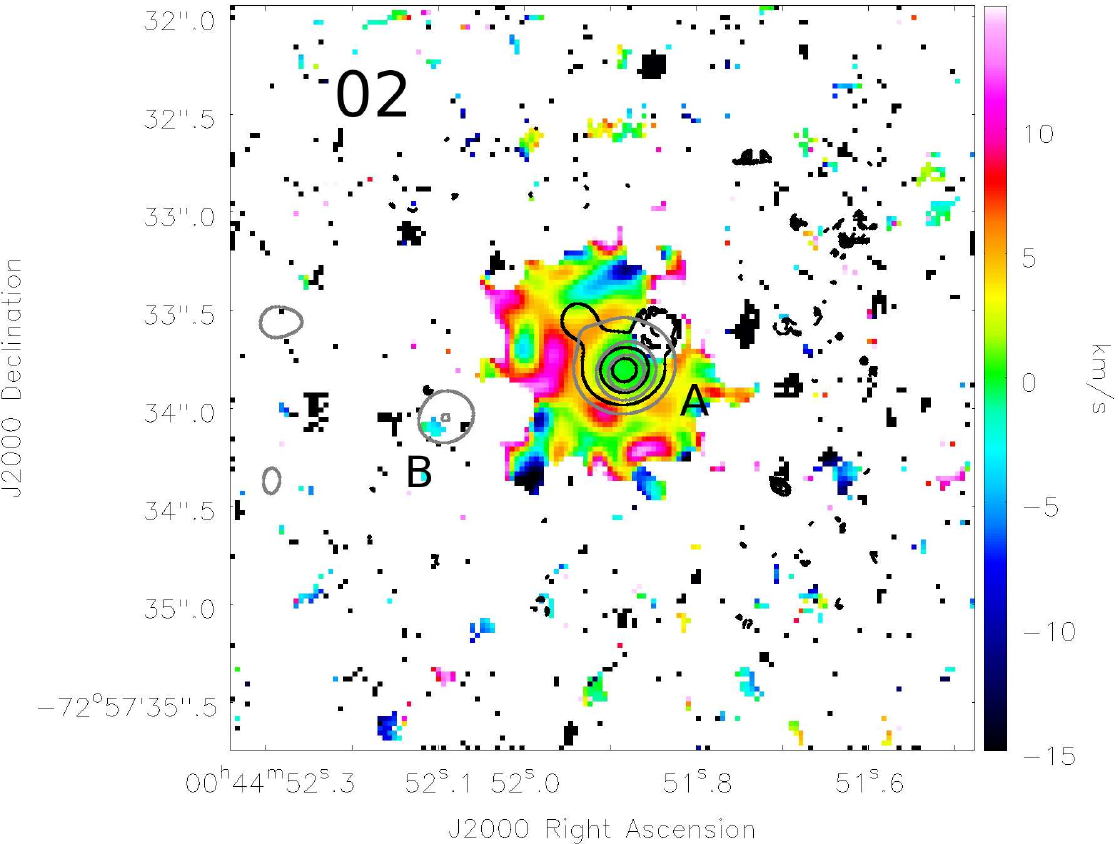}  
\includegraphics[width=0.48\linewidth]{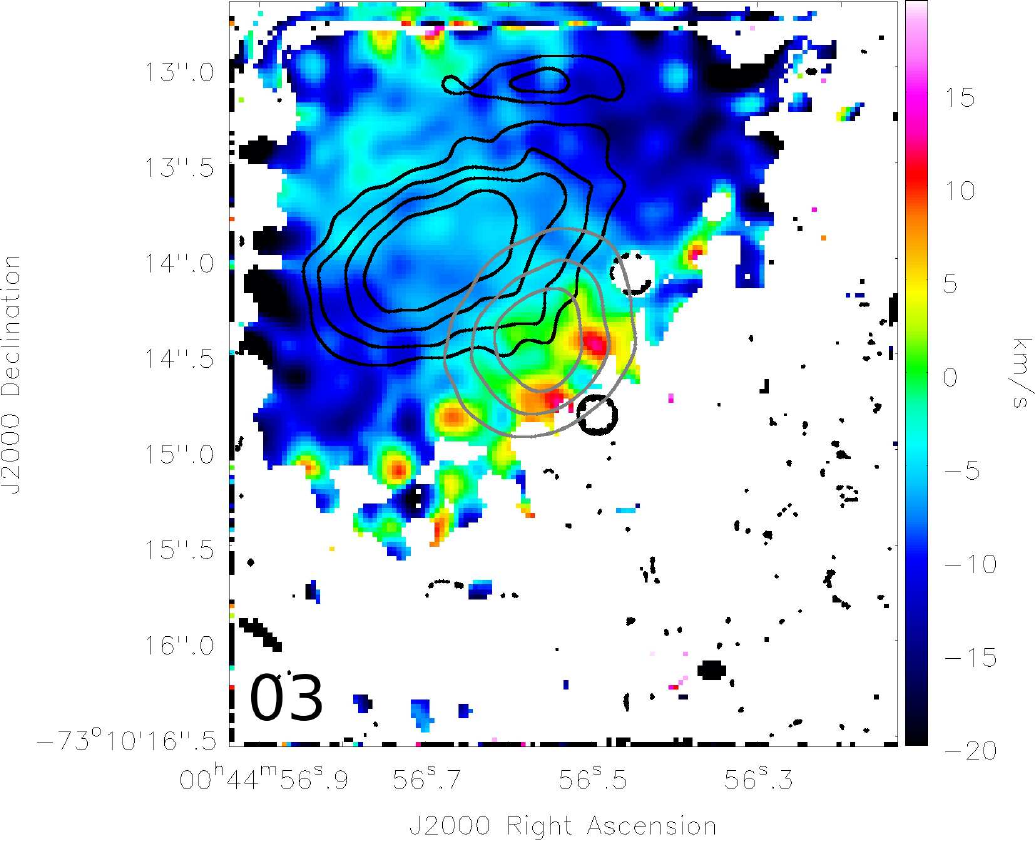}     
\includegraphics[width=0.48\linewidth]{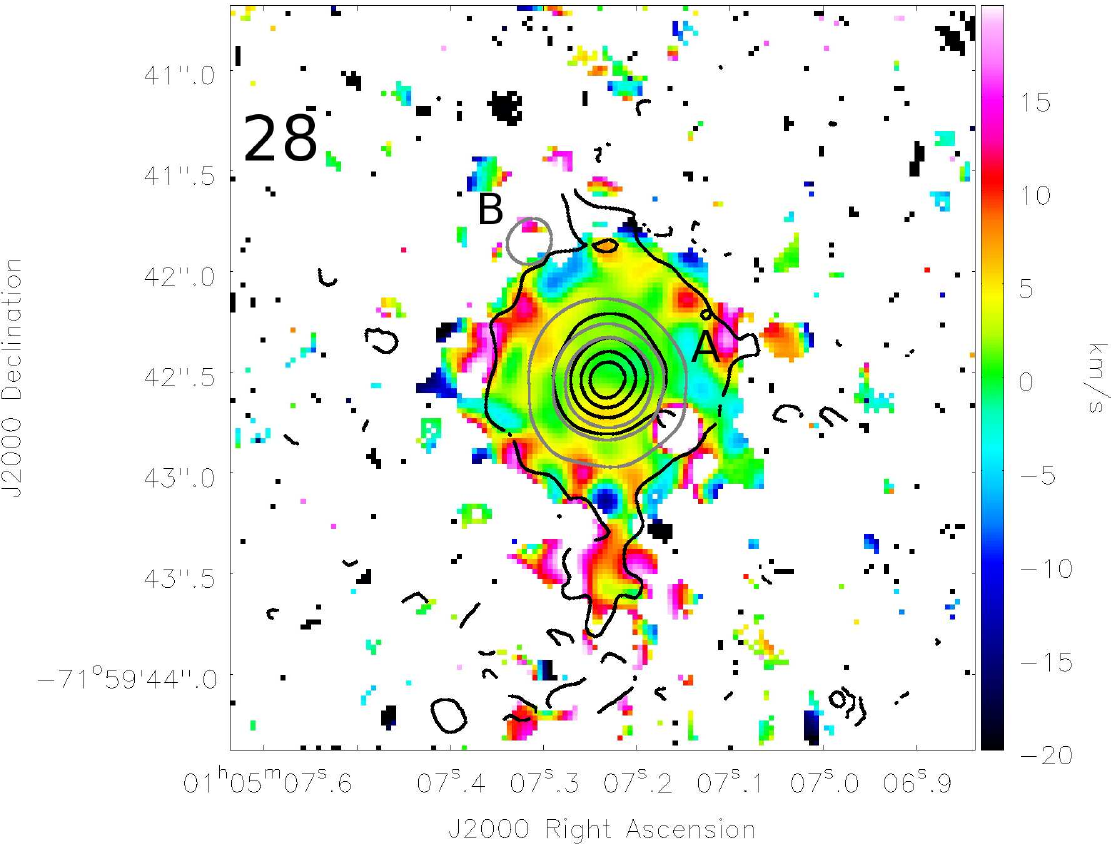}   
\includegraphics[width=0.48\linewidth]{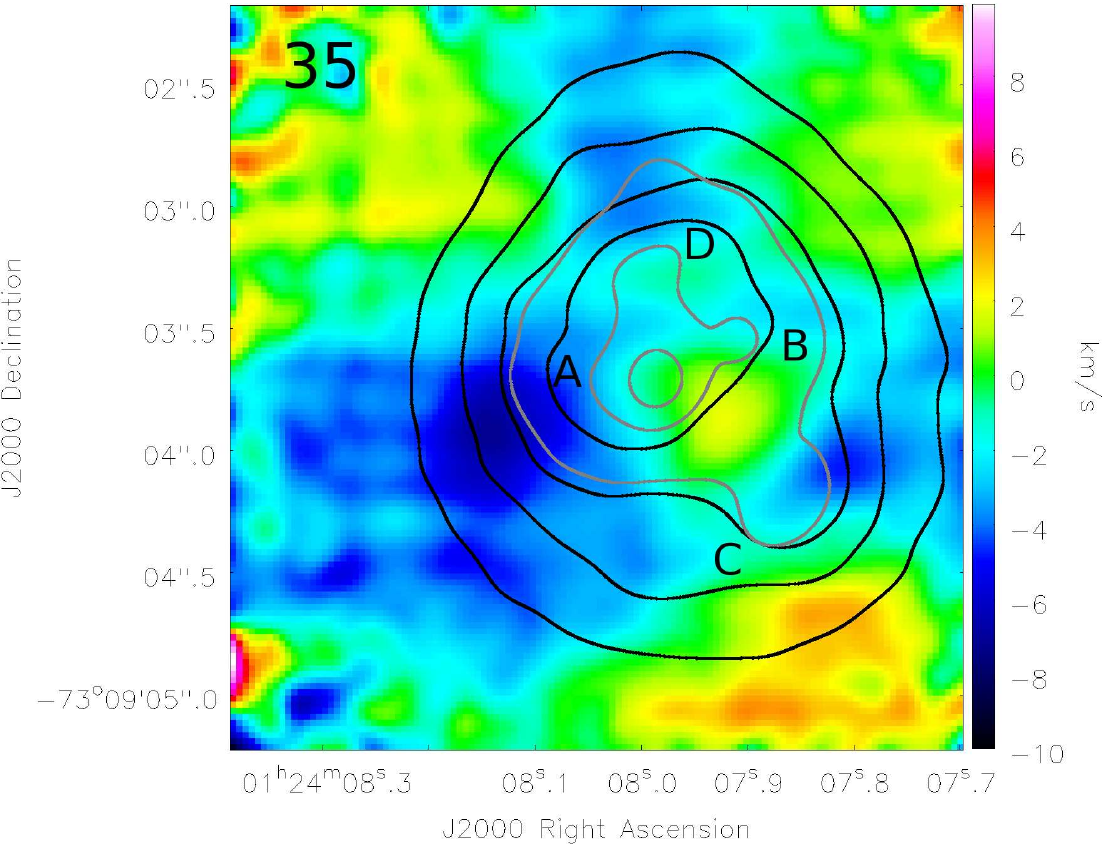}   
\includegraphics[width=0.48\linewidth]{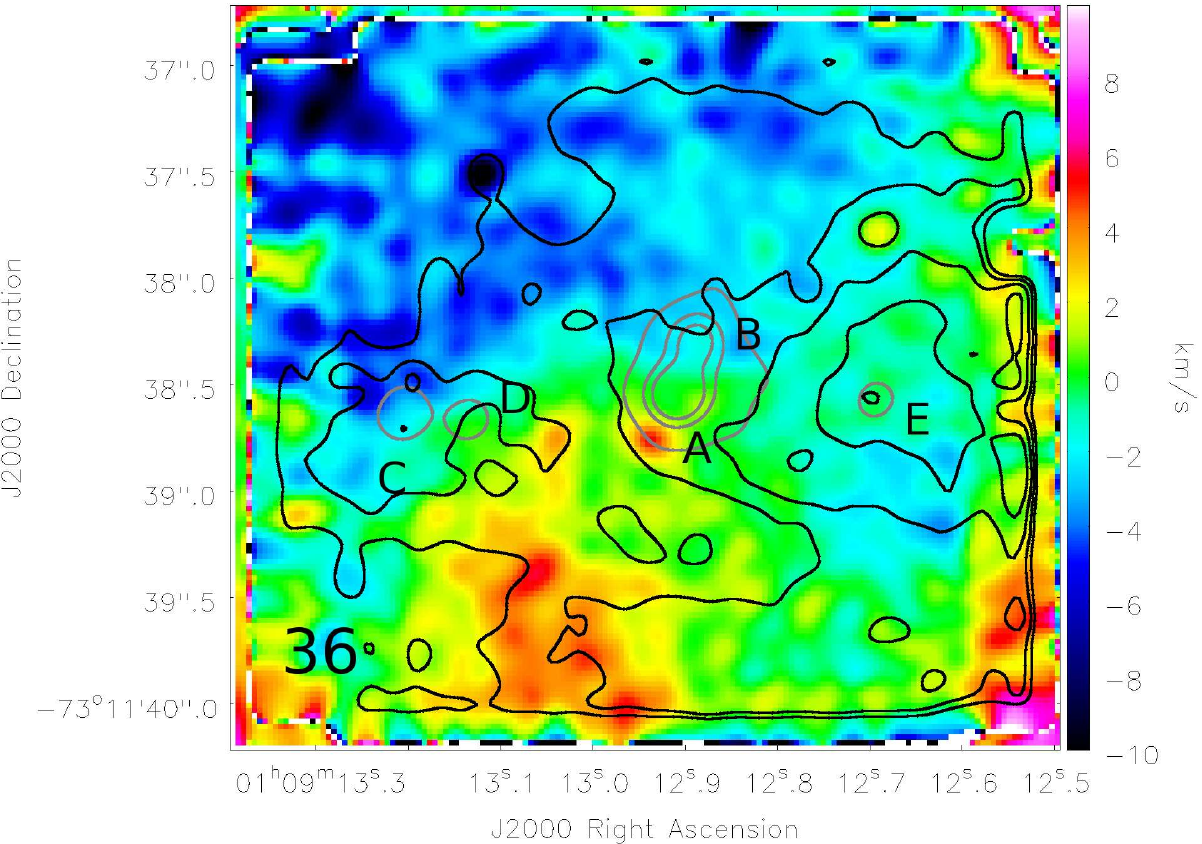}  
\caption{Br$\gamma$ centroid velocity maps for sources \#01, 02, 03, 28, 35 and 36. Black contours - Br$\gamma$ 2.1661\,$\mu$m integrated flux, grey contours - 
continuum flux (see Fig. 3). The black contours in source \#28 trace the same emission as the black and the red contours in Fig. 3.
Velocities are relative to that measured towards the brightest continuum source in the FOV.}
  \end{center}
 \end{minipage}
\end{figure*}

In addition to analysing Br$\gamma$ emission morphologies and fluxes, the centroid velocity of the emission has also been calculated using the 
centroid of the Gaussian profile fitted to the emission line at every spaxel
to create velocity maps. Figure 4 shows the Br$\gamma$ centroid velocity maps for sources \#01, 02, 03, 28, 35 and 36 relative to the brightest 
K-band continuum source in each field.
Sources \#02 and 28 exhibit similar velocity fields surrounding the continuum source with no clear gradient in Br$\gamma$ emission velocity; 
to the south of source \#28 the weak, elongated Br$\gamma$ component (red contour in Fig. 3 and black contour in Fig. 4) appears to be red-shifted with respect to the continuum source by 5--10 km s$^{-1}$.
For sources \#01 and 35 the velocity maps show velocity gradients of 5--10 km s$^{-1}$ which are consistent with an expanding ionized medium surrounding the continuum sources.
Source \#03 exhibits a wide extended and blue shifted region of Br$\gamma$ emission. 
The map of source \#36 shows a significant velocity gradient across the FOV; however, it remains 
unclear whether this is directly influenced by the continuum sources or whether it is more representative of larger scale motions of 
ionized gas in the region. 
Sources \#36\,C and 36\,B appear in regions of Br$\gamma$ emission which are slightly blue shifted with respect to the emission towards 36\,A.
Although Br$\gamma$ emission is extended in source \#31, the signal-to-noise ratio is not sufficient to map the velocity field of the emitting gas.

The Pfund series (between 2.37\,$\mu$m and 2.44\,$\mu$m) has only been detected and measured towards sources in \#35, the compact H\,{\sc ii} region N88\,A. 
However due to the relatively large uncertainties in extinction towards N88\,A and the 
poor signal-to-noise (S/N) in this region, physical properties cannot be determined from the Pfund emission.

\subsubsection{He\,{\sc i} emission}

He\,{\sc i} emission is produced from the ionization and subsequent recombination of helium atoms, a significant process at the ionization boundary of H\,{\sc ii}
regions
and potentially also in areas of collisional excitation \citep{Porter1998}. The strongest and most commonly detected helium emission line 
in the \textit{K}-band towards massive YSOs is the
2.059\,$\mu$m line produced by the 2$^{1}$P$^{0}$--2$^{1}$S transition. This emission line has been detected towards 15 of our \textit{K-}band continuum sources, nine of 
which lie in the FOVs of targets \#35 and 36 (N88\,A and N81, respectively).
\begin{figure}
  \begin{center}
   \includegraphics[width=0.98\linewidth]{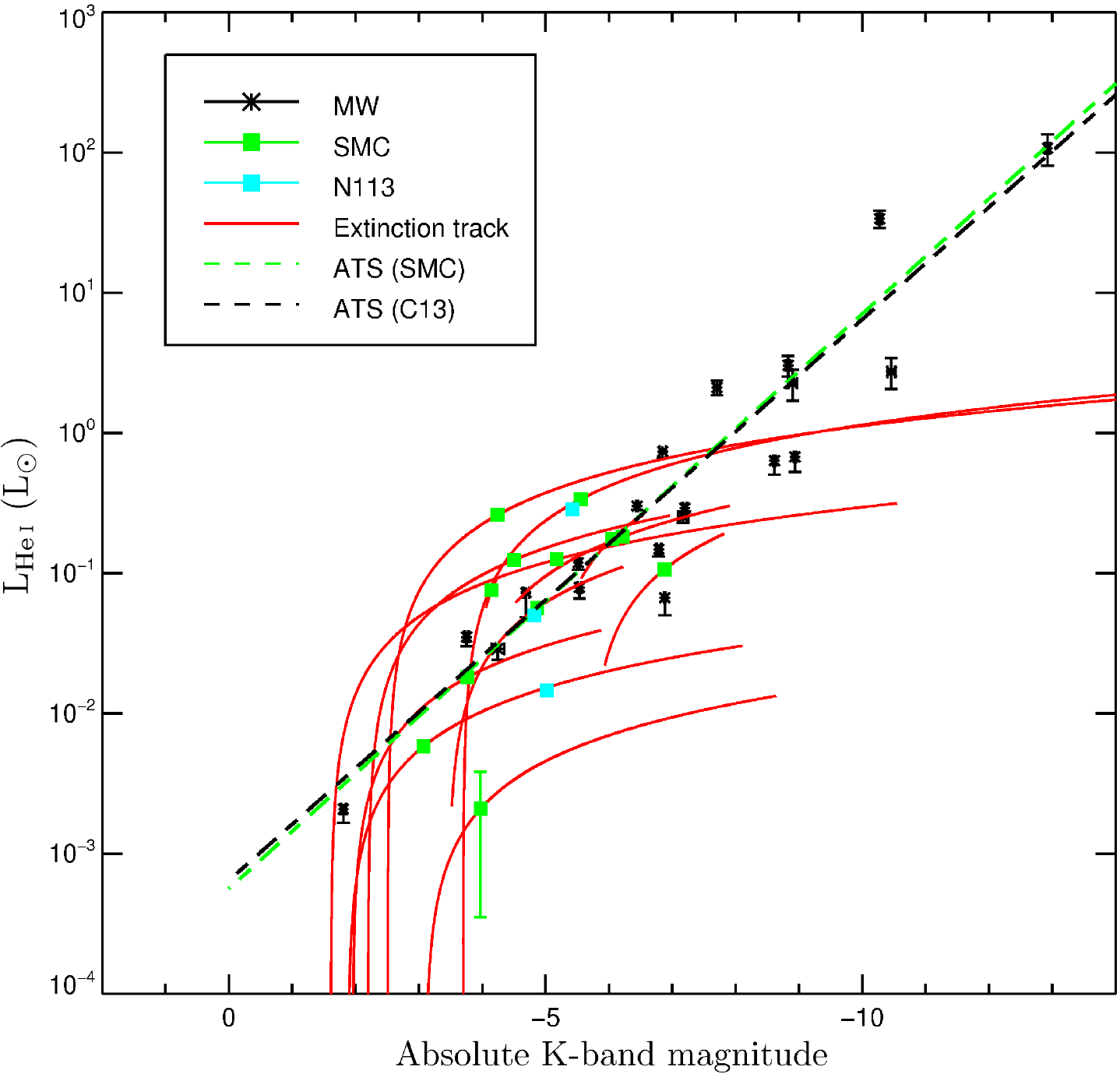}
\caption{He\,{\sc i} luminosity versus absolute \textit{K-}band magnitude. See also caption for Fig. 2.}
  \end{center}
\end{figure}

\begin{figure*}
 \begin{minipage}{175mm}
  \begin{center}
\includegraphics[width=0.325\linewidth]{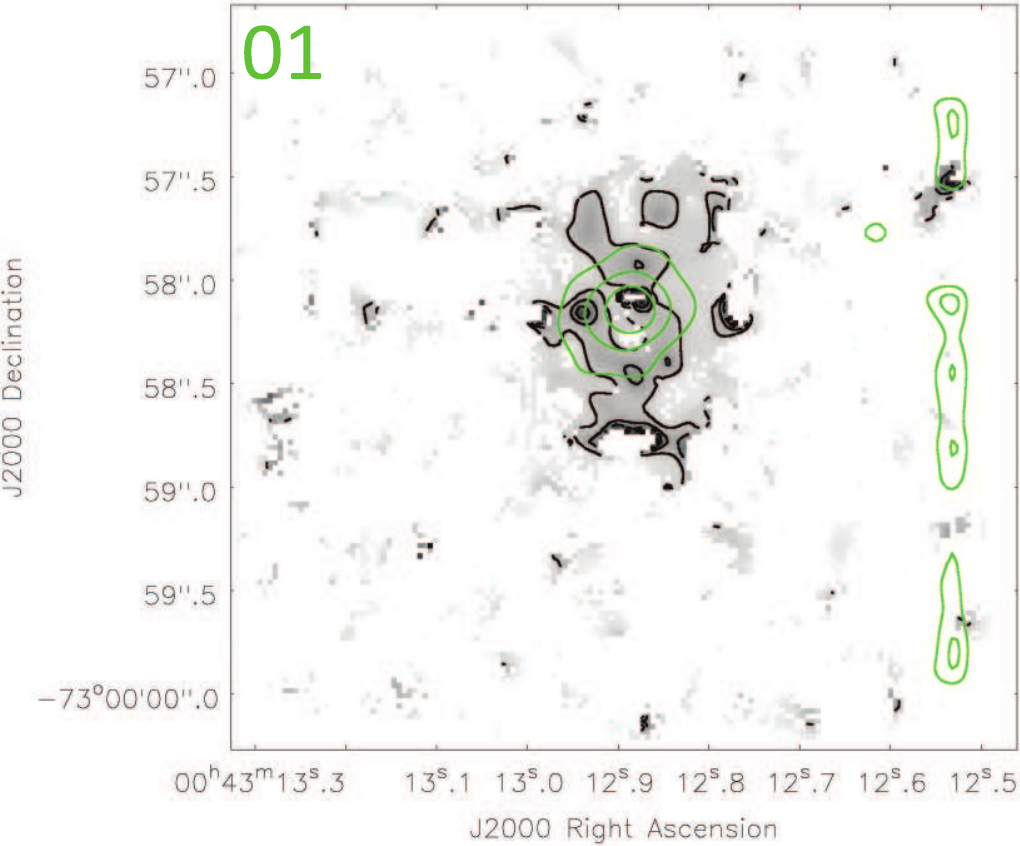}
\includegraphics[width=0.325\linewidth]{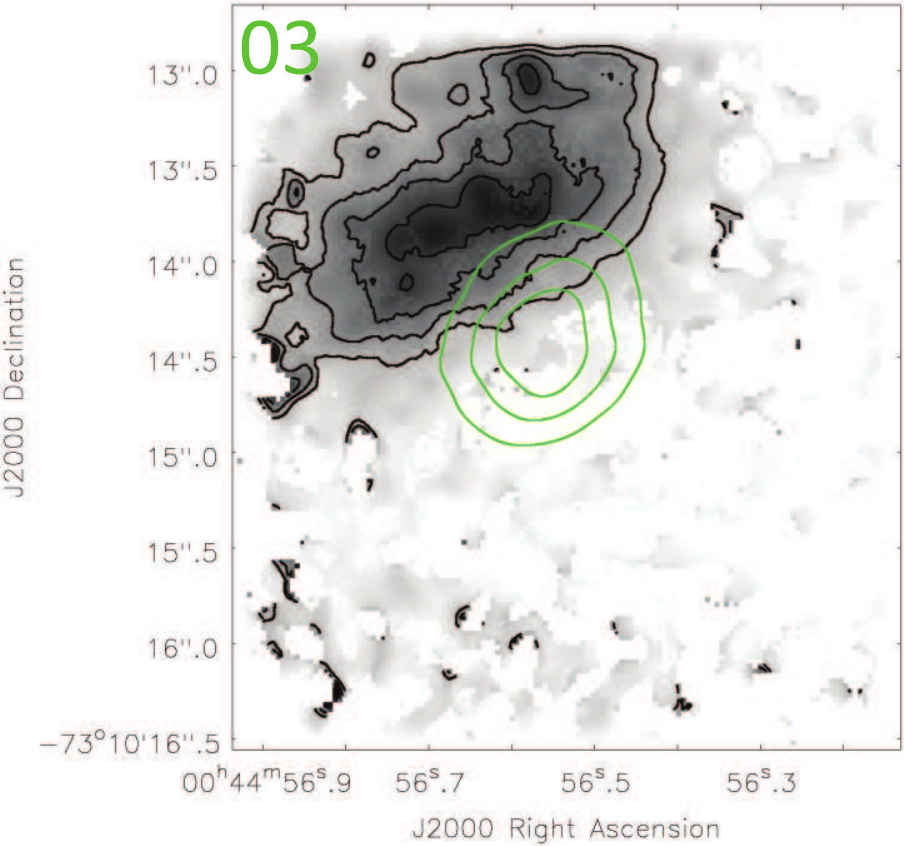}
\includegraphics[width=0.325\linewidth]{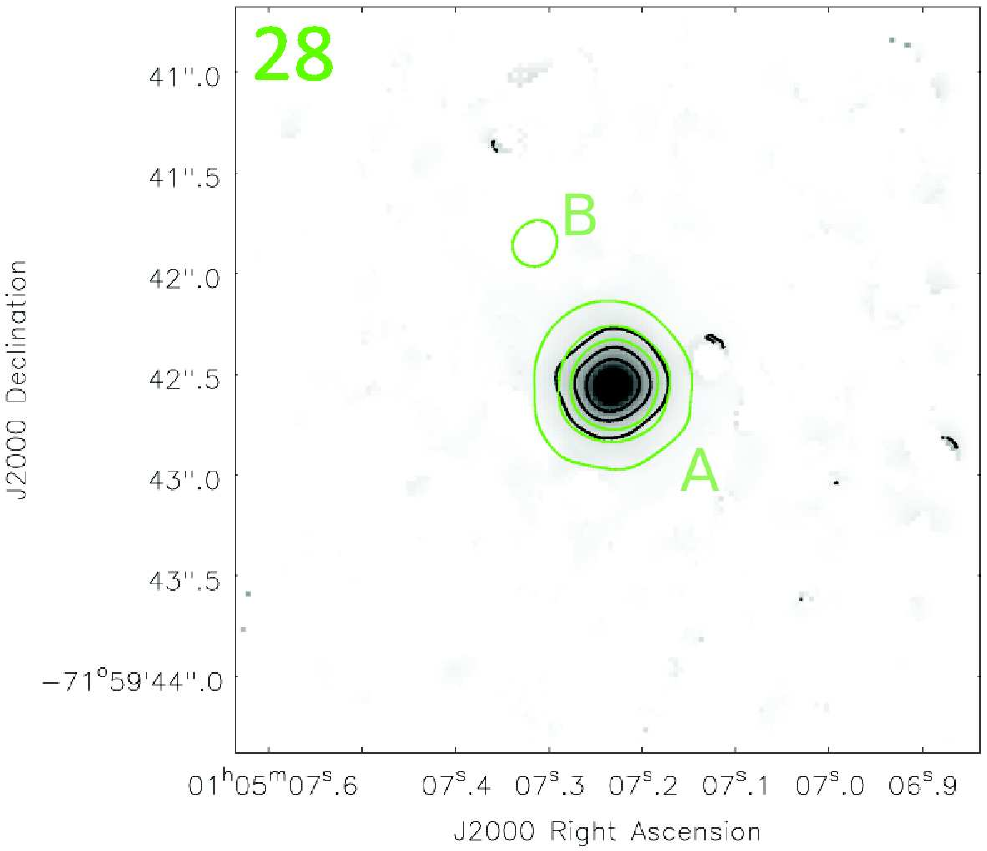}
\includegraphics[width=0.325\linewidth]{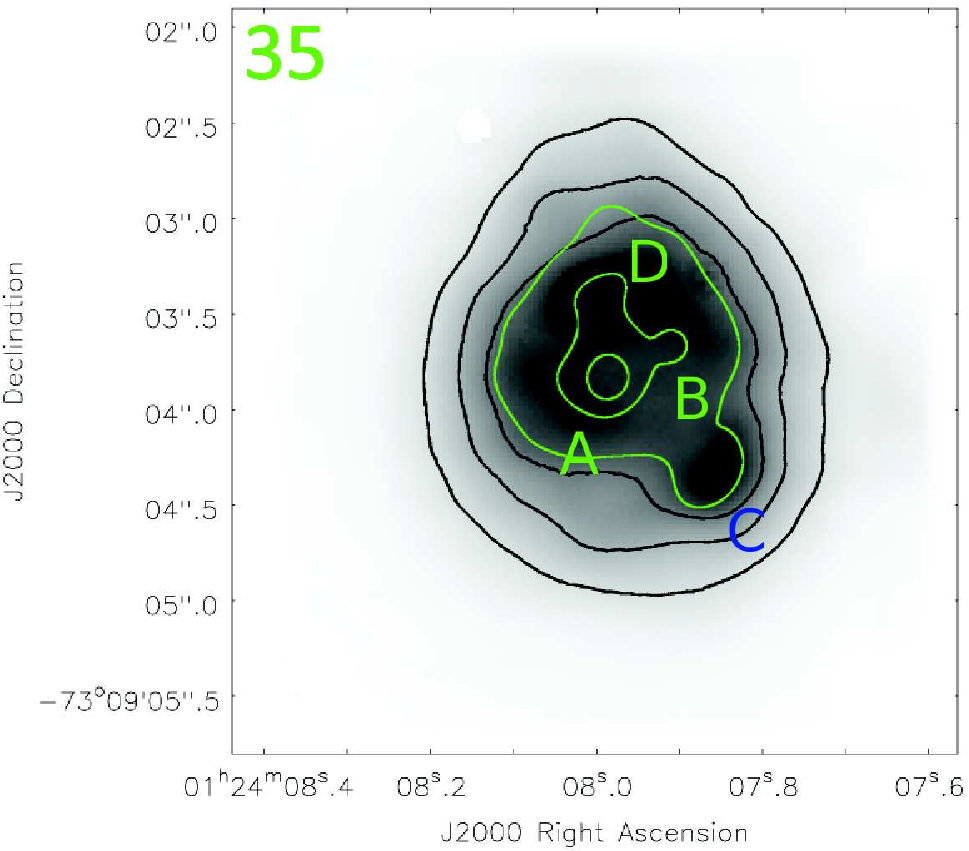}
\includegraphics[width=0.325\linewidth]{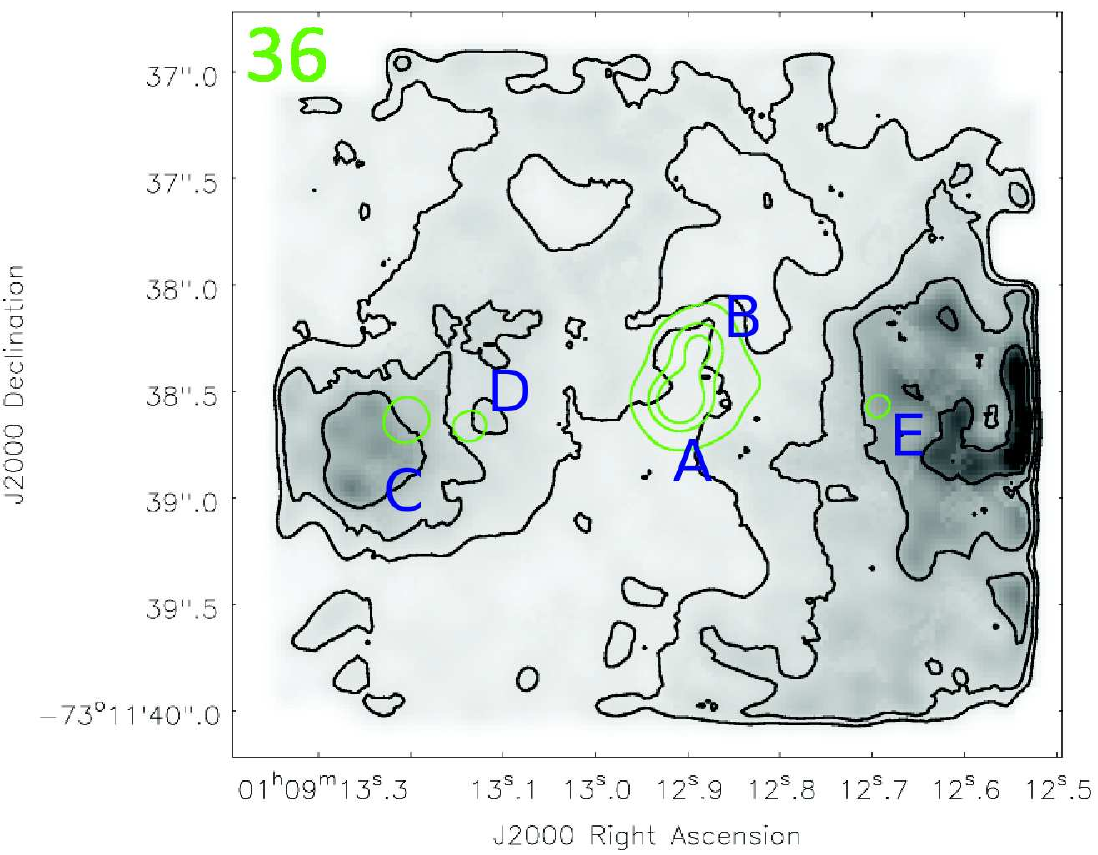}
   \caption{He\,{\sc i} emission line morphologies for sources \#01, 03, 28, 35 and 36. 
Black contours - [0.2,0.4,0.6,0.8] $\times$ maximum He\,{\sc i} 2.059$\mu$m integrated flux, green contours - continuum flux. 
The continuum contour levels are as in Fig. 3.}
  \end{center}
 \end{minipage}
\end{figure*}

Figure 5 plots the He\,{\sc i} 2.059\,$\mu$m emission line luminosity against the absolute \textit{K-}band luminosity, analogous with Fig. 2.
Also included are the results for Galactic sources from \citet{Cooper2013} and those for sources in N113 in the LMC from \citet{Ward2016}.
Despite the large uncertainties originating in the extinction calculation, 
it certainly appears that all of the He\,{\sc i} fluxes fall within the range observed in the Milky Way and in N113 in the LMC.
This suggests that the mechanism causing the increased Br$\gamma$ emission in the SMC does not appear to affect the He\,{\sc i} emission in 
the same way.
Furthermore the ATS regressions fitted to these data also suggest no significant increase is seen in the He\,{\sc i} fluxes towards the SMC compared
to the Galactic sample.
Using the form $y = 10^{ax + b}$, we find values of $a_{\text{SMC}} = -0.41$ and $b_{\text{SMC}} = -3.25$ for the SMC
data and $a_{\text{MW}} = -0.40$ and $b_{\text{MW}} = -3.19$ for the sample of \citet{Cooper2013}.
Again, we find a stronger correlation for the Galactic sources than the SMC sources with $\tau_{\text{SMC}} = -0.48$ whilst $\tau_{\text{MW}} = -0.78$.
Similarly the probability of a correlation is higher in the Galactic sample with values of $p_{\text{SMC}} = 0.03$ and $p_{\text{MW}} = 9 \times 10^{-7}$, indicating probabilities
that the magnitude and He\,{\sc i} luminosities are correlated of 97\% and $>$99\% for the SMC and the Milky Way, respectively.

\begin{figure*}
 \begin{minipage}{175mm}
  \begin{center}
\includegraphics[width=0.48\linewidth]{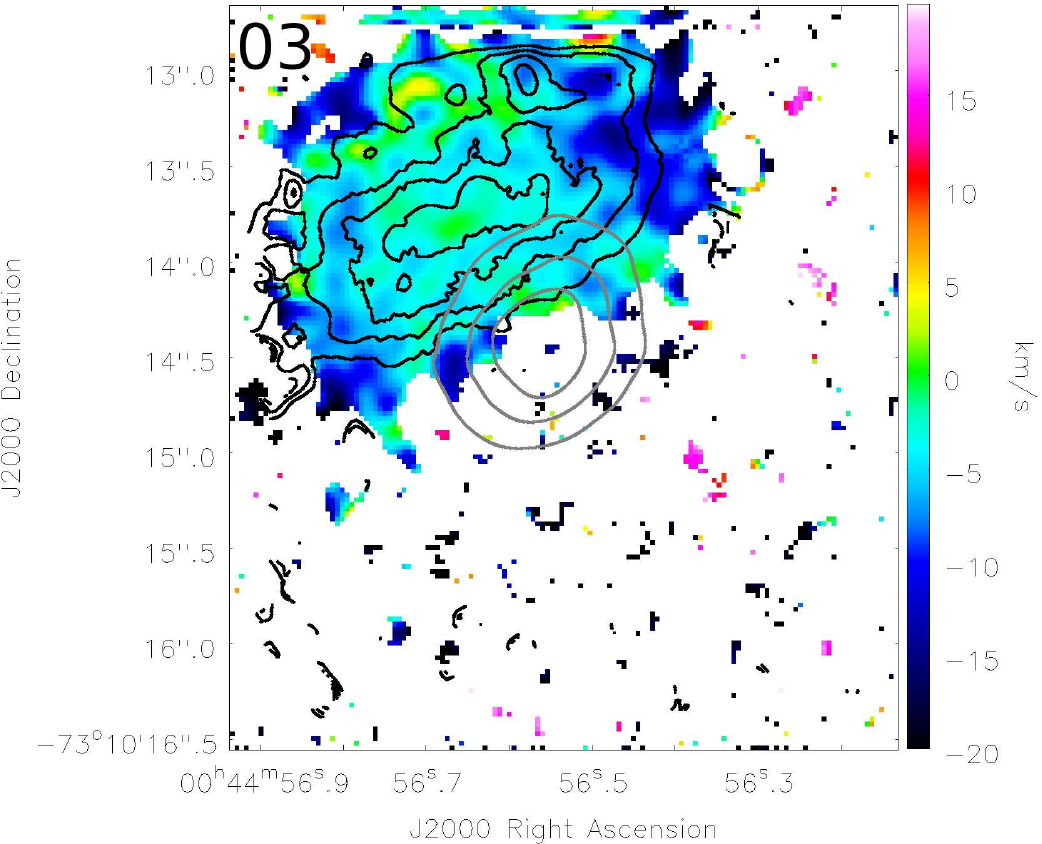}   
\includegraphics[width=0.48\linewidth]{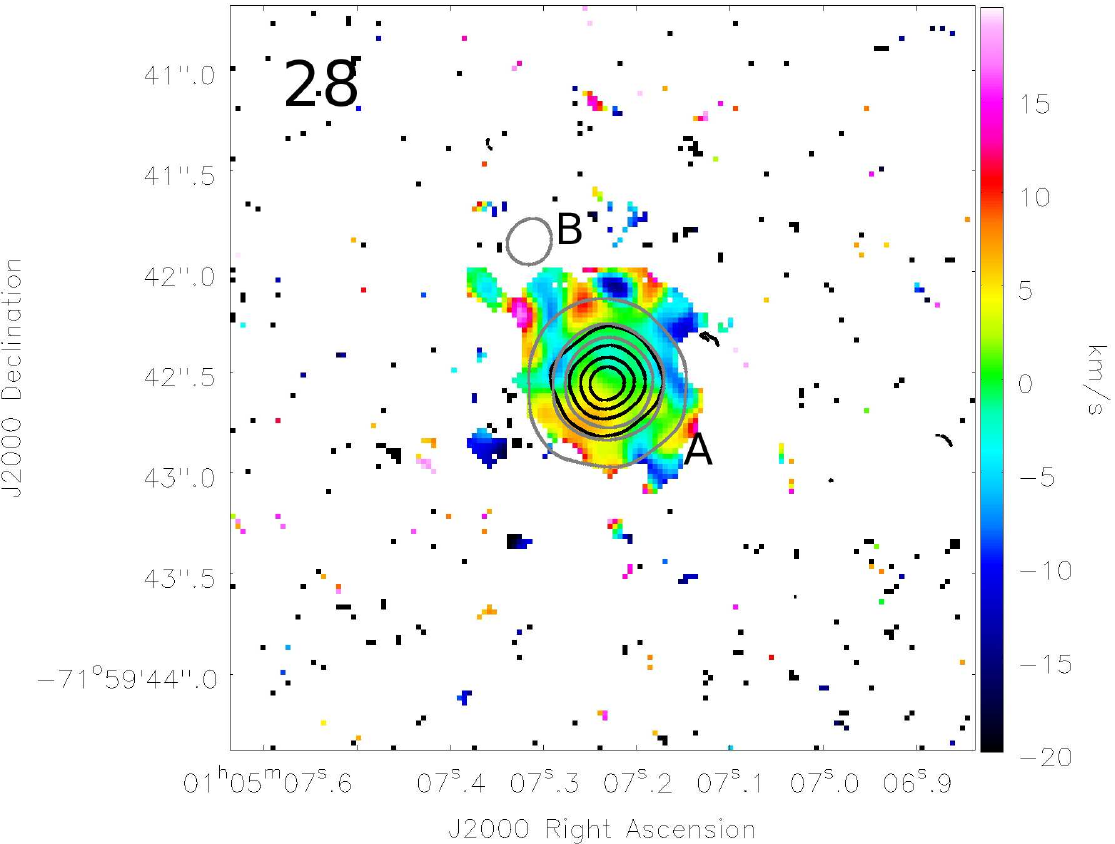}   
\includegraphics[width=0.48\linewidth]{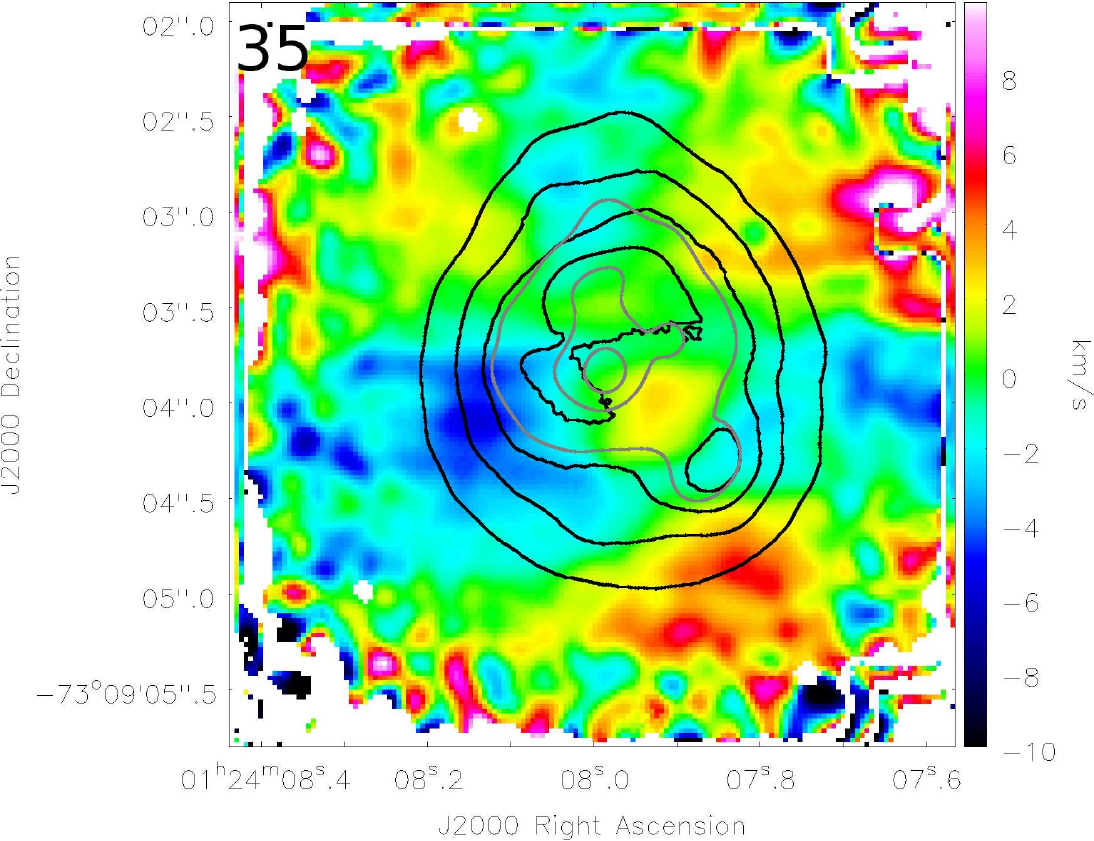}  
\includegraphics[width=0.48\linewidth]{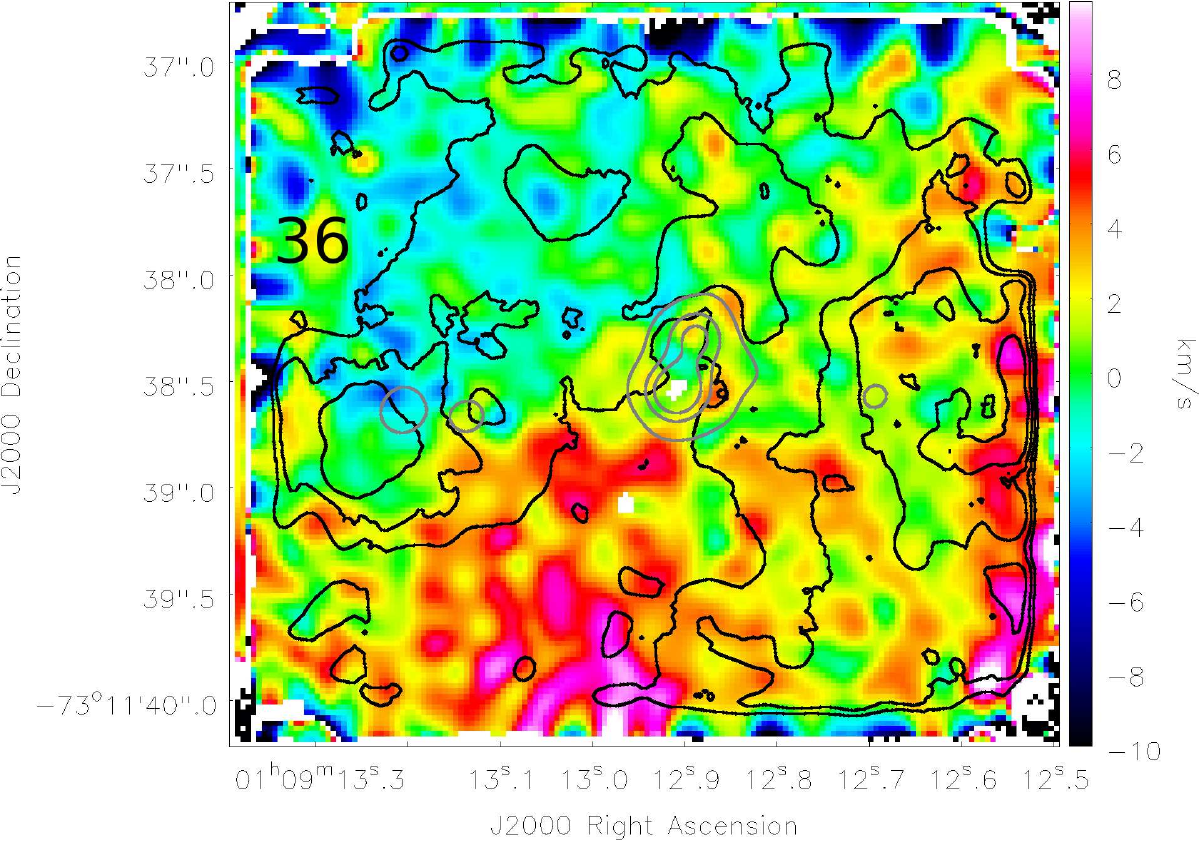}  
\caption{He\,{\sc i} centroid velocity maps for sources \#03, 28, 35 and 36. Black contours - He{\sc i} 2.059\,$\mu$m integrated flux, grey contours - 
continuum flux. See also the caption of Fig. 4.
Velocities are relative to that measured towards the brightest continuum source in the FOV.}
  \end{center}
 \end{minipage}
\end{figure*}

The morphology and kinematics of the He\,{\sc i} 2.059\,$\mu$m emission has been mapped for five sources for which the S/N was adequate to do so (Figs. 6 and 7).
For the most part the He\,{\sc i} emission traces the same morphological structures as the Br$\gamma$ emission.
For sources where line velocities could be reliably measured (\#03, 28, 35, 36),
the He\,{\sc i} velocity maps trace the same velocity structures as the Br$\gamma$ maps.
Both the morphologies and kinematics measured support a common physical origin of Br$\gamma$ and He\,{\sc i} emission for these four sources.
The exception is source \#28\,A which does not exhibit the same southwards narrow 
jet-like structure in He\,{\sc i} emission as in the Br$\gamma$ emission.

\subsubsection{H$_2$ emission}

Molecular hydrogen emission lines in the \textit{K}-band are a commonly used tracer of both shocked emission from outflows and of photo-dissociation regions (PDRs).
We detect the H$_2$ 1-0S(1) and 1-0Q(3) emission lines towards all but one of our sources (\#04) whilst the remaining detected H$_2$ emission
 lines (1-0S(0), 1-0S(2), 1-0S(3), 2-1S(0), 2-1S(1), 2-1S(2), 1-0Q(1), 1-0Q(2); wavelengths are listed in Table 3) have lower rates of detection. 
The full list of H$_2$ emission line fluxes is given in Table C1,
and in Fig. 8 we have plotted H$_2$ 2.1218\,$\mu$m luminosity against that of Br$\gamma$ emission.
We find that the H$_2$ emission dominates (significantly exceeds the Br$\gamma$ luminosity, H$_2$/Br$\gamma > 1.3$) in seven of our sources (\#06, 18, 22\,A, 22\,B, 26, 28\,B, 31)
and that six sources lie close to the H$_2$ $=$ Br$\gamma$ line (H$_2$/Br$\gamma > 0.6$; \#01, 02\,A, 03, 17, 28\,A, 34). The remaining sources
for which we measure both H$_2$ emission and Br$\gamma$ emission appear to be dominated by Br$\gamma$ emission (H$_2$/Br$\gamma < 0.35$).

\begin{figure*}
 \begin{minipage}{175mm}
\begin{center}
\includegraphics[width=0.85\linewidth]{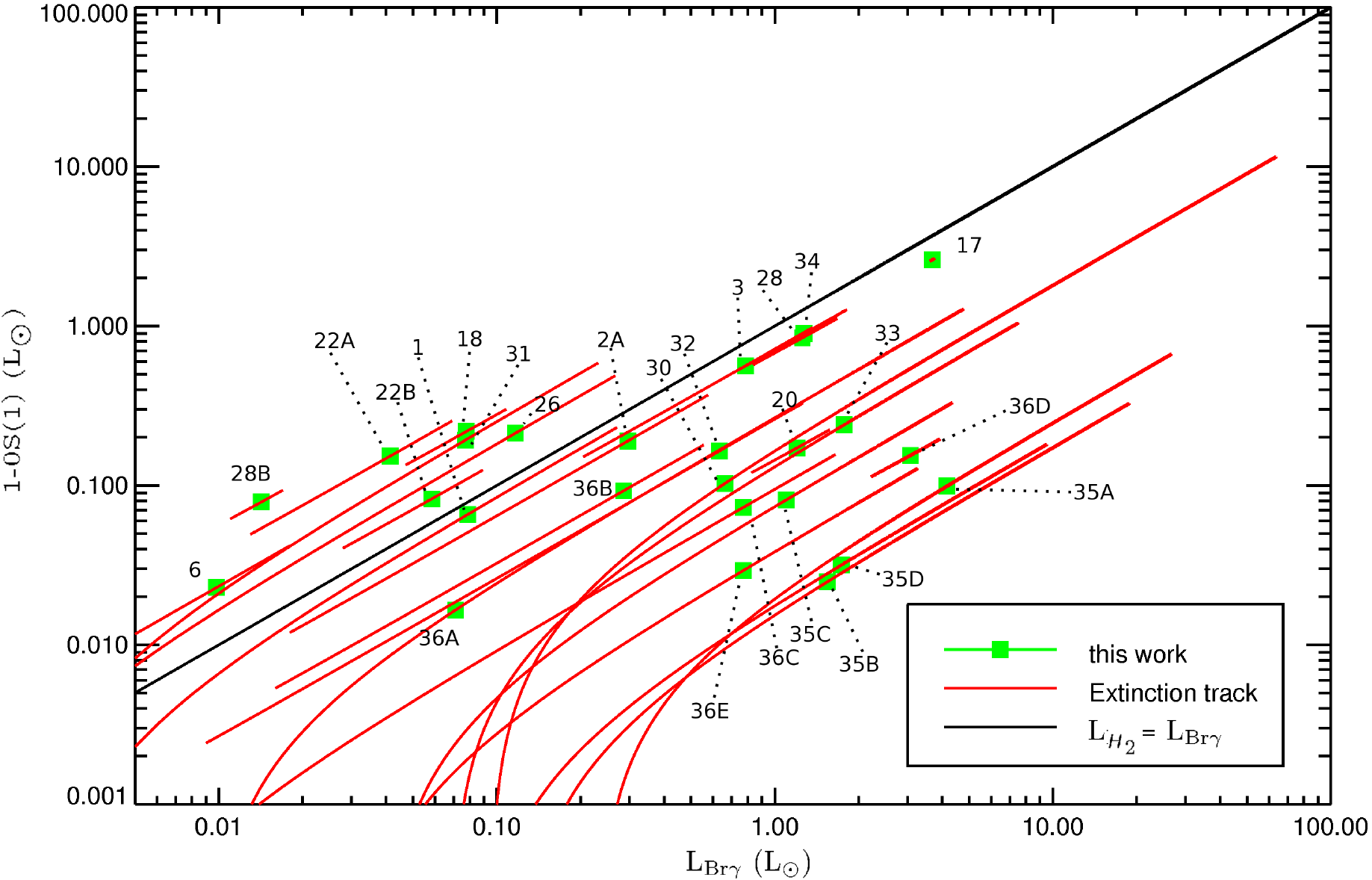}
\caption{H$_2$ 2.1218\,$\mu$m emission vs Br$\gamma$ emission. The solid black line shows where the Br$\gamma$ emission line luminosity and 
the H$_2$ emission line luminosity are equal. See also the caption of Fig. 2.}
\end{center}
 \end{minipage}
\end{figure*}

Through comparison with grids of models simulating H$_2$ emission from both shock excitation and photo-excitation, it is possible to distinguish 
between these excitation mechanisms using the flux ratios between different emission lines. 
Table 3 shows the extinction corrected intensity ratios of all of the measured H$_2$ emission lines relative to the 1-0S(1) line with the exception
of the Q-branch emission lines. Also included are the model line ratios for photo-excitation \citep{Black1987} and shocked emission \citep{Shull1978}
for comparison.
 Particularly useful are the 1--0S(0)/1--0S(1) and 2--1S(1)/1--0S(1) emission line ratios due to the relatively high fluxes of these lines and 
clear distinctions between the photo-excited and shock-excited ratios (see Table 3).

Using the 1-0S(0) / 1-0S(1) and 2-1S(1) / 1-0S(1) emission line ratios, sources \#03, 06, 18, 26 and 34 consistently appear to exhibit 
emission dominated by collisional excitation with ratio ranges of 0.14--0.27 and 0.11--0.25, respectively. 
Source \#25 is the only source which is clearly photo-excited whilst all other sources appear to exhibit contributions from both 
photo-excitation and shocked excitation.
For instance, whilst none of the 2-1S(1) / 1-0S(1) ratios fall within the photo-excitation range,
sources \#02\,B, 28\,B, 36\,B and 36\,E all have 1-0S(0) / 1-0S(1) ratios consistent with photo-excited gas (0.38--0.55) and 2-1S(1) / 1-0S(1) 
inconsistent with collisional excitation (0.33--0.46).
It is likely that the majority of measured H$_2$ emission has contributions from both photo-excitation and collisional excitation
as reflected in the low 2-1S(1) line strength compared to 1-0S(0) line strength which implies that the 2-1S(1) emission is primarily sourced
from shocked excitation whilst the 1-0S(0) emission has a larger contribution from photo-excitation.

\begin{table*}
\begin{minipage}{175mm}
 \begin{center}
\caption{Extinction corrected H$_2$ emission line ratios with respect to the 1-0S(1) emission line for all continuum sources. 
Also included are the expected 
H$_2$ line ratios for photo-excitation \citep{Black1987} and shocked emission \citep{Shull1978}.
The final column gives the origin of the emission for those sources for which we are confident of the diagnosis based on the H$_2$ emission line ratios.}
  \begin{tabular}{l c c c c c c c}
\hline
   Source & 1-0S(0) & 1-0S(2) & 1-0S(3) & 2-1S(1) & 2-1S(2) & 2-1S(3) & Origin\\
  & 2.2235\,$\mu$m & 2.0338\,$\mu$m & 1.9576\,$\mu$m & 2.2477\,$\mu$m & 2.1542\,$\mu$m & 2.0735\,$\mu$m \\
\hline
01	&	0.49	$\substack{+	0.01	 \\ -	0.02	}$ & 	0.44	$\substack{+	0.05	 \\ -	0.02	}$ & 	0.57	$\substack{+	0.13	 \\ -	0.06	}$ & 	0.193	$\substack{+	0.005	 \\ -	0.011	}$ & 	0.158	$\substack{+	0.005	 \\ -	0.002	}$ & 	0.35	$\substack{+	0.03	 \\ -	0.01	}$ 	\\
02\,A	&	0.36	$\substack{+	0.01	 \\ -	0.03	}$ & 	0.39	$\substack{+	0.07	 \\ -	0.01	}$ & 	2.09	$\substack{+	0.22	 \\ -	0.01	}$ & 	0.115	$\substack{+	0.010	 \\ -	0.001	}$ & 	\null & 	0.19	$\substack{+	0.21	 \\ -	0.01	}$ 	\\
02\,B	&	0.55	$\substack{+	0.02	 \\ -	0.07	}$ & 	0.35	$\substack{+	0.07	 \\ -	0.05	}$ & 	1.6	$\substack{+	0.7	 \\ -	0.5	}$ & 	\null & 	0.168	$\substack{+	0.010	 \\ -	0.008	}$ & 	\null 	\\
03	&	0.23	$\pm0.01$ & 	0.25	$\substack{+	0.05	 \\ -	0.01	}$ & 	2.23	$\substack{+	0.17	 \\ -	0.03	}$ & 	0.245	$\substack{+	0.003	 \\ -	0.019	}$ & 	0.104	$\substack{+	0.017	 \\ -	0.003	}$ & 	0.229	$\substack{+	0.025	 \\ -	0.004	}$ 	&	shock\\
04	&	\null	 & 		\null & 		\null & 		\null & 		\null & 		\null 	\\
06	&	0.14	$\pm$0.01 & 	0.37	$\substack{+	0.11	 \\ -	0.01	}$ & 	1.69	$\substack{+	0.01	 \\ -	0.22	}$ & 	0.113	$\substack{+	0.001	 \\ -	0.010	}$ & 	\null & 	0.089	$\substack{+	0.075	 \\ -	0.003	}$ 	&	shock\\
17	&	\null & 	0.48	$\substack{+	0.21	 \\ -	0.16	}$ & 	4.12	$\substack{+	0.18	 \\ -	0.14	}$ & 	0.138	$\pm$0.001 & 	\null & 	\null 	\\
18	&	0.25	$\pm$0.01 & 	0.38	$\substack{+	0.03	 \\ -	0.01	}$ & 	1.09	$\pm$0.01 & 	0.175	$\substack{+	0.004	 \\ -	0.010	}$ & 	0.070	$\substack{+	0.004	 \\ -	0.002	}$ & 	\null 	&	shock\\
20	&	0.23	$\pm$0.01 & 	0.45	$\substack{+	0.02	 \\ -	0.01	}$ & 	1.09	$\substack{+	0.22	 \\ -	0.11	}$ & 	\null & 	\null & 	\null 	\\
22\,A	&	0.38	$\substack{+	0.01	 \\ -	0.02	}$ & 	0.40	$\pm$0.01 & 	0.74	$\substack{+	0.01	 \\ -	0.08	}$ & 	0.182	$\substack{+	0.003	 \\ -	0.013	}$ & 	0.129	$\substack{+	0.010	 \\ -	0.002	}$ & 	0.117	$\substack{+	0.002	 \\ -	0.001	}$ 	\\
22\,B	&	0.59	$\substack{+	0.01	 \\ -	0.03	}$ & 	0.37	$\substack{+	0.03	 \\ -	0.01	}$ & 	1.87	$\substack{+	0.04	 \\ -	0.13	}$ & 	0.281	$\substack{+	0.006	 \\ -	0.018	}$ & 	0.265	$\substack{+	0.018	 \\ -	0.006	}$ & 	0.218	$\substack{+	0.038	 \\ -	0.012	}$ 	\\
25	&	0.44	$\substack{+	0.01	 \\ -	0.03	}$ & 	\null & 	\null & 	0.463	$\substack{+	0.001	 \\ -	0.041	}$ & 	\null & 	\null 	&	photo\\
26	&	0.22	$\substack{+	0.03	 \\ -	0.01	}$ & 	0.31	$\substack{+	0.01	 \\ -	0.10	}$ & 	1.14	$\substack{+	0.21	 \\ -	0.03	}$ & 	0.251	$\substack{+	0.038	 \\ -	0.003	}$ & 	0.103	$\substack{+	0.002	 \\ -	0.001	}$ & 	0.189	$\substack{+	0.001	 \\ -	0.011	}$ 	&	shock\\
28\,A	&	0.30	$\pm$0.01 & 	0.35	$\pm$0.01 & 	0.56	$\substack{+	0.01	 \\ -	0.03	}$ & 	0.151	$\substack{+	0.004	 \\ -	0.008	}$ & 	0.048	$\pm$0.001 & 	0.126	$\substack{+	0.007	 \\ -	0.003	}$ 	\\
28\,B	&	0.48	$\substack{+	0.01	 \\ -	0.02	}$ & 	0.54	$\substack{+	0.13	 \\ -	0.07	}$ & 	1.02	$\substack{+	0.18	 \\ -	0.11	}$ & 	0.375	$\substack{+	0.010	 \\ -	0.016	}$ & 	\null & 	0.308	$\substack{+	0.013	 \\ -	0.008	}$ 	\\
30	&	0.31	$\substack{+	0.03	 \\ -	0.09	}$ & 	0.39	$\pm$0.15 & 	1.2	$\pm$1.1 & 	\null & 	\null & 	\null 	\\
31	&	0.50	$\substack{+	0.01	 \\ -	0.02	}$ & 	0.48	$\substack{+	0.02	 \\ -	0.05	}$ & 	1.12	$\substack{+	0.08	 \\ -	0.24	}$ & 	0.242	$\substack{+	0.004	 \\ -	0.010	}$ & 	0.136	$\substack{+	0.003	 \\ -	0.001	}$ & 	0.296	$\substack{+	0.006	 \\ -	0.018	}$ 	\\
32	&	0.29	$\substack{+	0.01	 \\ -	0.02	}$ & 	$<$1.2	 & 	\null & 	0.337	$\substack{+	0.001	 \\ -	0.030	}$ & 	\null & 	\null 	\\
33	&	0.30	$\substack{+	0.01	 \\ -	0.02	}$ & 	0.23	$\substack{+	0.02	 \\ -	0.03	}$ & 	\null & 	0.270	$\substack{+	0.001	 \\ -	0.021	}$ & 	\null & 	\null 	\\
34	&	0.27	$\pm$0.01 & 	0.37	$\pm$0.01 & 	0.85	$\substack{+	0.02	 \\ -	0.05	}$ & 	0.112	$\substack{+	0.003	 \\ -	0.006	}$ & 	0.099	$\substack{+	0.016	 \\ -	0.006	}$ & 	\null 	&	shock\\
35\,A	&	0.37	$\substack{+	0.01	 \\ -	0.04	}$ & 	0.30	$\substack{+	0.03	 \\ -	0.05	}$ & 	0.92	$\substack{+	0.22	 \\ -	0.31	}$ & 	0.332	$\substack{+	0.004	 \\ -	0.041	}$ & 	0.116	$\substack{+	0.006	 \\ -	0.004	}$ & 	0.195	$\substack{+	0.012	 \\ -	0.016	}$ 	\\
35\,B	&	0.34	$\substack{+	0.01	 \\ -	0.05	}$ & 	0.23	$\substack{+	0.04	 \\ -	0.05	}$ & 	0.90	$\substack{+	0.34	 \\ -	0.41	}$ & 	0.318	$\substack{+	0.001	 \\ -	0.058	}$ & 	0.081	$\substack{+	0.005	 \\ -	0.004	}$ & 	\null 	\\
35\,C	&	0.39	$\substack{+	0.01	 \\ -	0.03	}$ & 	0.44	$\substack{+	0.04	 \\ -	0.07	}$ & 	1.25	$\substack{+	0.25	 \\ -	0.38	}$ & 	0.334	$\substack{+	0.006	 \\ -	0.035	}$ & 	0.095	$\substack{+	0.004	 \\ -	0.003	}$ & 	0.306	$\substack{+	0.016	 \\ -	0.024	}$ 	\\
35\,D	&	0.37	$\substack{+	0.01	 \\ -	0.02	}$ & 	0.28	$\substack{+	0.02	 \\ -	0.03	}$ & 	0.8	$\substack{+	0.1	 \\ -	0.2	}$ & 	0.173	$\substack{+	0.001	 \\ -	0.012	}$ & 	0.060	$\substack{+	0.002	 \\ -	0.001	}$ & 	0.125	$\substack{+	0.004	 \\ -	0.008	}$ 	\\
36\,A	&	0.29	$\substack{+	0.01	 \\ -	0.08	}$ & 	0.37	$\substack{+	0.13	 \\ -	0.14	}$ & 	1.2	$\substack{+	0.9	 \\ -	1.0	}$ & 	0.400	$\substack{+	0.001	 \\ -	0.126	}$ & 	\null & 	\null 	\\
36\,B	&	0.41	$\substack{+	0.01	 \\ -	0.03	}$ & 	0.16	$\substack{+	0.08	 \\ -	0.01	}$ & 	1.04	$\substack{+	0.43	 \\ -	0.01	}$ & 	0.388	$\substack{+	0.001	 \\ -	0.034	}$ & 	0.138	$\substack{+	0.087	 \\ -	0.002	}$ & 	0.337	$\substack{+	0.051	 \\ -	0.001	}$ 	\\
36\,C	&	\null & 	\null & 	\null & 	\null & 	\null & 	0.64	$\substack{+	0.02	 \\ -	0.04	}$ 	\\
36\,D	&	\null & 	\null & 	2.44	$\substack{+	0.04	 \\ -	0.02	}$ & 	\null & 	\null & 	\null 	\\
36\,E	&	0.38	$\substack{+	0.07	 \\ -	0.01	}$ & 	\null & 	1.13	$\substack{+	0.03	 \\ -	0.49	}$ & 	0.333	$\substack{+	0.072	 \\ -	0.002	}$ & 	\null & 	\null 	\\
\hline
Photo-excitation & 0.4--0.7 & 0.4--0.6 & & 0.5--0.6 & 0.2--0.4 & 0.2--0.3 \\
1000 K shock &  0.27 & 0.27 &  0.51 & 0.005 & 0.001 & 0.003 \\
2000 K shock & 0.21 & 0.37 & 1.02 &  0.083 & 0.031 & 0.084 \\
3000 K shock & 0.19 & 0.42 & 1.29 & 0.21 & 0.086 & 0.27 \\
4000 K shock & 0.19 &  0.44 & 1.45 & 0.33 & 0.14 & 0.47 \\
\hline
  \end{tabular}

 \end{center}
\end{minipage}
\end{table*}

Half of our YSO targets exhibit extended H$_2$ emission; the morphologies of the 2.1218\,$\mu$m 1--0S(1) emission line are shown in Fig. 9. 
Comparing Figs. 3 and 9, it is clear that the Br$\gamma$ and H$_2$ emission morphologies are very different.
Sources \#01, 02\,A, 28\,A and 30 exhibit H$_2$ morphologies extended along a single axis, asymmetric with respect to the continuum source. 
Source \#28\,A exhibits a v-shape morphology, indicative of a cone like physical structure. 
Source \#06 exhibits two knots of H$_2$ emission separated from the continuum source (0.6\,arcsec and 1.4\,arcsec for the southern and
north-western knots, respectively), although the weaker north-western knot 
appears to be linked by a tenuous tail of emission to the continuum source.
Sources \#03 and 31 both exhibit irregular elongated
emission across the continuum source with offset Br$\gamma$ emission.
Bipolar extended emission is observed centred on sources \#22\,A and 36\,B although that of 36\,B appears to be slightly off-centre.
Finally source \#35 exhibits an arc of H$_2$ emission as previously studied in \citet{Testor2010}, tracing the edge of the compact
H\,{\sc ii} region shown in the Br$\gamma$ emission.
The morphologies of the different emission lines are discussed further in Section 6.

\begin{figure*}
 \begin{minipage}{175mm}
  \begin{center}
\includegraphics[width=0.32\linewidth]{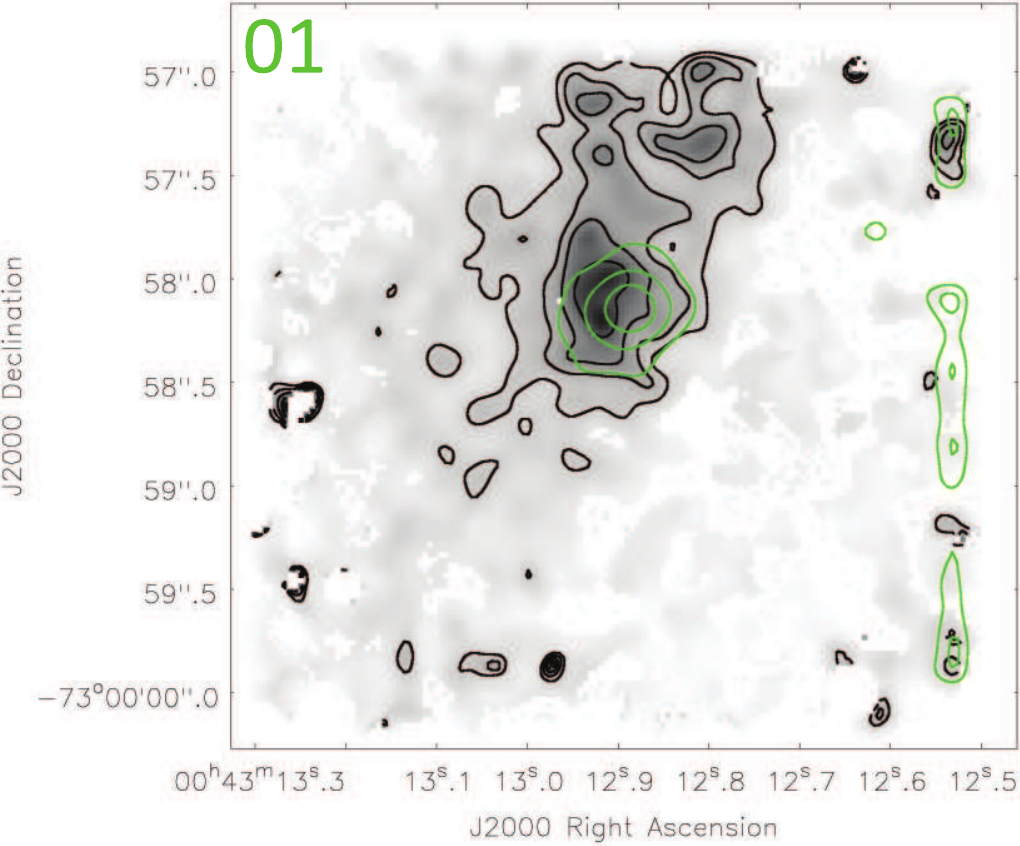}
\includegraphics[width=0.32\linewidth]{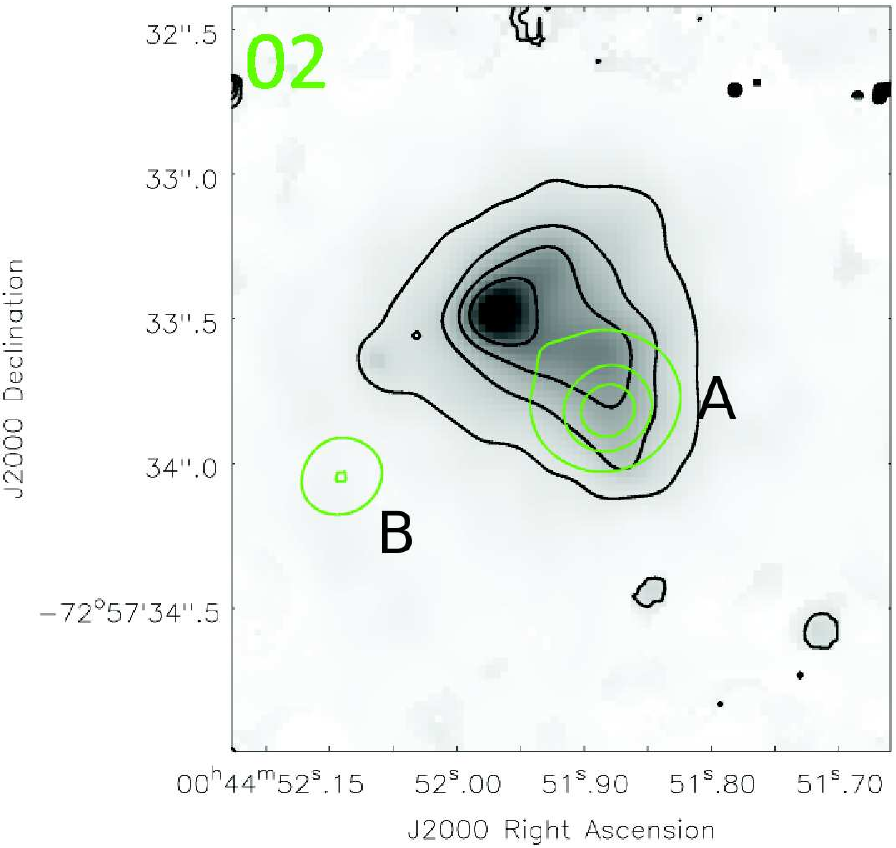}
\includegraphics[width=0.32\linewidth]{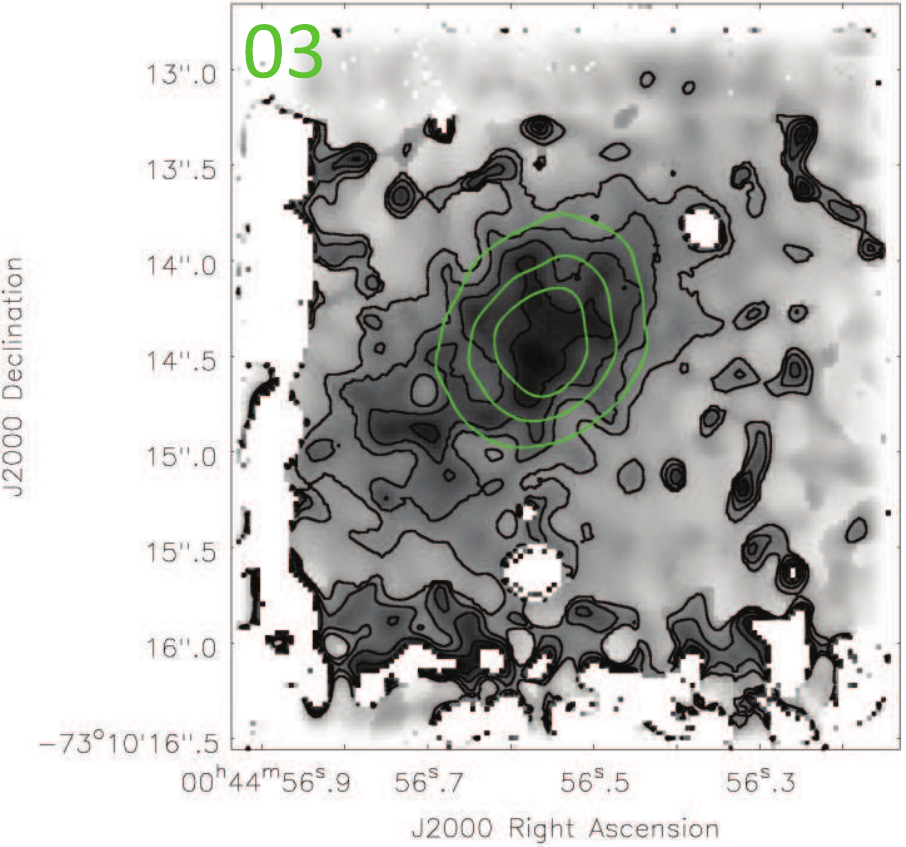}
\includegraphics[width=0.32\linewidth]{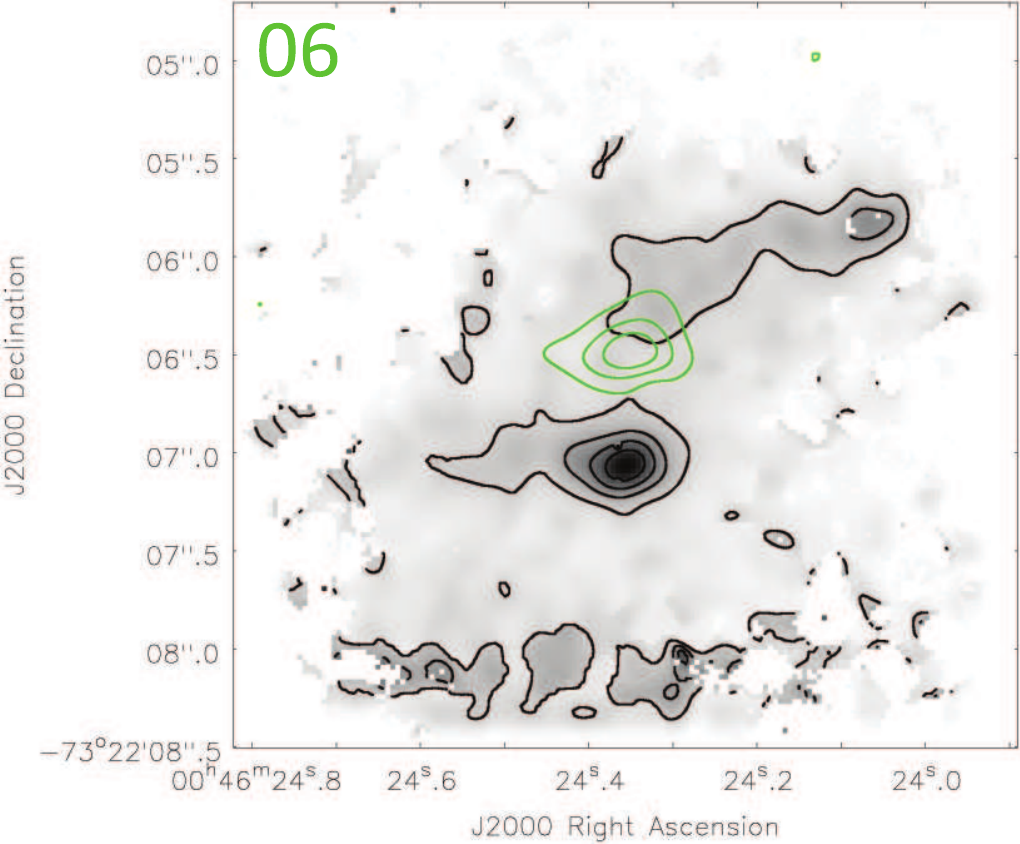}
\includegraphics[width=0.32\linewidth]{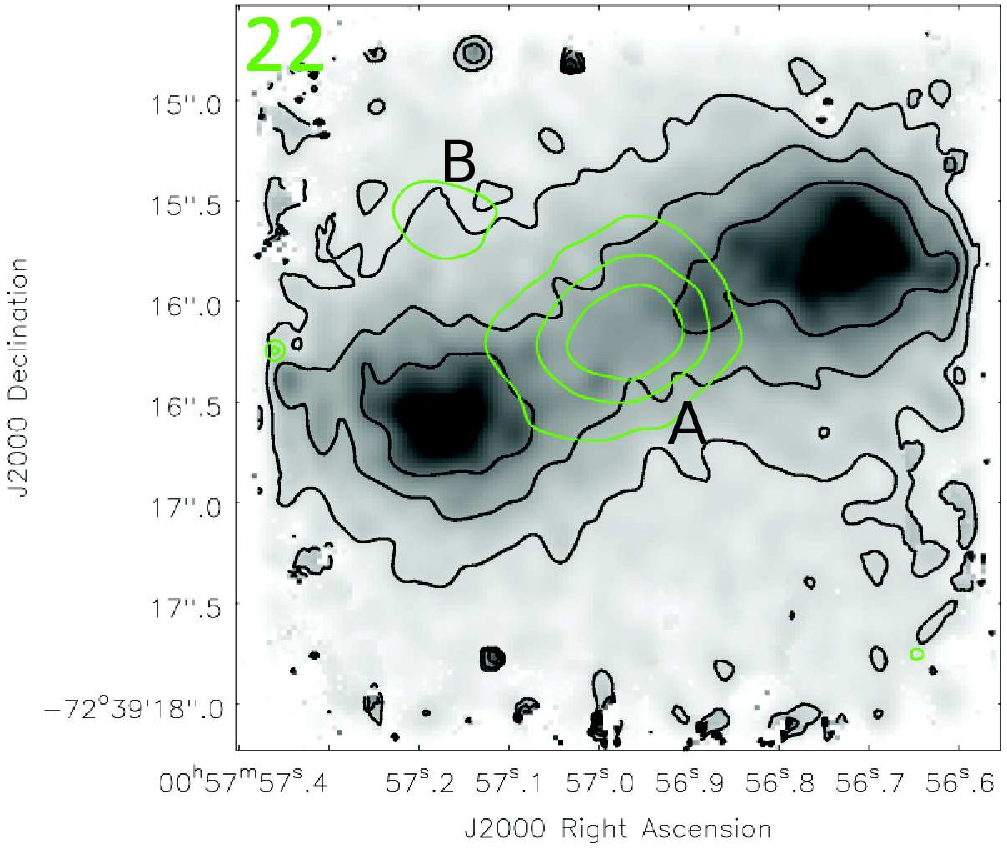}
\includegraphics[width=0.32\linewidth]{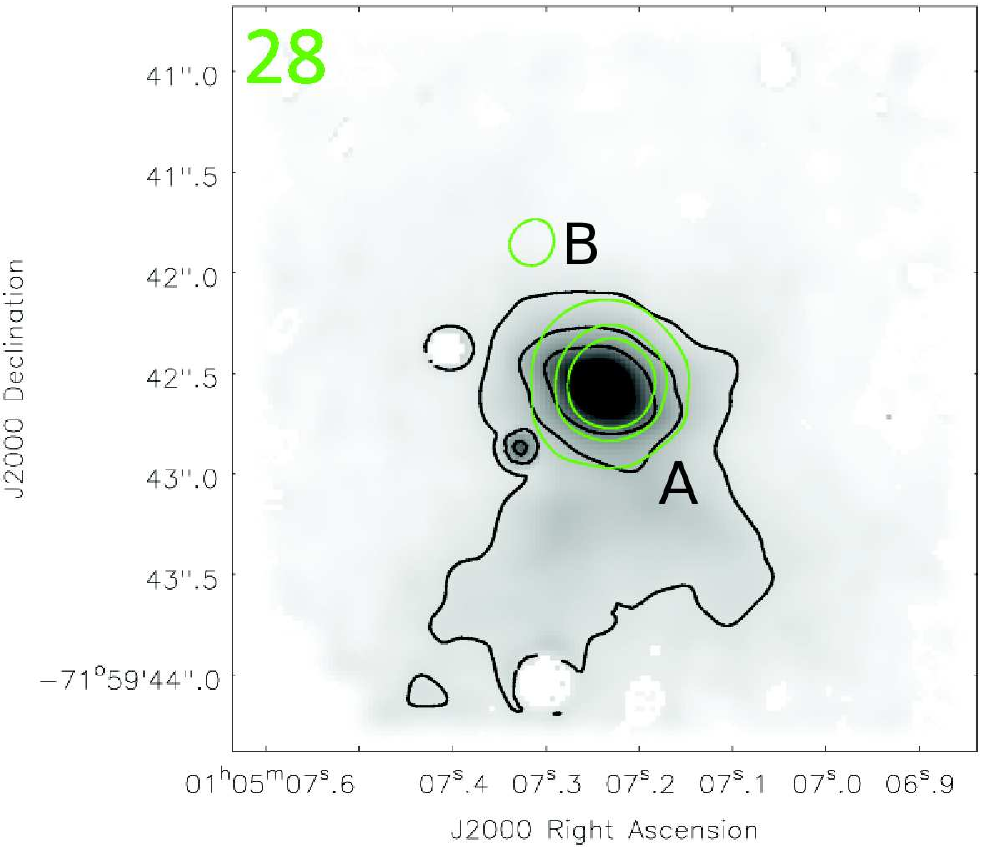}
\includegraphics[width=0.32\linewidth]{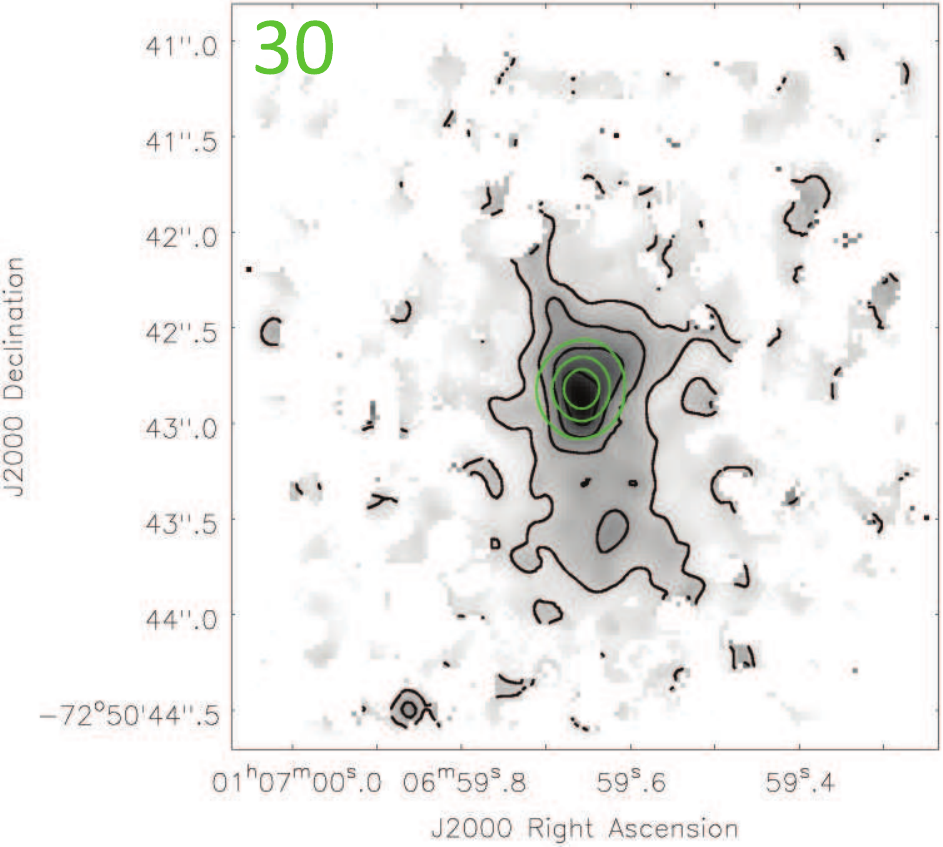}
\includegraphics[width=0.32\linewidth]{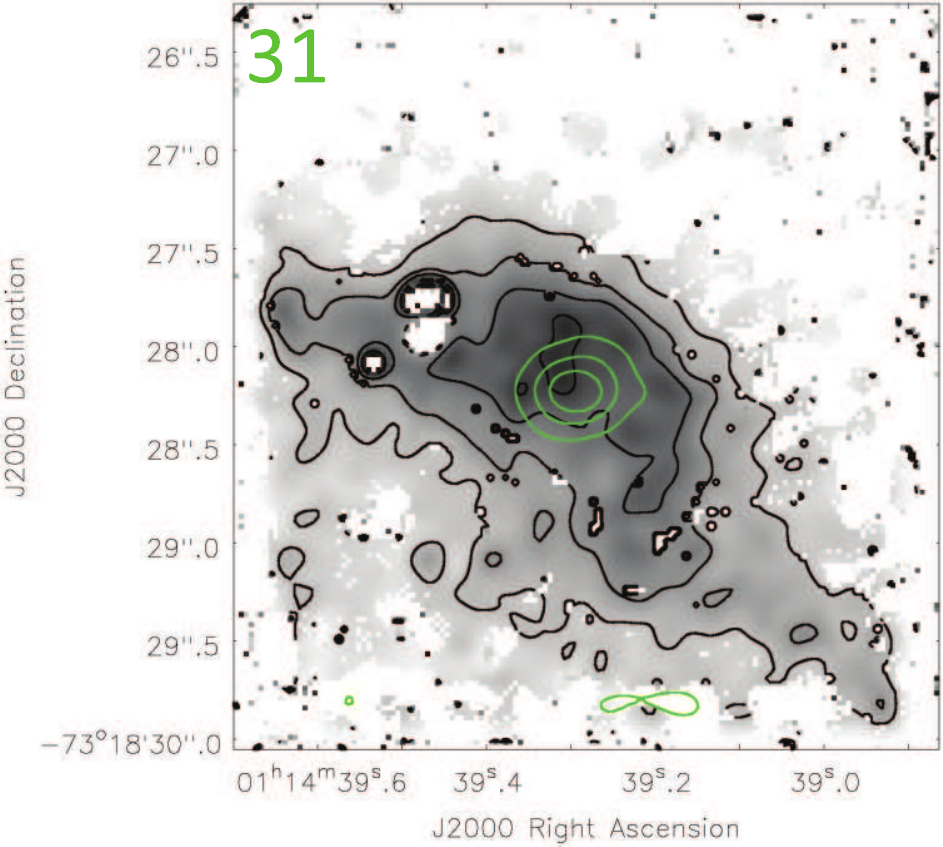}
\includegraphics[width=0.32\linewidth]{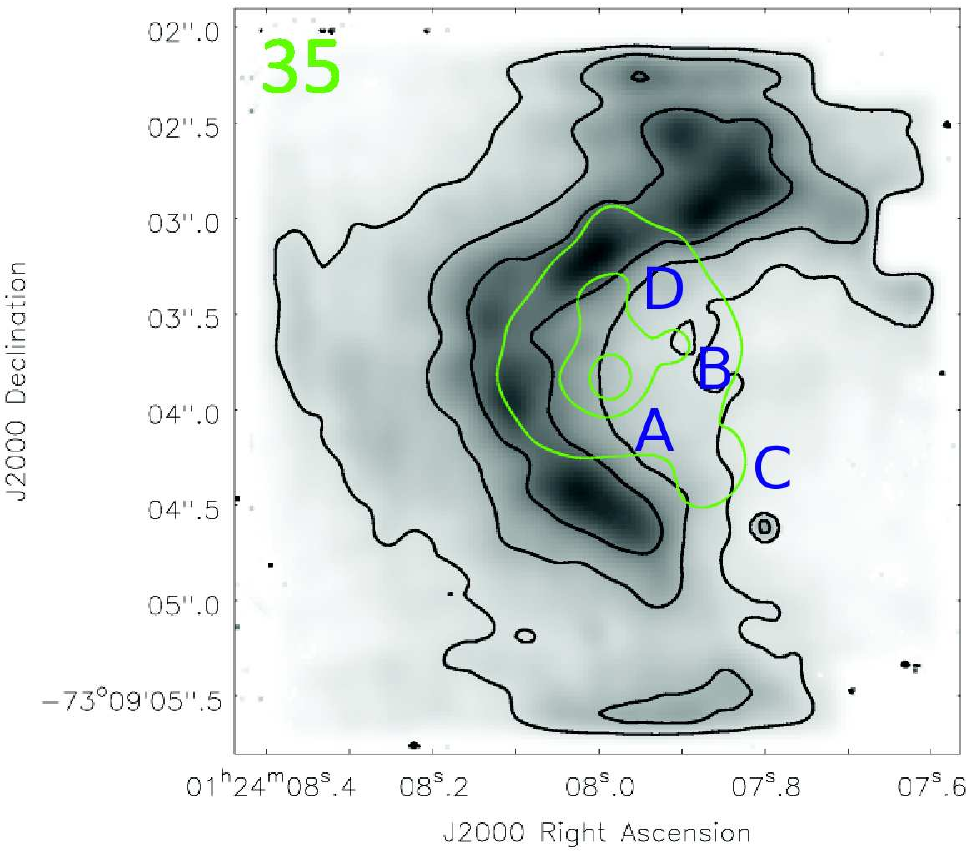}
\includegraphics[width=0.32\linewidth]{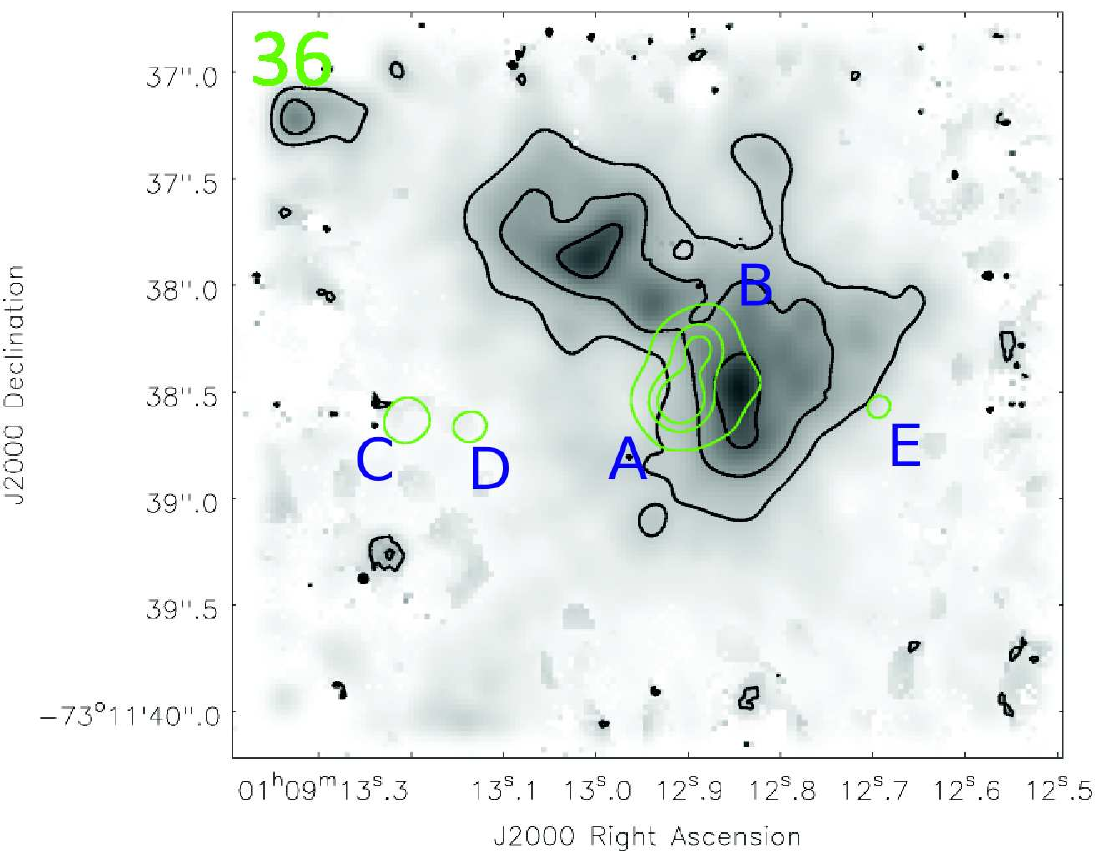}
   \caption{H$_{2}$ emission line morphologies for sources: \#01, 02, 03, 06, 22, 28, 30, 31, 35 and 36. 
Black contours - [0.2,0.4,0.6,0.8] $\times$ maximum H$_2$ 2.1218\,$\mu$m integrated flux, green contours - continuum flux. The continuum contour levels are as in Fig. 3.}
  \end{center}

 \end{minipage}

\end{figure*}

\begin{figure*}
 \begin{minipage}{175mm}
  \begin{center}
\includegraphics[width=0.48\linewidth]{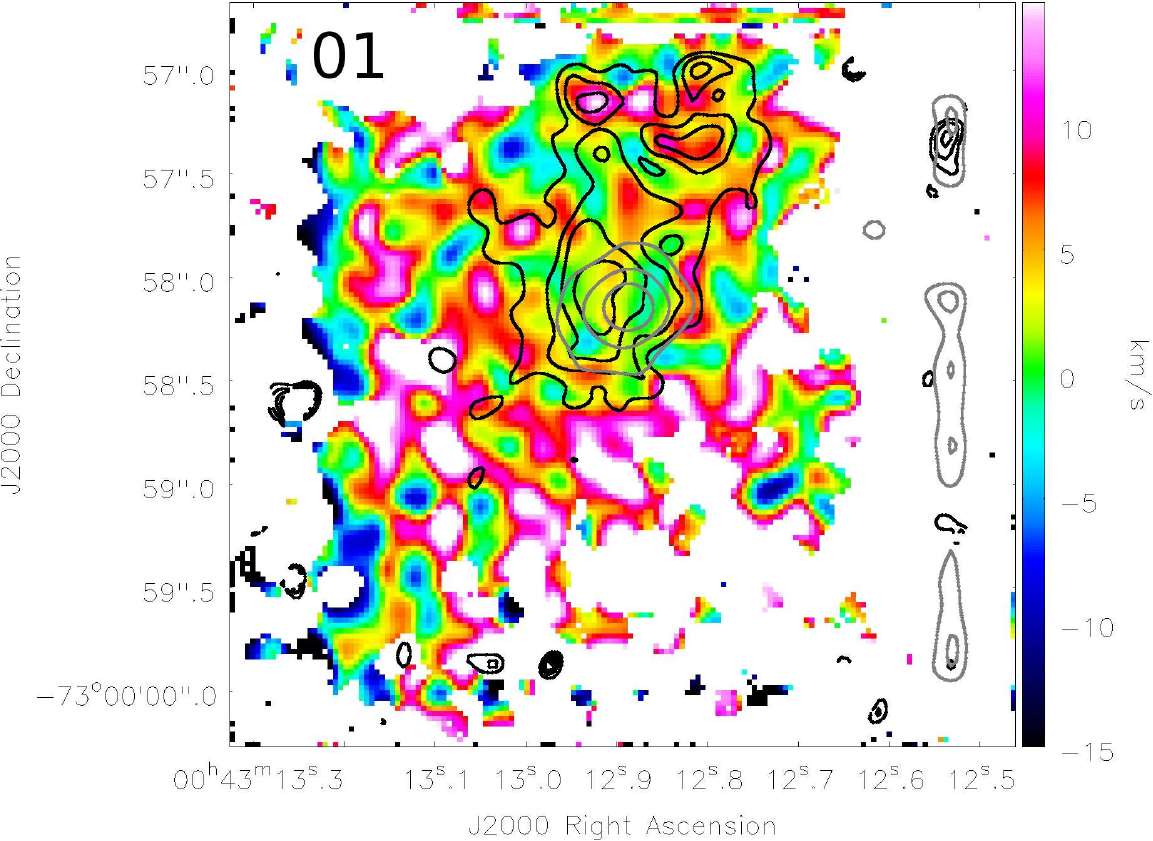}
\includegraphics[width=0.48\linewidth]{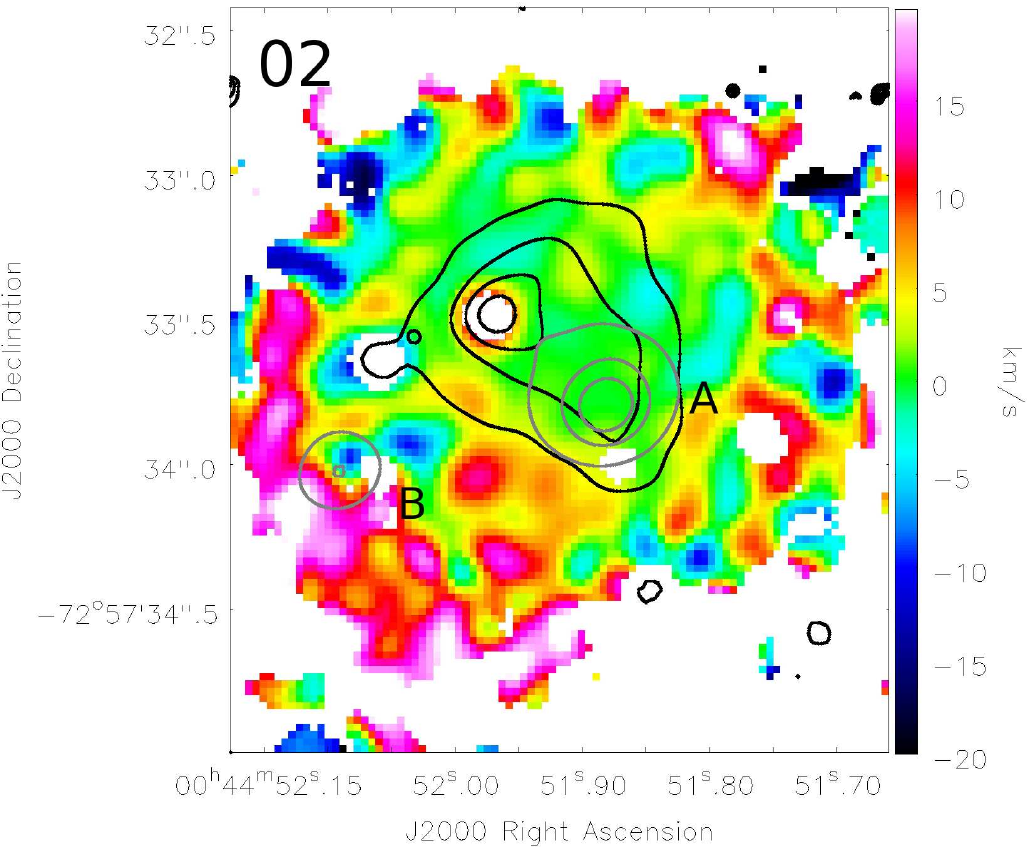}
\includegraphics[width=0.48\linewidth]{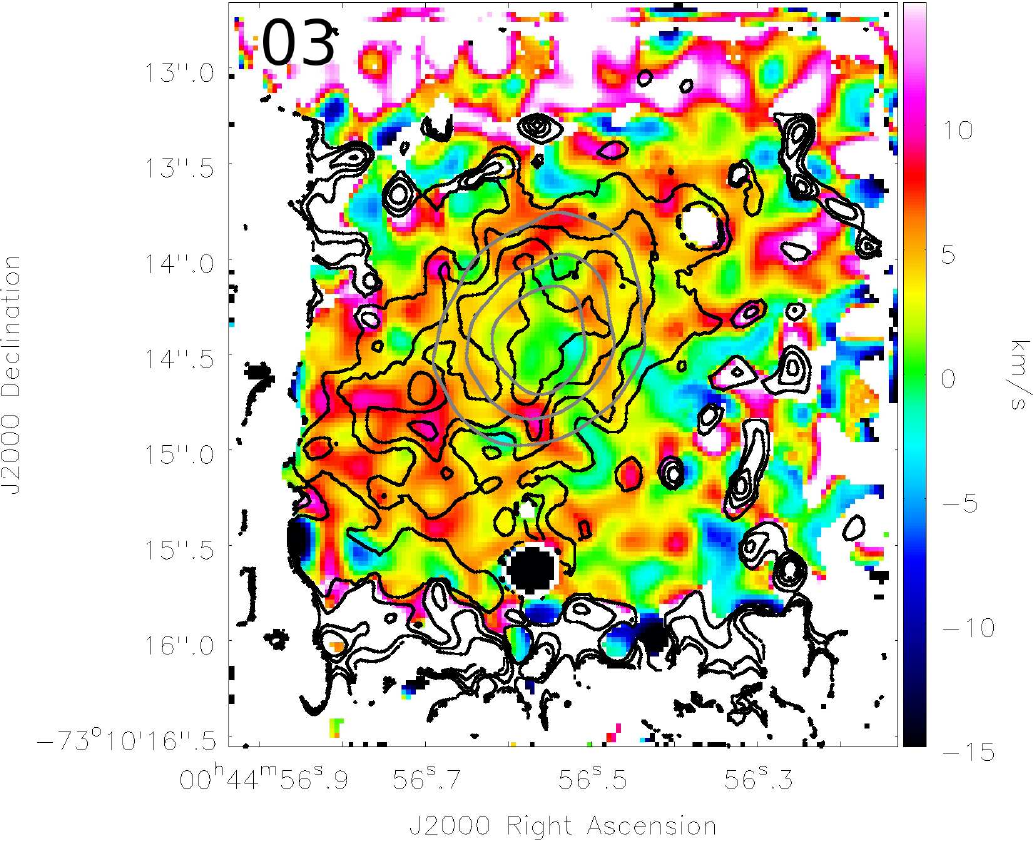}
\includegraphics[width=0.48\linewidth]{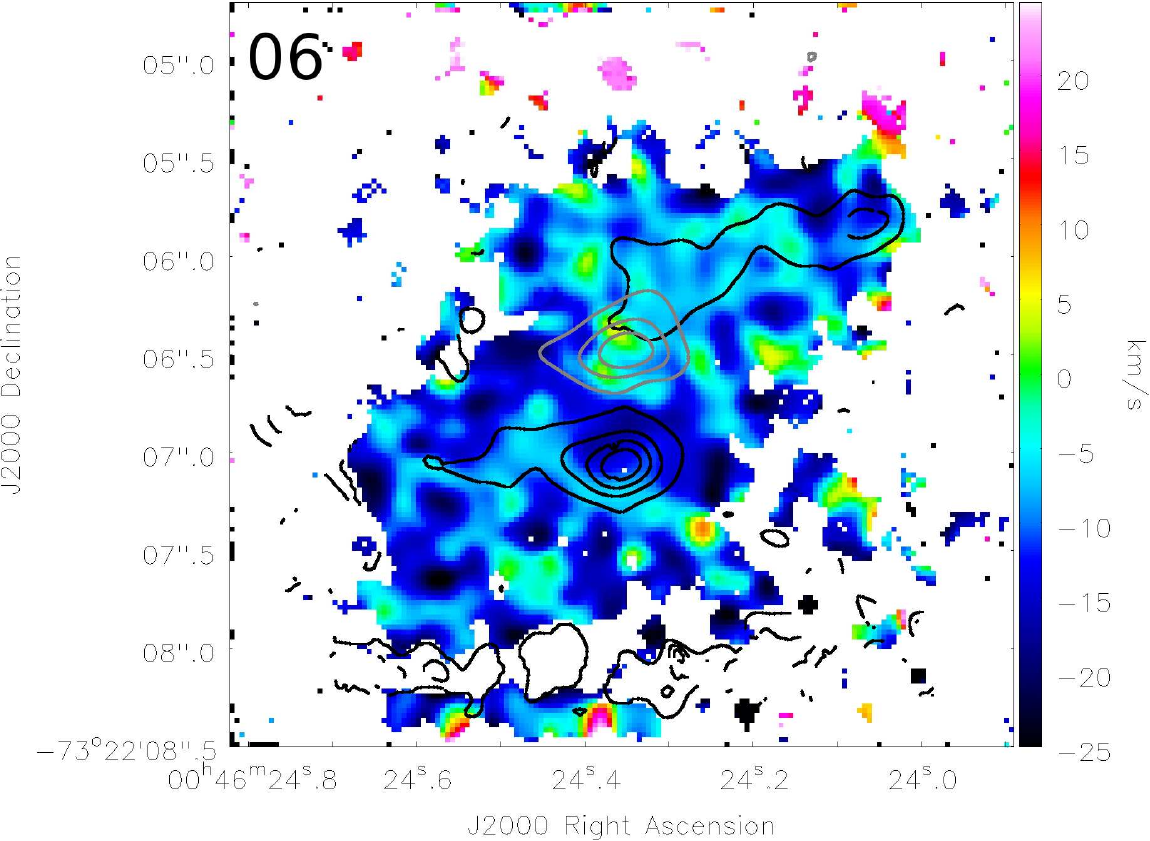}
\includegraphics[width=0.48\linewidth]{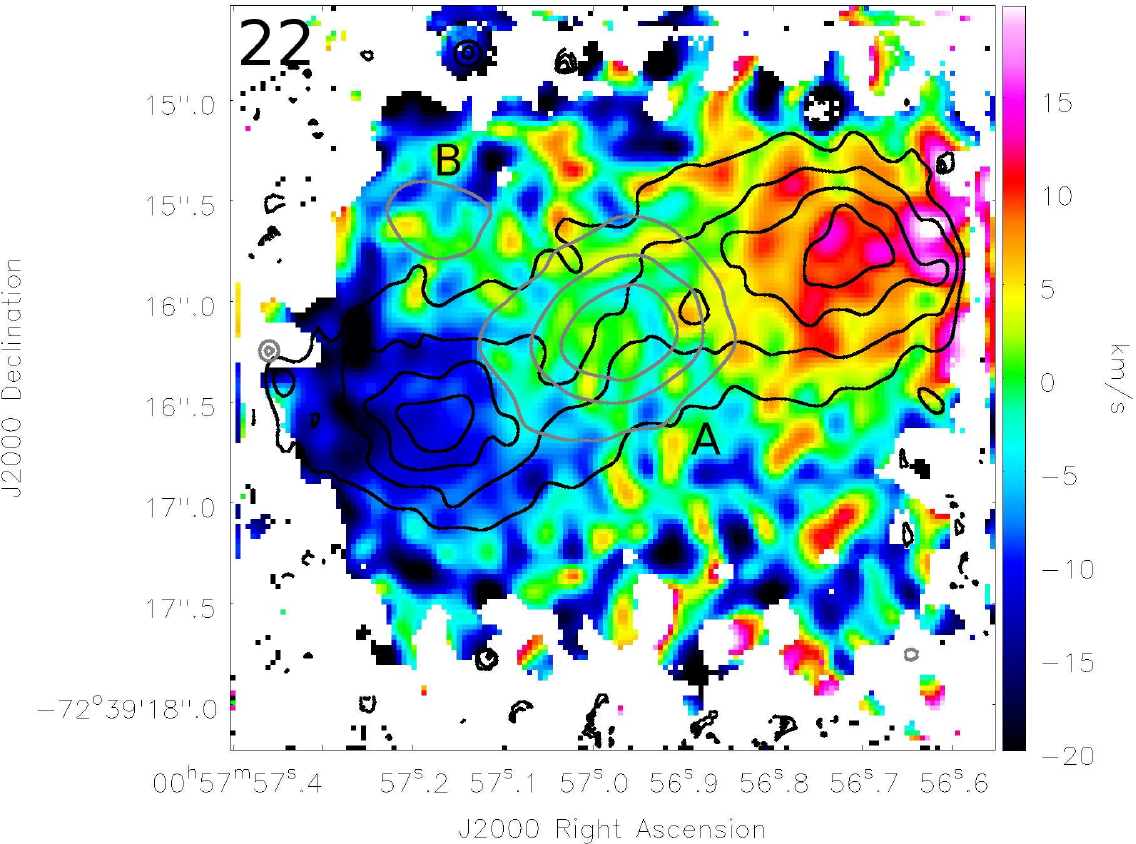}
\includegraphics[width=0.48\linewidth]{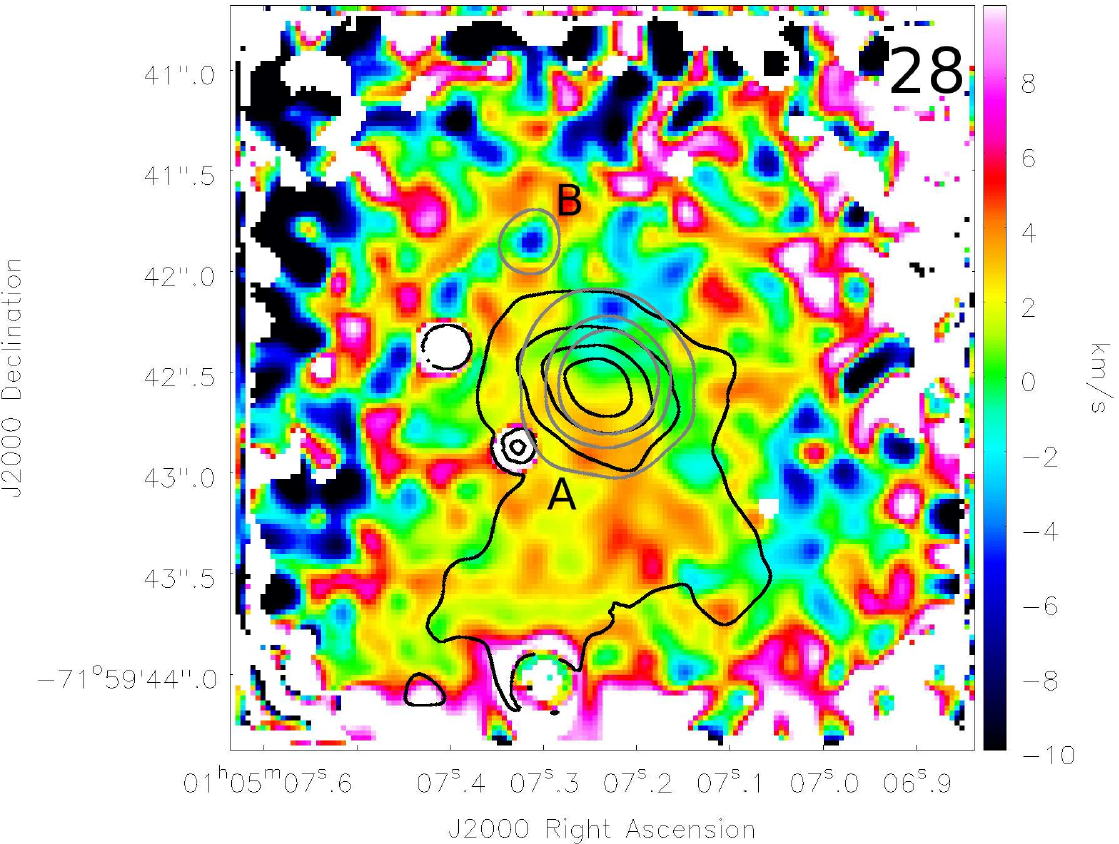}
   \caption{H$_{2}$ emission line velocity maps for sources \#01, 02, 03, 06, 22, 28. 
Black contours - [0.2,0.4,0.6,0.8] $\times$ maximum H$_2$ 2.1218\,$\mu$m integrated flux, grey contours - [0.25,0.5,0.75] $\times$ maximum continuum flux.}
  \end{center}
 \end{minipage}
\end{figure*}
\begin{figure*}
 \begin{minipage}{175mm}
  \begin{center}
\includegraphics[width=0.48\linewidth]{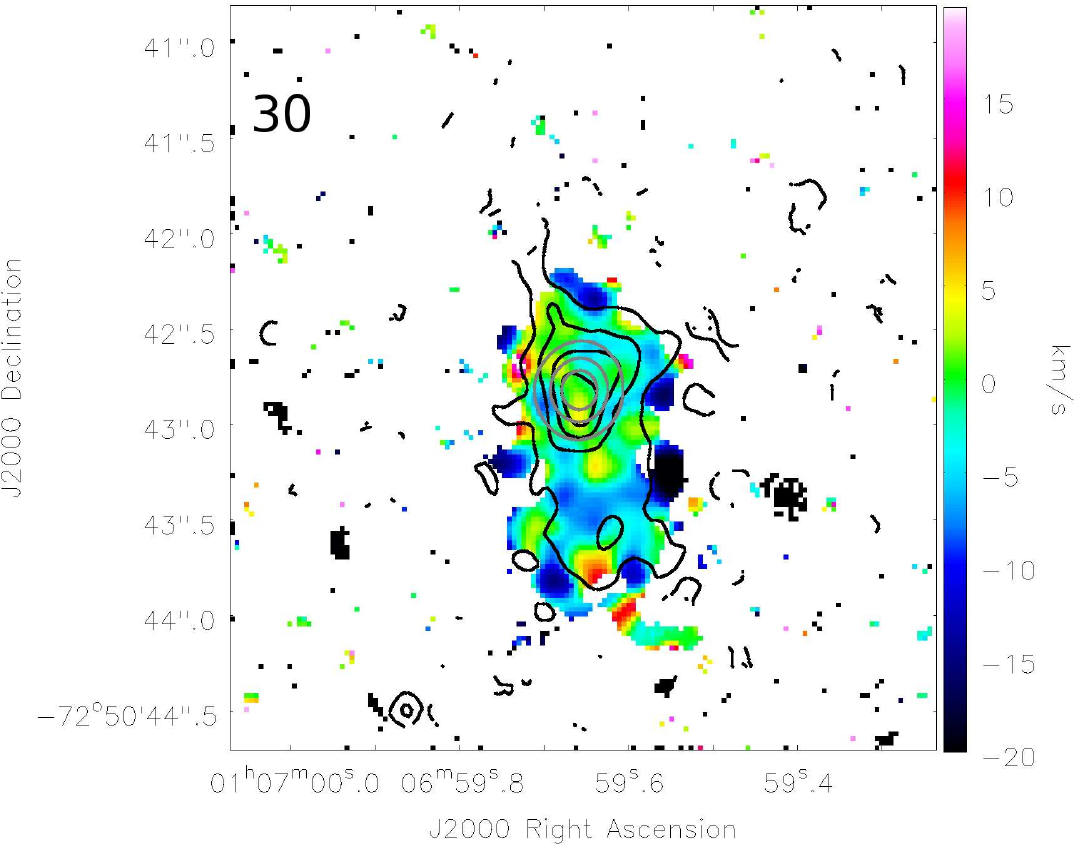}
\includegraphics[width=0.48\linewidth]{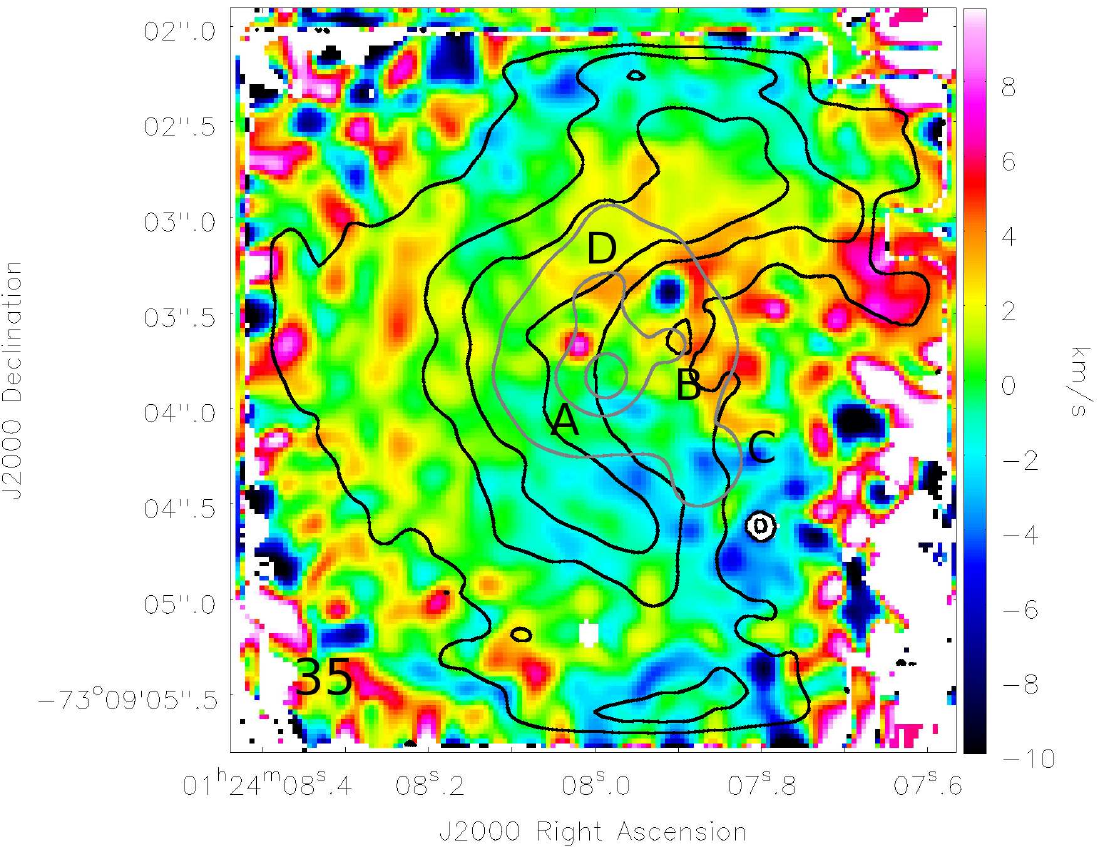}
\includegraphics[width=0.48\linewidth]{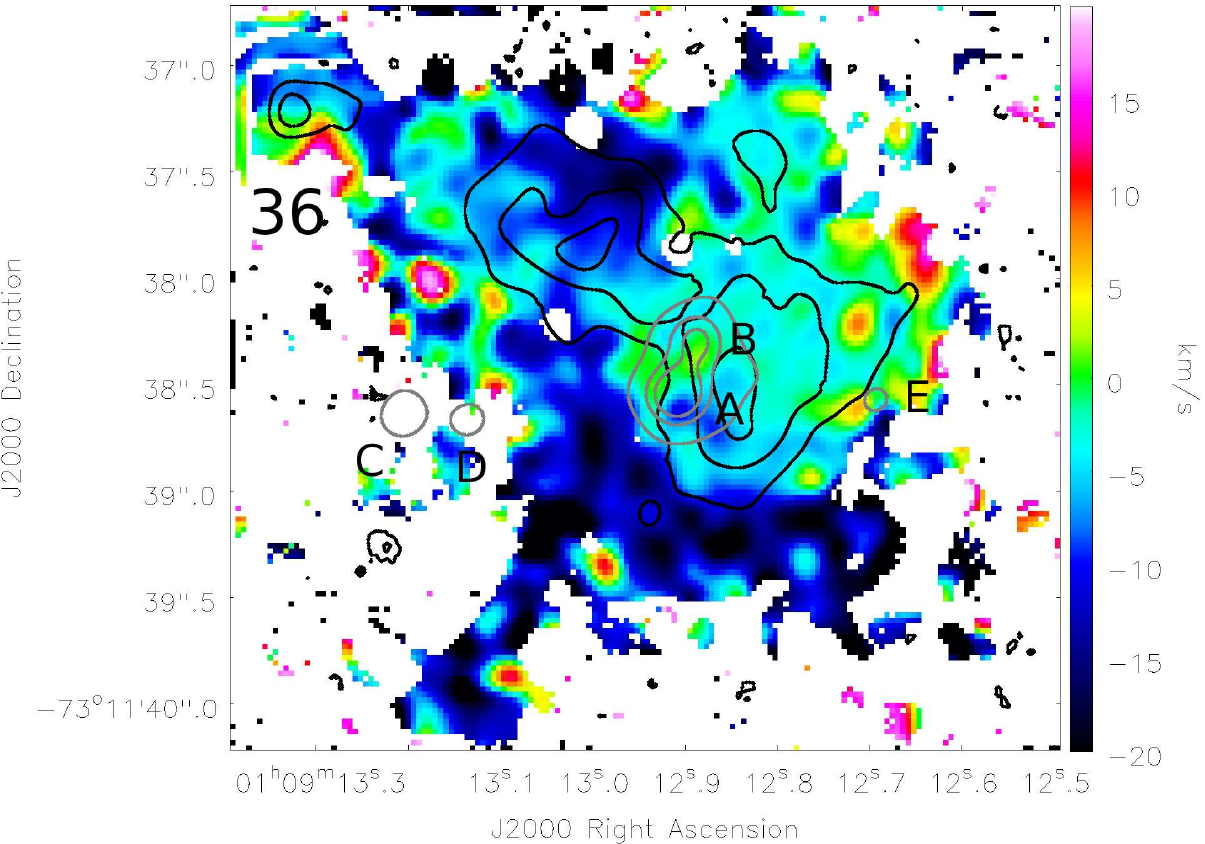}
  \end{center}
\textbf{Fig 10 cont.} H$_{2}$ emission line velocity maps for sources \#30, 35, 36. 
 \end{minipage}
\end{figure*}

The centroid velocity fields of the extended H$_2$ emission are shown in Fig. 10 with the exception of source \#31 for which the 
S/N was not sufficient to measure the centroid position at each spaxel.
The H$_2$ emission detected towards sources \#01, 03 and 28 all appears to be slightly red-shifted with respect to the continuum source whilst source \#02 appears to 
exhibit no significant velocity gradient. 
The H$_2$ emission in sources \#06 and 30 appear to be blue shifted relative to the continuum source.
Sources \#22\,A and 36\,B both appear to be at the centre of bipolar outflow 
structures and the velocity gradients across the length of the structures certainly support this. Finally in source \#35 (N88\,A) there appears to be a small
velocity gradient from south to north.

\subsubsection{CO bandhead emission/absorption}

The CO bandhead emission red-wards of 2.29\,$\mu$m in the \textit{K}-band is widely associated
with accretion discs in YSOs 
\citep{Davies2010, Wheelwright2010}.
Towards our targets in the SMC, only source \#03 exhibits detectable CO bandhead emission as shown in Fig. 11.
This emission is weak even for the v = 2--0 transition at $\sim$2.295\,$\mu$m which is not contaminated by low-$J$ CO absorption lines between 2.32 and 2.38\,$\mu$m.
Whilst the detection of CO bandhead emission originating in discs is strongly dependent on geometry \citep{Kraus2000, Barbosa2003},
the detection rate towards the SMC sources presented in this work is significantly lower than the 17\% of \citet{Cooper2013} and
is therefore suggestive of a physical difference between this sample and that of \citet{Cooper2013}. 

When we restrict our comparison to only the range of bolometric luminosities of our SMC targets (1.5$\times$10$^{3}$--1.7$\times$10$^{5}$ L$_{\sun}$)
and exclude any sources from \citet{Cooper2013} which exhibit P-cygni line profiles (as we have not observed any in the SMC), we find the CO bandhead
detection rate of \citet{Cooper2013} drops to 15\%. However this is still significantly higher than the detection rate found towards sources in the SMC
(5\%).

\begin{figure*}
\begin{minipage}{175mm}
\begin{center}
 \includegraphics[width=0.75\linewidth]{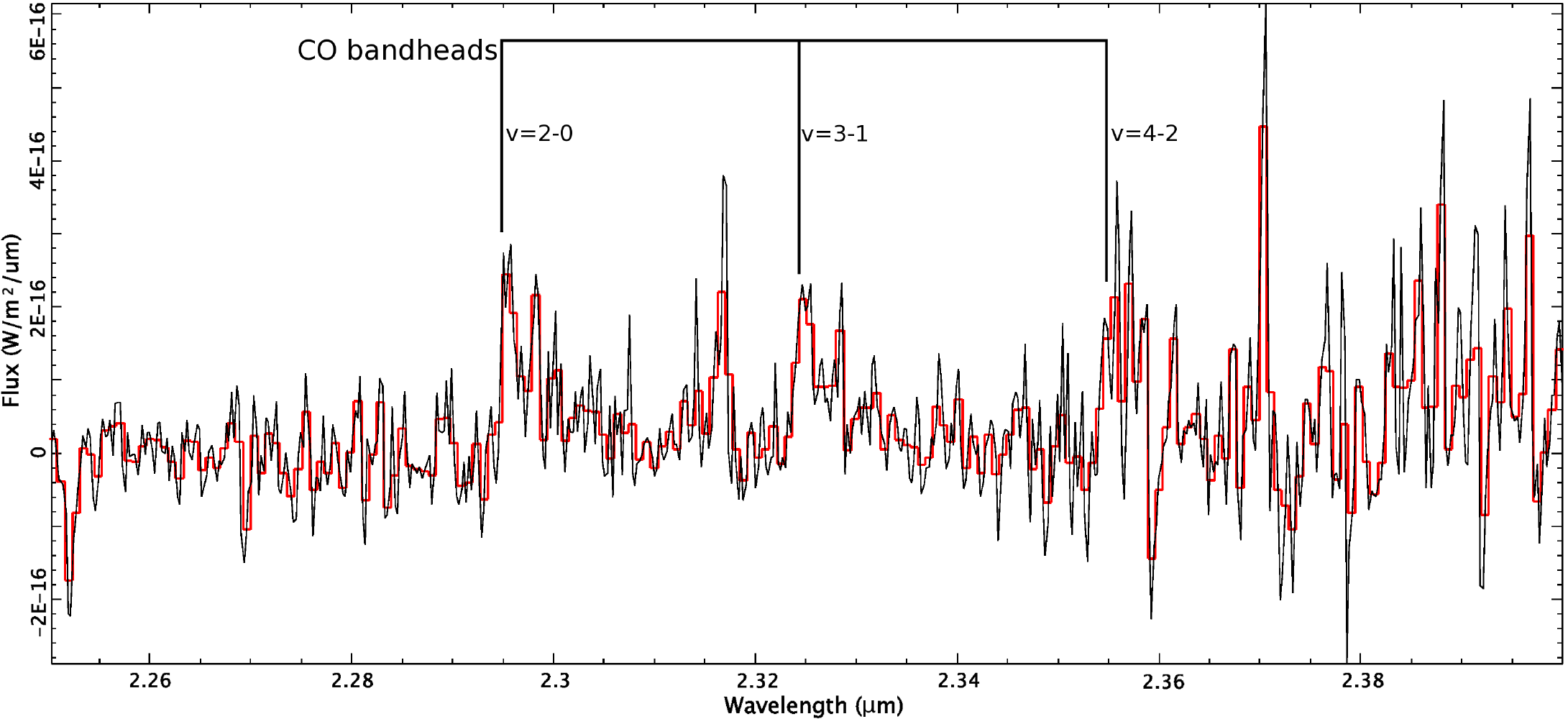}\\
 \includegraphics[width=0.75\linewidth]{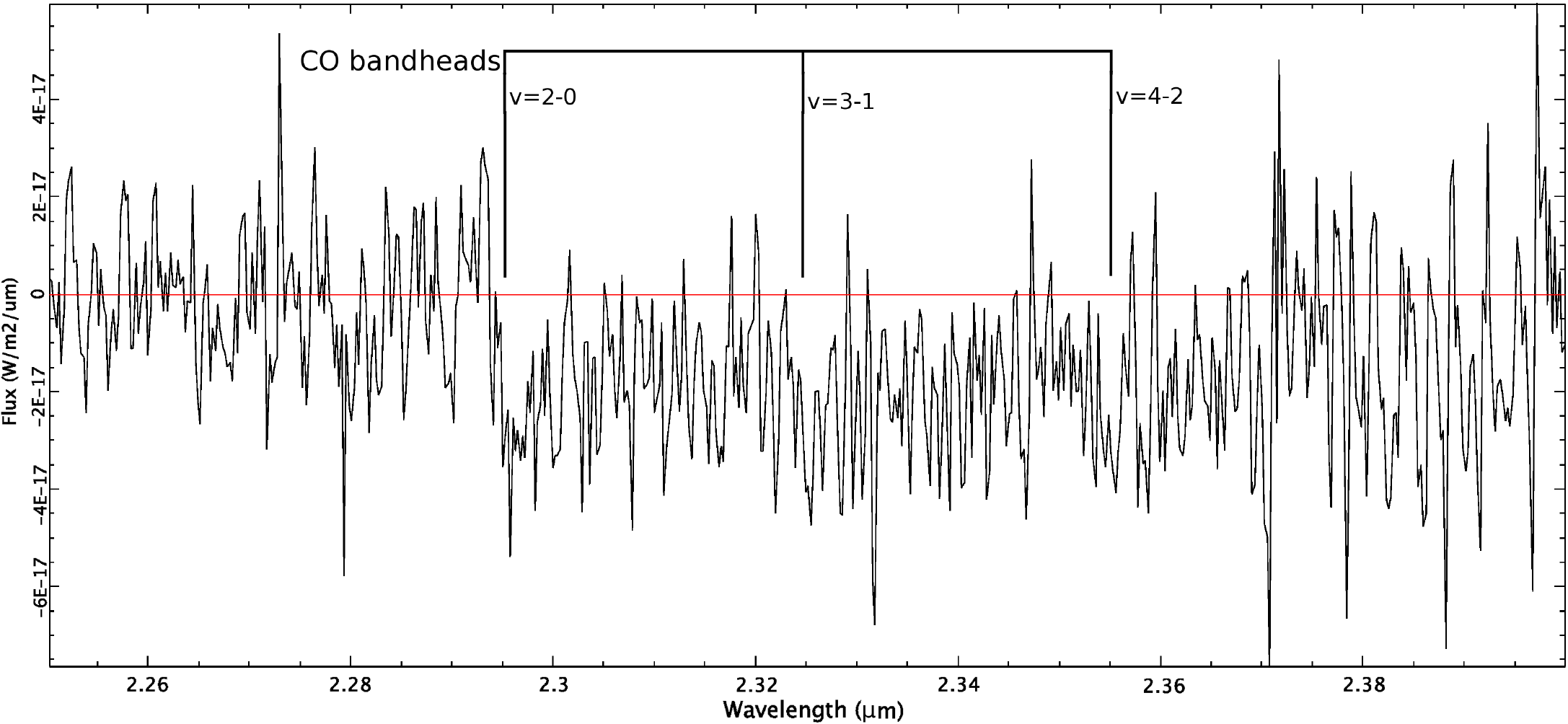}
\end{center}
\caption{Upper panel: The region of the continuum subtracted spectrum of source \#03 which contains CO bandhead emission. The 3-1 and 4-2 bandheads appear to be
contaminated by CO absorption lines.
Lower panel: Continuum subtracted spectrum of source \#02\,B exhibiting CO absorption features. The positions of the CO bandheads are marked on both spectra.}
\end{minipage}
\end{figure*}

The spectrum of source \#02\,B appears to exhibit CO in absorption (see lower panel of Fig. 11). Whilst uncommon in massive YSOs, CO absorption is 
commonly associated with lower mass YSOs (e.g. \citealt{Casali1996}). 
Although the S/N is very poor, using the blue edge of the v=2--0 bandhead, we estimate that the measured velocity towards the CO bandhead falls in the range 60--200 km s$^{-1}$.
This is consistent with the velocity measurements made towards source \#02\,A (see Section 4.4 and Table 4).
This detection is discussed in more detail in Section 6.3.1.

\subsection{Kinematics of molecular, atomic and excited hydrogen emission}

The Gaussian profiles used to fit the emission lines towards the continuum sources in this work yield centroid wavelengths which are converted into velocities. 
The heliocentric centroid velocities of the measured Br$\gamma$ and 
H$_2$ 2.1218\,$\mu$m emission lines (with the flexure wavelength shift component subtracted; see Section 3) are shown in Table 4.
Also shown in Table 4 are the most likely corresponding heliocentric velocities measured from the H\,{\sc i} column density data cube of the SMC 
\citep{Stanimirovic1999}. Apertures of 2 arcmin radius centred on the coordinates of each of the
\textit{Spitzer} YSOs were used to measure the H\,{\sc i} 21 cm velocity towards each source.

\begin{table}
 \caption{Emission line centroid velocities for Br$\gamma$ and  H$_2$ 2.1218\,$\mu$m emission.
Also included are the local H\,{\sc i} 21\,cm line velocities closest to those velocities measured with SINFONI.
Where it is unclear which 21\,cm velocity component is associated with the \textit{K}-band emission, the closest two components are
listed.
The H\,{\sc i} 21\,cm data is from \citet{Stanimirovic1999}, measured from a radius of 2$^{\prime}$ surrounding the YSO source.}
\small
\begin{tabular}{c  c c c}
\hline
& \multicolumn{3}{c}{Centroid velocity (km s$^{-1}$)} \\
\# & Br$\gamma$ & H$_2$ 2.12\,$\mu$m & H\,{\sc i} 21\,cm \\
\hline
01 & 139$\pm$8 & 133$\pm$8 & 130$\pm$1 \\
02\,A & 152$\pm$7 & 162$\pm$7 & 151.3$\pm$0.2\\
02\,B & & 166$\pm$7 &  151.3$\pm$0.2\\
03 & 156$\pm$5 & 143$\pm$5 & 167.7$\pm$0.1\\
04 & 137$\pm$5 & & 133.4$\pm$0.1 \\
06 & 124$\pm$13 & 152$\pm$6 & 130.4$\pm$0.6, 159.6$\pm$0.4\\
17 & 171$\pm$6 & 179$\pm$5 & 172.0$\pm$0.1\\
18 & 170$\pm$10 & 183$\pm$8 & 171.1$\pm$0.2\\
20 & 165$\pm$6 & 168$\pm$7 & 170.0$\pm$0.3\\
22\,A & 182$\pm$8 & 187$\pm$6 & 175.2$\pm$0.3\\
22\,B & 191$\pm$8 & 182$\pm$6 & 175.2$\pm$0.3\\
25 & 167$\pm$8 & 174$\pm$7 & 166.8$\pm$0.6 \\
26 & 184$\pm$8 & 174$\pm$8 & 175.7$\pm$0.7\\
28\,A & 190$\pm$7 & 202$\pm$7 & 186.9$\pm$0.1\\
28\,B & 189$\pm$10 & 200$\pm$6 & 186.9$\pm$0.1\\
30 & 185$\pm$7 & 186$\pm$7 & 179.8$\pm$0.5\\
31 & 185$\pm$7 & 189$\pm$7 & 182.0$\pm$0.3 \\
32 & 136$\pm$8 & 145$\pm$8 & 140.8$\pm$0.1 \\
33 & 177$\pm$4 & 171$\pm$3 & 165.0$\pm$0.1, 196.2$\pm$0.3\\
34 & 205$\pm$8 & 205$\pm$7 & 196.9$\pm$0.4\\
35\,A & 156$\pm$7 & 166$\pm$8 & 161.0$\pm$0.4\\
35\,B & 155$\pm$7 & 168$\pm$9 & 161.0$\pm$0.4\\
35\,C & 153$\pm$7 & 162$\pm$8 & 161.0$\pm$0.4\\
35\,D & 154$\pm$7 & 167$\pm$8 & 161.0$\pm$0.4\\
36\,A & 167$\pm$8 & 180$\pm$9 & 170.0$\pm$0.3\\
36\,B & 166$\pm$8 & 180$\pm$9 & 170.0$\pm$0.3\\
36\,C & 166$\pm$8 & 219$\pm$13 & 170.0$\pm$0.3\\
36\,D & 167$\pm$8 & 184$\pm$9 & 170.0$\pm$0.3\\
36\,E & 167$\pm$9 & 183$\pm$8 & 170.0$\pm$0.3\\
\hline
\end{tabular}

\end{table}

Figure 12 shows the Br$\gamma$ velocity plotted against H\,{\sc i} 21 cm velocity and H$_2$ 2.1218\,$\mu$m velocity against 
Br$\gamma$ velocity. For all but five cases (sources \#01, 03, 22\,B, 26 and 33) the atomic hydrogen emission velocities agree within 
1$\sigma$ uncertainty. All but two of the H$_2$ emission velocities are consistent with the Br$\gamma$ velocities within
2$\sigma$; however, on the whole the H$_2$ emission does appear to be slightly red-shifted with respect to the Br$\gamma$ emission.
The linear fits to the data have the equations v$_{\text{Br}\gamma} = 1.04\text{v}_{21\text{cm}} -5 $ km s$^{-1}$ and
v$_{\text{H}_{2}} = 0.80\text{v}_{\text{Br}\gamma} + 43$ km s$^{-1}$. This indicates a strong correlation of both the excited atomic 
gas velocities and the molecular gas velocities with the bulk motions of atomic gas in the ISM of the SMC.

\begin{figure*}
\begin{minipage}{175mm}
 \includegraphics[width=0.48\linewidth]{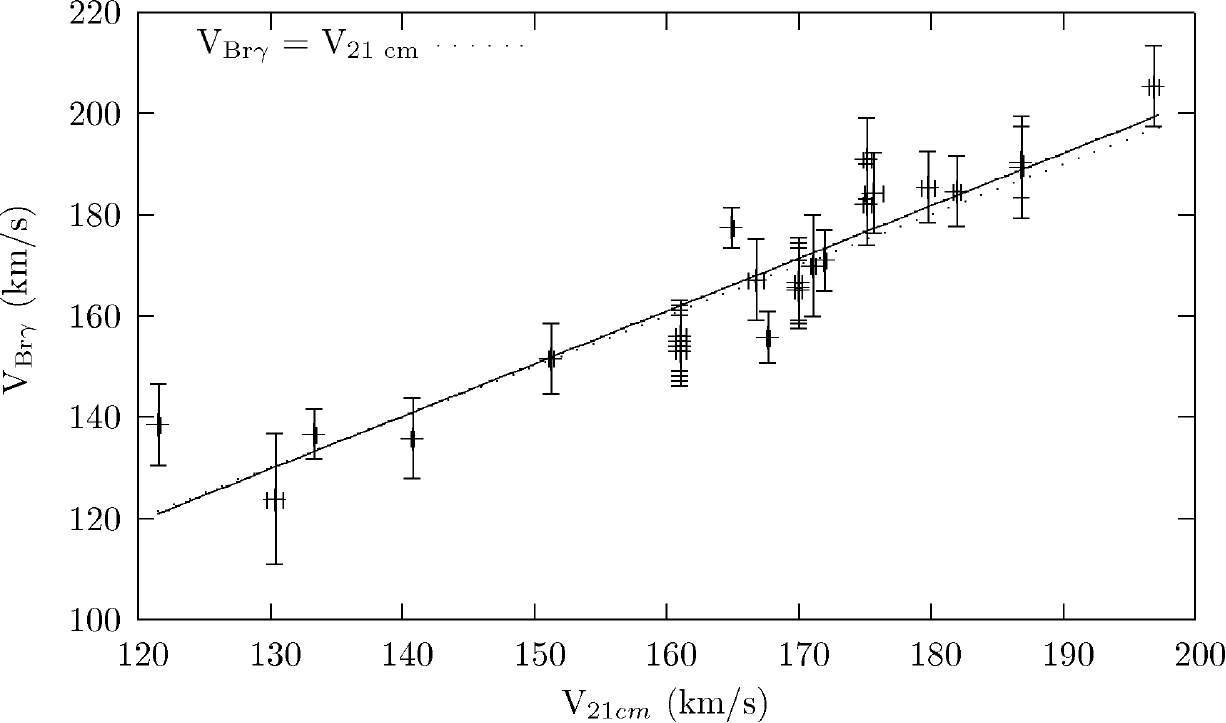}
 \includegraphics[width=0.48\linewidth]{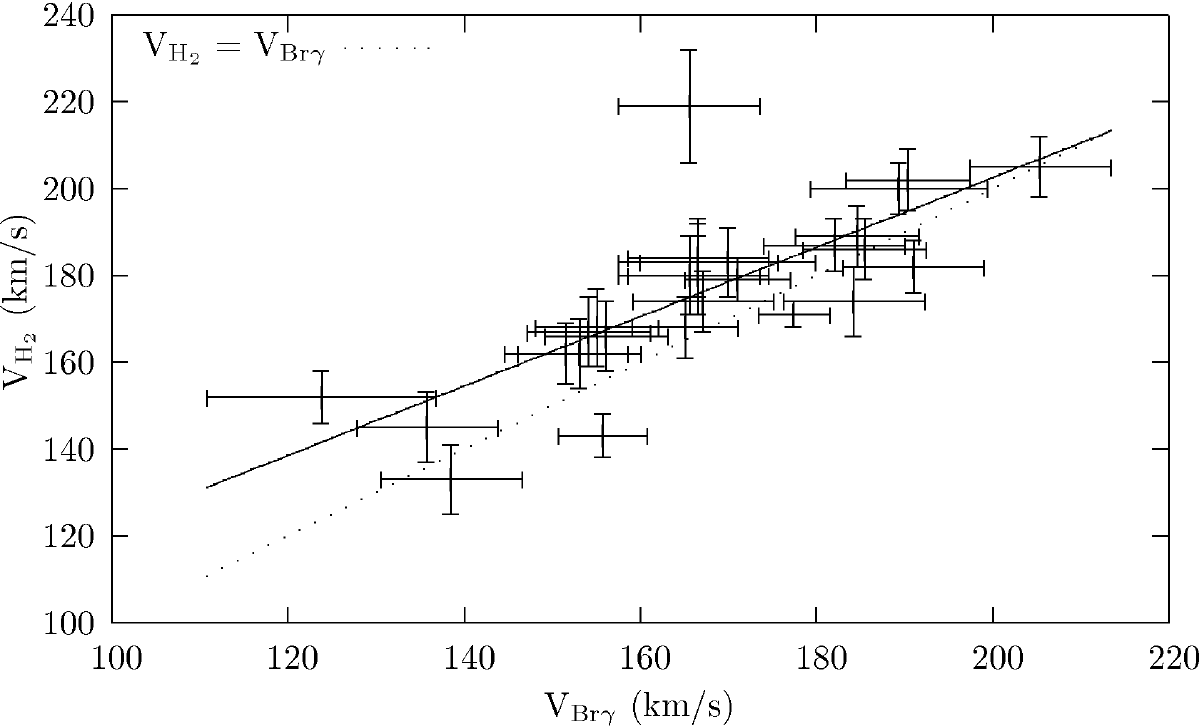}
\caption{Left: Br$\gamma$ centroid velocity against H\,{\sc i} velocities as measured in data cube from \citet{Stanimirovic1999}.
H\,{\sc i} 21 cm centroid velocity taken from an aperture of 2\,arcmin centred on the YSO position.
Right: H$_2$ 2.1218\,$\mu$m emission centroid velocity against Br$\gamma$ centroid velocity for all sources for which both can be measured.
The solid lines represent a linear fit to the data whilst the dotted lines show equal velocities.}
\end{minipage}
\end{figure*}

\section{Optical spectra}

Optical spectra have been previously presented for all but one (\#31) of these sources in \citet{Oliveira2013} and we will analyse these in more detail here. 
We also present the results of the new RSS spectroscopy carried out towards sources \#35 and 36 (see Section 3.2). Note that for seven sources
with optical spectra, SINFONI observations are not available.

\subsection{Extinction towards optical emission}

Although extinction significantly impacts optical observations, it also
gives an indication of where the emission originates. This is a particularly important consideration for massive YSOs because they are embedded objects
and it is therefore unexpected for them to be observable in the optical regime. 

We calculate values of extinction using H\,{\sc i} emission lines comparing the attenuated ratio
 $[H_{\alpha}/H_{\beta}]_{a}$, with the expected intrinsic ratio, $[H_{\alpha}/H_{\beta}]_{i}$;
\begin{equation}
 E(B-V) = \frac{\log[H_{\alpha} / H_{\beta}]_{a} - \log[H_{\alpha} / H_{\beta}]_{i}}{0.4 (K_{H\beta}-K_{H\beta})}
\end{equation}
where $K_{H\alpha}$ and $K_{H\beta}$ are 2.535 and 3.609, respectively \citep{Calzetti2001}. 
The Case B recombination intrinsic line ratio is 
normally $H_{\alpha}/H_{\beta} \approx 2.87$ at 10\,000\,K; however, this ratio can vary by 5--10\%  over a temperature range of
5000--20\,000 K \citep{Osterbrock2006}. For Case A recombination the ratio is 2.86 at 10\,000\,K \citep{Osterbrock2006} with a similar temperature dependent
variance making this ratio largely independent of whether the emission is optically thick or thin.
Assuming a Milky Way-like extinction curve with R=3.1, we obtain the extinction estimates shown in Table 5. We find that these are significantly lower than those
calculated from the \textit{K}-band emission (see Section 4.2) although we are only able to calculate extinction values using both methods for 9 of our targets. 
For these 9 targets, the 
mean extinction values are $A_V =$ 12.4$\pm$2.4\,mag in the \textit{K}-band whilst only $A_V =$ 1.0$\pm$0.3\,mag in the optical emission.
The median values for extinction obtained for these
9 sources are $A_V =$ 13.3 and 0.81 mag for the \textit{K}-band and optical measurements, respectively.
We use upper limits as $A_\text{V}$ values in the optical to provide an adequate sample
of extinctions derived from optical emission
whilst excluding sources with limits in the \textit{K-}band.
The disparity between the extinction values calculated for the optical and near-infrared emission
 suggests that the optical emission is in fact sampled in a much shallower region of the YSO environment rather than at the YSOs themselves and most 
likely close to the outer edges of the molecular clouds.

\begin{table}
\caption{Extinction values calculated from H{\sc i} line ratios. 
With the exception of source \#04 the sources without visual extinctions derived from \textit{K-}band measurements have not been
observed with SINFONI.}
 \begin{tabular}{c c c}
\hline
  source & A$_V$ (optical) & A$_V$ (\textit{K}-band)\\
\hline
1 & 1.51$\pm$0.05 & 2.6$\pm$10.9 \\
3 & 0.98$\pm$0.18 & 13.7$\pm$10.8 \\
4 & 3.10$\pm$0.18 & \\
7 & 1.95$\pm$0.57 & \\
8 & 1.15$\pm$0.21 & \\
9 & 0.54$\pm$0.13 & \\
12 & 0.91$\pm$0.99 & \\
13 & 2.65$\pm$0.28 & \\
15 & 0.86$\pm$0.19 & \\
20 & 2.47$\pm$1.16 & 24.1$\pm$13.2 \\
22 & $<$0.24 & 9.9$\pm$6.4 \\
25 & 0.47$\pm$0.18 & $<$12 \\
26 & $<$0.30 & 14.2$\pm$16.6 \\
28 & 0.81$\pm$0.35 & 19.5$\pm$9.3 \\
30 & 2.61$\pm$0.34 & 13$\substack{+	55	 \\ -	25	}$ \\
35 & $<$0.09 & 1.2$\pm$17.0 \\
36 & $<$0.08 & 13.3$\pm$12.8 \\
mean & 1.5$\pm$0.4 & 12.3$\pm$2.7 \\
\hline
 \end{tabular}
\end{table}

\subsection{Nature of emission}

An important consideration in the analysis of the optical emission lines we have measured is the physical process driving the emission, in particular whether the emission is 
photo-excited or shock excited. In order to constrain this we compare the ratios of emission lines measured in our data with those predicted by models.
To investigate the source of the optical emission 
we use the MAPPINGS III pre-run photoionization grids \citep{Kewley2001} and shock grids \citep{Allen2008} to constrain the source of the emission.
In Fig. 13 we plot $\log$([O\,{\sc iii}]/H$\beta$) against $\log$([S\,{\sc ii}]/H$\alpha$) (upper panel) and 
$\log$([O\,{\sc iii}]/H$\beta$) against $\log$([N\,{\sc ii}]/H$\alpha$) (lower panel) for all of the sources for which these emission lines could be measured,
as well as predictions from the photoionization and collisional excitation grids.
\,Also shown in both panels is the maximum starburst line from \citet{Kewley2001}, a theoretical limit above which photoionization alone is not sufficient to produce the 
observed emission.

\begin{figure*}
 \begin{minipage}{175mm}
\begin{center}
  \includegraphics[width=0.99\linewidth]{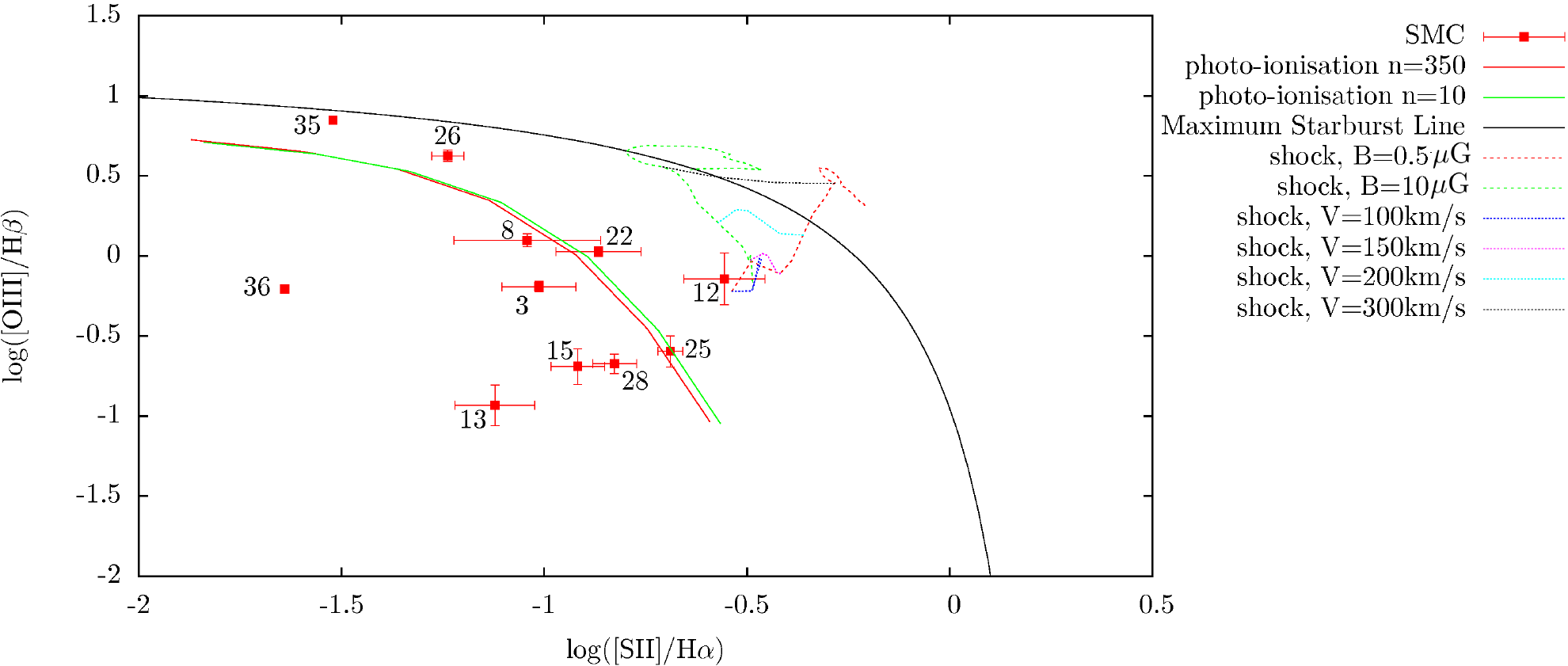}
  \includegraphics[width=0.99\linewidth]{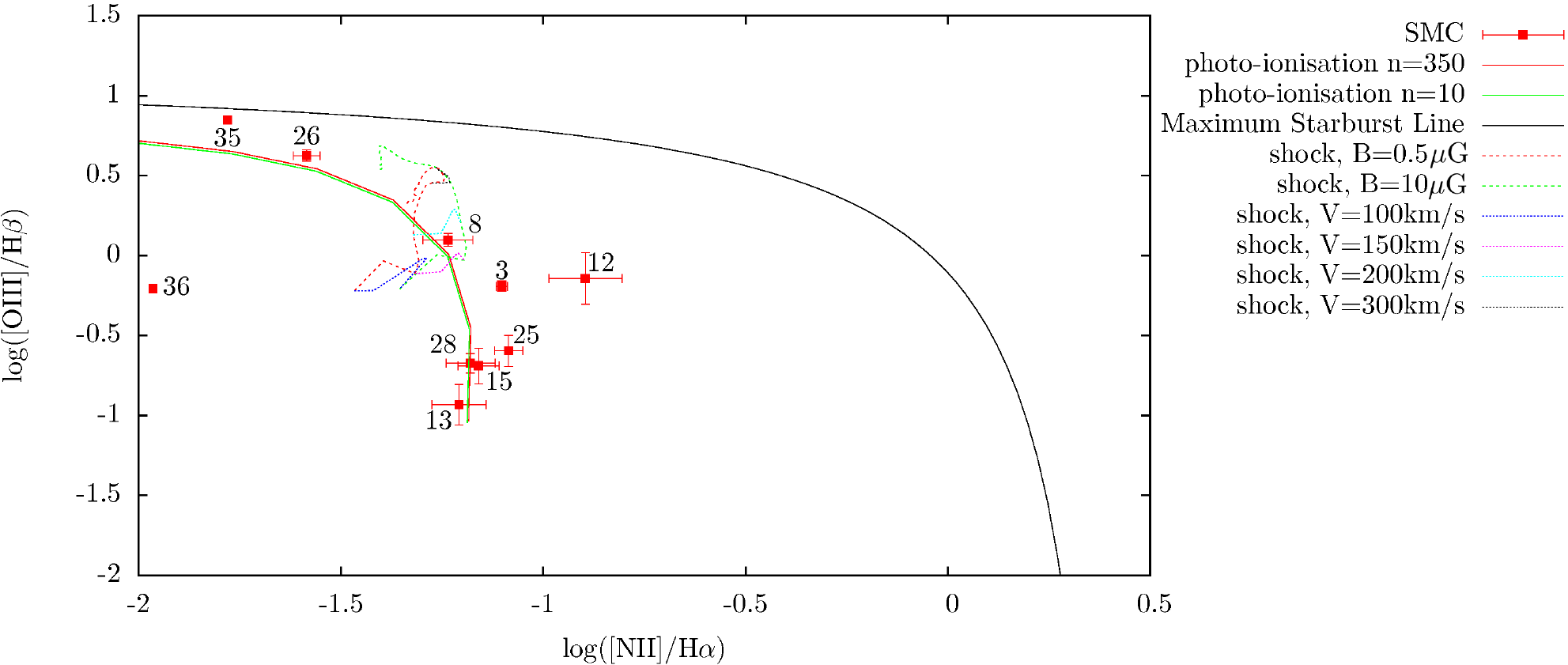}
 \end{center}
\caption{Upper panel: 
log([O\,{\sc iii}]/H$\beta$) vs log([S\,{\sc ii}]/H$\alpha$);
lower panel: log([O\,{\sc iii}/H$\beta$) vs log([N\,{\sc ii}]/H$\alpha$).
Also shown are the MAPPINGS III photoionization grids \citep{Kewley2001} and shock excitation 
grids \citep{Allen2008} and the maximum starburst line (see text, \citealt{Kewley2001}).
\,A metallicity of 0.2 Z$_{\sun}$ was adopted for 
the model grids.
} 
 \end{minipage}

\end{figure*}

We find that the optical emission is not consistent with shocked excitation towards any of our sources.
The majority of the sources
fall close to the photoionization models with some scatter.
It seems reasonable to assert that, as the emission is caused by photo-excitation and the optical extinctions are significantly lower than 
those measured in the \textit{K}-band, the optical emission arises at a significant distance from the SINFONI continuum sources, in a region 
closer to the surface of the molecular cloud. It is therefore also reasonable to assume that the optical emission may have
contributions from stars which are outside of our SINFONI FOVs.

\subsection{{\sc cloudy} optimization}

Where multiple emission lines were detected in the optical spectra we have used the \textsc{cloudy} photoionization code to fit parameters
using the Tlusty grid of OB-type stellar atmosphere models and assuming abundances of \citet{Dufour1984} where available and a metallicity of 0.2 Z$_{\sun}$
for species where measured abundances are not available. 
When interpreting these results it should be noted that the elemental abundances of the cloud used in the {\sc cloudy} line optimisation
process have a significant impact on the results of the analysis.

For the SALT spectra
we have treated the red and blue spectral regions separately, using the pg1800 grating observations for N88\,A in order to resolve the [N\,{\sc ii}]
emission lines at 6548 \AA{} and 6584 \AA{}
which are blended with the broad H$\alpha$ emission line in the pg900 grating spectrum.
The resulting stellar temperatures and hydrogen densities of the {\sc cloudy} line optimization are shown in Table 6. Also given are the values of the average deviation between the observed
extinction corrected line flux relative to H$\beta$ (or H$\alpha$ where H$\beta$ is not used as an input) and the model line flux. This is expressed as
a multiple of the uncertainty in the observed relative line flux and thus a number significantly less than one is desirable. The penultimate column shows the 
average line flux as a multiple of the uncertainty in that line flux and hence values greater than three represent 3$\sigma$ measurements.

\begin{table*}
\begin{minipage}{175mm}
 \caption{Results of the \textsc{cloudy} emission line optimization for optical spectra based on the Tlusty grid of OB star model atmospheres \citep{Lanz2003, Lanz2007}.
Shown are the emission lines used as inputs and the output stellar temperature and cloud density.
Also shown are values representing the variation between the model emission line fluxes and the observed fluxes and the average line flux for 
each source (\textquoteleft avg $\lvert$mod--obs$\rvert$\textquoteright\, and \textquoteleft avg flux\textquoteright\,, respectively), 
both expressed as multiples of the uncertainty ($\sigma$). 
The final column gives the spectral type based on ZAMS stars of equivalent effective temperature from \citet{Hanson1997}. 
Sources \#07--21 were not observed with SINFONI.
}
\setlength{\tabcolsep}{4pt}
\begin{tabular}{c c c c c c c}
\hline
Source & emission lines used & $\log$($T_{\text{eff}}$) & $\log$($n_{\text{H}}$) & avg $\lvert$mod--obs$\rvert$ & avg flux  & Spectral \\
& & (K) & (cm$^{-3}$) & ($\sigma$) & ($\sigma$) & type \\
\hline
01& H$\alpha$, H$\beta$, [S\,{\sc ii}]6716,6731, [N\,{\sc ii}]6548,6584  & 4.511 & 3.286 & 1.921 & 11.736 & B0 \\
03& H$\alpha$, H$\beta$, [S\,{\sc ii}]6716,6731, [N\,{\sc ii}]6548,6584, [O\,{\sc iii}]4959,5007 & 4.505 & 3.162 & 2.234 & 3.329 & B0 \\
04& H$\alpha$, H$\beta$, [S\,{\sc ii}]6716,6731 & 4.495 & 2.670 & 0.540 & 3.425 &  B0 \\
07& H$\alpha$, H$\beta$, [S\,{\sc ii}]6716,6731, [N\,{\sc ii}]6584 & 4.517 & 2.586 & 0.143 & 0.964 & B0 / O9.5 \\
08& H$\alpha$, H$\beta$, [S\,{\sc ii}]6716,6731, [N\,{\sc ii}]6548,6584, [O\,{\sc iii}]4959,5007 & 4.539 & 2.369 & 0.575 & 2.477 & O9.5  \\
09& H$\alpha$, H$\beta$, [S\,{\sc ii}]6716,6731, [N\,{\sc ii}]6548,6584 & 4.513 & 3.029 & 0.958 & 4.631 & B0 \\
12& H$\alpha$, H$\beta$, [S\,{\sc ii}]6716,6731, [N\,{\sc ii}]6584, [O\,{\sc iii}]4959,5007 & 4.740 & 1.343 & 0.104 & 0.419 & $<$O3 \\
13& H$\alpha$, H$\beta$, [S\,{\sc ii}]6716,6731, [N\,{\sc ii}]6548,6584, [O\,{\sc iii}]4959,5007 & 4.521 & 0.106 & 0.544 & 1.981 & B0 / O9.5 \\
15& H$\alpha$, H$\beta$, [S\,{\sc ii}]6716,6731, [N\,{\sc ii}]6548,6584, [O\,{\sc iii}]4959,5007 & 4.533 & 0.998 & 0.658 & 2.829 &  O9.5 \\
21& H$\alpha$, H$\beta$, [S\,{\sc ii}]6716,6731 & 4.457 & 3.125 & 0.069 & 1.701 & B0.5 \\
22& H$\alpha$, H$\beta$, [S\,{\sc ii}]6716,6731, [O\,{\sc iii}]4959,5007 & 4.585 & 2.368 & 0.062 & 1.724 & O8 \\
25& H$\alpha$, H$\beta$, [S\,{\sc ii}]6716,6731, [N\,{\sc ii}]6548,6584, [O\,{\sc iii}]4959,5007 & 4.534 & 0.009 & 0.991 & 3.263 & O9.5 \\
26& H$\alpha$, H$\beta$, [S\,{\sc ii}]6716,6731, [N\,{\sc ii}]6584, [O\,{\sc iii}]4959,5007 & 4.551 & 1.171 & 0.445 & 1.935 & O9 \\
28& H$\alpha$, H$\beta$, [S\,{\sc ii}]6716,6731, [N\,{\sc ii}]6548,6584, [O\,{\sc iii}]4959,5007 & 4.531 & 2.098 & 0.407 & 1.665 & O9.5 \\
35 (N88\,A) & H$\beta$, [O\,{\sc iii}]4959,5007 & 4.63 & 0.46 & 0.015 & 8.854 & O6 \\
35 (N88\,A) & H$\alpha$, [S\,{\sc ii}]6716,6731, [N\,{\sc ii}]6548,6584 & 4.54 & 2.82 & 2.833 & 24.52 & O9.5 \\
36 (N81) & H$\beta$, [O\,{\sc iii}]4959,5007  & 4.58 & 2.96 & 0.670 & 43.167 & O8 \\
36 (N81) & H$\alpha$, [S\,{\sc ii}]6716,6731, [N\,{\sc ii}]6548,6584  & 4.56 & 2.64 & 1.417 & 29.0 & O8.5 \\
\hline
\end{tabular}
\end{minipage}
\end{table*}

The effective temperatures obtained using {\sc cloudy} for the majority of sources are consistent with late O-/early B-type stars.
For sources \#07 and 12, the uncertainties in the line flux ratios exceed the values themselves and thus these results are highly 
uncertain. 
The fits for sources \#01, 03, 09 and 25 are poor: the average difference between the observed and model line fluxes 
is close to (\#09, 25) or greater than (\#01, 03) the average uncertainty in the measured line fluxes
 despite having relatively well constrained emission line fluxes available.
This may be an indication that multiple unresolved sources contribute to the excitation of the optical lines.
The red sections of the spectra for sources \#35 and 36 have proven particularly challenging to fit with {\sc cloudy}. For these spectra 
multiple bright sources of differing spectral types fall within the 1 arcsec width of the slit. It is therefore unsurprising that it is challenging to fit a 
single star model to these spectra. Nevertheless these fits are consistent with excitation by O-stars as expected (see Section 2).

\section{Discussion}

SINFONI IFU observations have revealed a total of 29 \textit{K}-band sources (with obtained resolutions ranging from $\sim$0$\rlap{.}^{\prime\prime}$2 to 
$\sim$1$\rlap{.}^{\prime\prime}$0) within 19 \textit{Spitzer} YSO candidates 
in the SMC. 
Of these $\sim$50\% exhibit line emission which extends beyond the full-width-half-maximum (FWHM) of the 
continuum emission. The emission line properties are summarised in Table 7.

\subsection{Br$\gamma$ emission and accretion rates}

In Section 4.3.1 we showed that the Br$\gamma$ emission line fluxes for our SMC sources are typically higher than those of Galactic massive YSOs
with comparable \textit{K-}band magnitudes. 
The ATS regressions fitted to the SMC and Galactic data sets yield the following relations between \textit{K}-band magnitude and 
Br$\gamma$ emission line luminosity.
\begin{equation}
 \log(L_{Br\gamma(SMC)}) = -0.39 K -2.46
\end{equation}
\begin{equation}
 \log(L_{Br\gamma(MW)}) = -0.37 K -2.69
\end{equation}
If we assume that the same empirically derived relation between Br$\gamma$ emission and accretion luminosity (Eqn. 4) holds true for both data sets
and take, for example, a  massive YSO of $-$6 mag then we arrive at accretion luminosities of $\sim$620\,L$_{\sun}$ and $\sim$300\,L$_{\sun}$ 
for the SMC and Milky Way, respectively. Assuming these hypothetical sources are of the same mass and evolutionary state this suggests an accretion rate
in the SMC which is double that of an equivalent object in the Milky Way.
Similarly, for the range of absolute \textit{K-}band magnitudes from 0 to --12 the L$_{\text{acc}}$ ratio between the SMC and the Milky Way ranges from 1.6 to 2.6.
The Br$\gamma$ emission measurements towards sources in N113 in the LMC ($Z_{\text{SMC}} \sim $0.4\,Z$_{\sun}$, \citealt{Bernard2008}) are on average even higher than those of 
the SMC sample. However, at least two of the three N113 sources exhibit 
compact H\,{\sc ii} regions surrounding the continuum source, contributing significant levels of Br$\gamma$ emission, likely produced by stellar photons rather 
than as a result of the accretion shock.

It should be noted that the \textit{K-}band magnitude for massive YSOs may not be the same in the SMC as in the Milky Way
for a given mass and age. Additionally the uncertainties resulting from our extinction corrections for the SMC sources are large
(see Fig. 2).

Nevertheless our results appear to be consistent with
observations of $\sim$1000 Pre-Main-Sequence (PMS) stars in the LMC by 
\citet{Spezzi2012} and 680 PMS stars in NGC 346 in the SMC by \citet{DeMarchi2011} which
suggest significantly higher accretion rates in PMS stars in the lower metallicity Magellanic Clouds. 
However, \citet{Kalari2015} analyse the mass accretion rates of PMS stars in the low metallicity ($Z_{\text{Sh2-284}} \sim $0.2\,Z$_{\sun}$) Galactic star forming region
Sh 2-284 and find no evidence of a systematic change in the mass accretion rate with metallicity.
It has been suggested that both detection issues towards the Magellanic Clouds sample and physical factors can contribute
towards this discrepancy.

Whilst the results in this work allude to higher accretion rates in massive YSOs in the SMC, it is far from certain whether this
effect is the result of the lower metallicity environment presented by the SMC.
Addressing the question of how accretion is affected by 
elemental composition will require significantly improved evolutionary models for 
low metallicity stars for a wide variety of elemental abundances, masses and ages, as well as more extensive Galactic and Magellanic Clouds observations.

\subsection{Optical spectra}

From our analysis of the H$\alpha$ and H$\beta$ emission lines we determine values of visual extinction towards the 
optical emission which are significantly lower than the extinction values calculated from the \textit{K-}band emission features.
We have also determined that, without exception, the optical emission we have measured is dominated by photo-excitation
rather than shocked emission.
These findings indicate that the optical emission is sourced in a much shallower region of the molecular cloud
and that it is produced via the interaction of UV photons with the gas in the outer regions of the cloud. 
This therefore implies a large mean free path of UV photons which in turn is likely caused by a porous, possibly clumpy
ISM which is consistent with the findings of \citet{Madden2006}, \citet{Cormier2015} and \citet{Dimaratos2015}.

Using the photoionization code \textsc{cloudy}, we have been able to fit excitation parameters to the optical emission based on a single
OB type star in a molecular cloud with elemental abundances similar to those of the SMC.
We find that the majority of optical spectra are consistent with late-O/early-B type
zero-age-main sequence (ZAMS) stars, indicating mass ranges of 10--30 M$_{\sun}$ \citep{Hanson1997}.
In the majority of cases, however, the objects in question are not straightforward systems comprising of a single ZAMS star illuminating a 
molecular cloud and the accretion process produces a significant amount of ionizing photons as well. As previously mentioned 
 it is possible that multiple objects contribute to the optical emission (although the emission is still likely to 
be dominated by the most massive star).

\subsection{Properties of massive YSOs in the SMC}

Notwithstanding resolution effects (see Section 6.3.1) we expect the \textit{K}-band emission features to reflect the
 spectral types outlined in \citet{Woods2011}, loosely describing an evolutionary sequence.
The ice-feature rich G1 type sources are predicted to be the earliest stage YSOs and as such are expected to be dominated by
 H$_2$ emission produced in the shock fronts of strongly collimated outflows.
The G2 (silicate absorption) sources are expected to exhibit significant Br$\gamma$ and H$_2$ emission resulting from UV photons originating at 
the accretion shock front and outflows, respectively.
The later stage YSOs and compact H\,{\sc ii} regions are expected to be strong PAH emitters, falling into the G3 type. 
They are also both expected to exhibit strong Br$\gamma$ and He\,{\sc i} emission as a result of the strongly ionizing radiation field 
of the young massive star (and the shock front of any still ongoing accretion). Consistently the H$_2$ emission is expected to be largely 
photo-excited due to the broadening of outflows into lower velocity, uncollimated stellar winds and higher fluxes of energetic photons from the star itself.
Compact H\,{\sc ii} regions may be distinguished by the presence of free--free radio emission produced by the large volume of ionized gas.
Shock excited H$_2$ emission may also be present at the boundaries of compact H\,{\sc ii} regions during the expansion of the Str\"{o}mgren sphere.
Finally, G4 type sources are dominated by silicate emission which is observed in Galactic Herbig AeBe stars and could be expected in lower mass, later evolutionary
stage objects.

Of the nine sources in the sample that exhibit infrared ice absorption features (G1 type; \citealt{Woods2011}), 
all nine exhibit H$_2$ emission, indicative of outflows in early stage YSOs, with extended H$_2$ emission in five cases.
Only two G1 type sources show significant He\,{\sc i} emission and only two exhibit 
free-free radio emission, more commonly found in later stage YSOs and H\,{\sc ii} regions.
Nevertheless our results are largely consistent with the G1 classification representing younger sources.
Likewise, of the five G3 type sources (PAH emission), four exhibit radio emission and all of them exhibit He\,{\sc i} emission, indicating that these are
indeed the more evolved sources.
The two G2 type sources (silicate absorption) both exhibit H$_2$ emission and Br$\gamma$ emission. 
The following equivalence can be established between \citet{Seale2009} and \citet{Woods2011}: S and SE types are equivalent to G2 types,
P and PE types are equivalent to G3 types and O types are equivalent to G4 types. G1 type sources from the \citet{Woods2011} scheme
do not have an analogue in the \citet{Seale2009} classifications.

It must be noted that the spatial scales probed by the \textit{Spitzer}-IRS spectra and the radio data differ significantly from
the observations presented in this work. 
The radio emission column in Table 1 indicates that such emission is present within a distance of 4 arcsec
from the source and thus in reality may be not be directly associated with the \textit{K}-band continuum sources discussed here. 
Additionally the slit width of \textit{Spitzer}-IRS varies from 3.6--11.1 arcsec depending on the spectral range, sampling a large area
in complex star forming environments and making it difficult to separate the contributions of the YSO and its environment to the measured flux. 
It should also be noted that some of the IRS sources are resolved into multiple components in this study,
demonstrating the obvious limitations of a classification scheme based on low spatial resolution observations.

\begin{table*}
 \begin{minipage}{175mm}
\caption{Observed spectral properties of all sources. For sources marked with an \textquotedblleft E\textquotedblright\, 
the relevant line emission is extended beyond the FWHM of the 
continuum emission. \textquotedblleft A\textquotedblright\, signifies that the emission measured is likely to be ambient whilst ? is indicative of a high degree of uncertainty in either the measurement
or the source of the emission. 
Also included are the mid-IR classifications (W11 type; \citealt{Woods2011}) and whether radio emission is detected towards the sources.
The 9th column gives the optical spectrum type for each source (see Table 1) where H$\alpha$ 
indicates that only H$\alpha$ emission is detected.
The final column is a type based on the \textit{K-}band morphology observed in this study for each source. Compact sources (C) are discussed in
Section 6.3.2 and extended outflow sources (O) are discussed in Section 6.3.3. the outflow sources \#03 and 31 (O$^{\ast}$) in which the Br$\gamma$ and He\,{\sc i} emission
line morphologies are distinctly offset from the continuum and H$_2$ emission are discussed separately in Section 6.3.4.
The H\,{\sc ii} morphological classification indicates that the source is the major ionizing source of a compact H\,{\sc ii} region.
  The compact H\,{\sc ii} region N88\,A is discussed in Section 6.3.6. and N81 is discussed in 6.3.5.}
\setlength{\tabcolsep}{5pt}
\begin{tabular}{c c c c c c c c c c}
\hline
 & H$_2$  & Maser  & Br$\gamma$  & He{\sc i}  &  CO  &  & Radio  &  & Morphological \\
Source & emission & emission & emission & emission & bandhead & W11 type & emission & Optical type & type \\
\hline
01 & \checkmark E & & \checkmark E & \checkmark E & & G3 & Y & IV/V & O\\
02\,A & \checkmark E & & \checkmark & & & G1 & Y & H$\alpha$& O\\
02\,B & \checkmark & & & & &  &  & &C \\
03 & \checkmark E & H$_2$O & \checkmark E & \checkmark E & \checkmark & G1 & Y & V & O$^{\ast}$\\
04 & & & \checkmark & \checkmark & & G3 & N & II & C \\
06 & \checkmark E & & \checkmark & ? & & G1 & N & & O\\
17 & \checkmark & & \checkmark &  & & G1 & N & H$\alpha$ & C\\
18 & \checkmark & & \checkmark & \checkmark & & G1 & N & H$\alpha$ & C\\
20 & \checkmark & & \checkmark & & & G4 & N & I & C\\
22\,A & \checkmark E & & \checkmark & & & G1 & N & IV/V & O\\
22\,B & \checkmark & & \checkmark & & &  &  &  & C\\
25 & \checkmark ? & & \checkmark ? & ? && G3 & Y & IV & \\
26 & \checkmark & & \checkmark & \checkmark & & G3 & Y & V & C\\
28\,A & \checkmark E & & \checkmark E & \checkmark & & G2 & Y & IV/V & O\\
28\,B & \checkmark & & & \checkmark & & & & & C\\
30 & \checkmark E & & \checkmark & & & G1 & N & I & O\\
31 & \checkmark E & & \checkmark E & \checkmark &  & G3 & Y$^{\ast}$ & & O$^{\ast}$\\
32 & \checkmark & & \checkmark & & & G1 & N & I/II& C\\
33 & \checkmark & & \checkmark & & & G2 & N & I/II & C\\
34 & \checkmark & & \checkmark & & & G1 & N & H$\alpha$ & C\\
35\,A & \checkmark E & & \checkmark E & \checkmark E & && & V & C/H\,{\sc ii}\\
35\,B & \checkmark  & & \checkmark A & \checkmark A & & & & & C\\
35\,C & \checkmark  & & \checkmark A & \checkmark A & && & & C\\
35\,D & \checkmark  & & \checkmark A & \checkmark A & && & &C\\
36\,A & \checkmark A & & \checkmark A & \checkmark A & && & V& C\\
36\,B & \checkmark E & & \checkmark A & \checkmark A & && & &O\\
36\,C & & & \checkmark A & \checkmark A & && & &C\\
36\,D & & & \checkmark A & \checkmark A & && & &C\\
36\,E & \checkmark A & & \checkmark A & \checkmark A & && & &C\\
\hline
\end{tabular}

\end{minipage}
\end{table*}

\subsubsection{Compact \textit{K}-band sources}

The majority (19/29) of the \textit{K}-band continuum sources identified in this work fall within the ultra-compact regime (diameter $\leq$0.1 pc)
with little or no line emission extending beyond the FWHM of the continuum source.
In this section we discuss these ultra-compact sources with the exception of those in the N88\,A (\#35) and N81 (\#36)
fields which are discussed in detail in Sections 6.3.5 and 6.3.6.
In the remaining FOVs we classify 11 sources as compact: \#02\,B, 04, 17, 18, 20, 22\,B, 26, 28\,B, 32, 33 and 34.

The five compact sources classified as Seale et al. (2009) S-type (silicate absorption dominated), \#17, 18, 32, 33 and 34, all exhibit either H$\alpha$ only in the optical spectrum or type I/II optical
spectra. None are sources of radio emission, nor do they exhibit He\,{\sc i} emission, indicating that these are early stage massive YSOs. 
Furthermore all but one of these sources (\#33, a G2 type) are G1 types, exhibiting ice absorption features indicative of early-stage YSOs.
Two of these sources (\#18 and 34) exhibit H$_2$ emission line ratios consistent with shocked excitation linked to outflows whilst none exhibit photo-excitation
dominated H$_2$ emission.
They exhibit a wide range
of luminosities, \textit{K}-band magnitudes and extinctions, suggesting a wide range of physical conditions and/or masses but what seems apparent is that 
these make up our embedded, early stage massive YSOs.

Source \#04 is a compact P-type/G3-type (Table 1) source, exhibiting a bright \textit{K}-band continuum despite a relatively low luminosity.
Br$\gamma$ and He\,{\sc i} emission are detected towards the source but no H$_2$ emission, so it is unlikely to be a
very early stage YSO due to the lack of outflow indicators. On the other hand, there is no radio emission
detected towards this source so identification as an ultra-compact H{\sc ii} region is unlikely. 
With a spectral type of B0 determined from the \textsc{cloudy} optimization, this source is likely to fall at the lower end 
of our mass range.

The only compact PE type (dominated by PAH emission with fine-structure emission), source \#26, 
is also a G3 \citet{Woods2011} type exhibiting a fairly high luminosity (1.2$\times$10$^{4}$ L$_{\sun}$).
 It has a type V optical spectrum with a log([O\,{\sc iii}]/H$\beta$) value comparable to that of N88\,A (see Fig. 13) and exhibits Br$\gamma$, He\,{\sc i} and 
H$_2$ emission in the \textit{K}-band, indicative of a strongly ionizing source.
It also exhibits free--free radio emission, typical of an UCH{\sc ii} but the H$_2$ emission in the \textit{K}-band is more consistent with shocked emission
than photo-excitation. It is therefore possible that the H$_2$ emission originates in a shock front of an expanding UCH{\sc ii} region.

Source \#20 is the only YSO in this work that exhibits silicate emission (O type and 
 G4 type classifications of \citet{Seale2009} and \citet{Woods2011} respectively). 
It is therefore likely to be lower mass than the majority of sources in this study and possibly in a later evolutionary state. 
The bolometric luminosity is low (1.5 $\times$ 10$^{3}$ L$_{\sun}$), consistent with a Herbig Be type object and the optical type of the star (I) 
suggests relatively low levels of ionizing radiation. The extinction 
appears to be unusually high (24$\pm$13\,mag; c.f. \citealt{Fairlamb2015}) suggesting that it is more embedded than would be expected for a Herbig Be star.
The high level of extinction measured in the \textit{K}-band could be a geometric effect, for example an edge on dusty disc; however, because the 
object is unresolved this is only speculation. 

Source \#02\,B exhibits a spectrum which is featureless except for H$_2$ emission which appears to be largely ambient (see Fig. 9) and CO absorption red-wards
of 2.9 $\mu$m. CO in absorption is not a common feature of 
massive YSOs but it has been detected towards two high confidence massive YSOs in the Milky Way (G023.6566$-$00.1273 and G032.0518$-$00.0902; \citealt{Cooper2013}).
CO absorption is commonplace in lower mass YSOs (e.g., \citealt{Casali1996}); however, source \#02\,B is relatively bright ($K = 15.75\pm0.1$\,mag)
and it is therefore unlikely to be a low mass star.
The velocity range estimated for the CO absorption towards source \#02\,B is consistent with the line centroid measurements towards
the massive YSO \#02\,A meaning it cannot be dismissed as an unrelated background source.

The remaining compact sources (\#22\,B and 28\,B) exist in close proximity to a brighter \textit{K}-band source and are unlikely to be the 
dominant source of the spectral properties examined by \citet{Oliveira2013}. The line emission towards each of these sources  
appears to be ambient (see Figs 2, 6 and 9) and as such the nature of these sources remains unclear; if they are YSOs then 
they could be early stage and/or low mass objects.

\subsubsection{Extended outflow sources}

Seven sources exhibit extended H$_2$ emission morphologies indicative of molecular outflows: \#01, 02A, 06, 22A, 28\,A, 30 and 36\,B.
Two of these sources (\#22\,A and 36\,B) exhibit clear bipolar, relatively collimated structures with a blue- and a red-shifted component (see Figs. 9 \& 10)
consistent with a bipolar outflow perpendicular to a disc. Source \#22\,A exhibits weak, compact Br$\gamma$ emission whilst the Br$\gamma$
emission detected towards source \#36\,B appears to be largely ambient to the region.
Source \#28\,A (see Fig. 14) exhibits a slightly red-shifted v-shaped region of H$_2$ emission to the south, most likely tracing cone shaped shocked and PDR emission 
at the edges of an outflow. In the centre of the outflow, a low density region is formed allowing the gas there to be ionized which causes the Br$\gamma$ recombination
emission (see Fig. 14).
Sources \#01, 02\,A and 30 exhibit extended H$_2$ emission in a single direction from the continuum source.
The last of the outflow-like morphology sources, source \#06 exhibits a knot of H$_2$ emission to the south and a more extended emission structure to 
the north west. Both of these structures appear slightly blue shifted with respect to the continuum source.
\begin{figure}
 \includegraphics[width=0.95\linewidth]{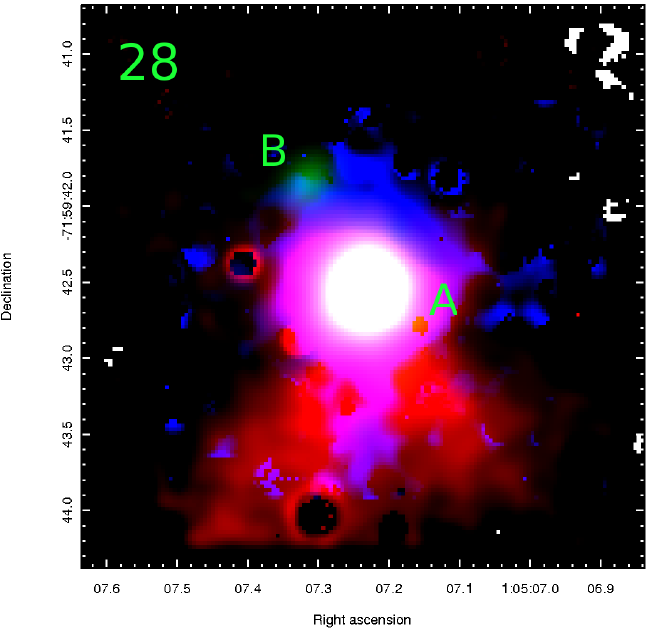}
\caption{Three colour image of source \#28. Red - H$_2$ 1-0S(1) emission, green - \textit{K}-band continuum emission, blue - Br$\gamma$ emission.}
\end{figure}

Only a single source exhibits a 1-0S(0) / 1-0S(1) emission line ratio consistent with purely shocked emission with the 
majority of sources exhibiting ratios which lie between the expected ranges for shocked and photo-excited emission.
On the other hand all but one 2-1S(1) / 1-0S(1) ratios (where the 2-1S(1) line is measured) are consistent with 
a shock excitation mechanism.
We deduce therefore that, since the upper energy level of the 2-1S(1) transition (12\,550 K) is almost double that of the 1-0S(0) transition (6471 K),
the 1-0S H$_2$ transitions are stimulated by a combination of photo-excitation from the central (proto-)star and shocks from outflows/winds whilst
2-1S transitions are dominated by shocked excitation from outflows.

As discussed previously, the observed ice features towards sources \#02, 06, 22 and 30 are consistent with early-stage YSOs
 (sources \#02 and 22 are both resolved into multiple components in the \textit{K-}band so the origin of the ice absorption
 in these fields remains uncertain). 
Source \#02 
also exhibits radio emission but the difference in spatial resolution means that the radio emission may not originate from the 
\textit{K-}band sources detected in this work.
Source \#01 is dominated by PAH emission indicating a later stage massive YSO 
whilst source \#28 is dominated by silicate absorption and likely is at an evolutionary state between
the ice feature dominated sources and source \#01.

Target \#28 (see Fig. 14) is by far the most luminous of our targets with a bolometric luminosity of 1.4$\times$10$^{5}$ L$_{\sun}$
and source \#28\,A is the dominant source of \textit{K-}band emission in the field. Source \#28\,A also exhibits relatively high level of 
extinction in the \textit{K}-band (for the SMC) and a low level of extinction in the optical. The results of the {\sc cloudy} optimization for the optical
spectrum of source \#28 indicates excitation consistent with an O9.5 type star. 

\subsubsection{Sources \#03 and 31}

Although the H$_2$ emission line morphologies of sources \#03 and 31 are consistent with the extended outflow sources,
the Br$\gamma$ and He\,{\sc i} morphologies (see Figs. 3 and 6) clearly distinguish them from the rest of the sources
so we discuss these sources separately in this section.
In both cases the atomic emission line structures are offset significantly from both the continuum emission and 
the H$_2$ emission.

Originally classified as a planetary nebula by \citet{Lindsay1961} and reclassified as a YSO by \citet{vanLoon2010SMC}, 
source \#03 presents a challenging morphology to interpret. 
A three colour image of source \#03 is given in Fig. 15, showing the displacement of the Br$\gamma$ emission (shown in blue) with respect to both the 
continuum source and the H$_2$ emission (in green and red, respectively).
As this is the only source to exhibit CO bandhead emission, the presence of a disc is probable.
The most likely origin of the extended H$_2$ emission and offset extended Br$\gamma$ emission is a wide, relatively uncollimated outflow
which is bound by the presence of a disc and has created a low density cavity along the axis of rotation in which the remaining gas is ionized. 
\begin{figure}
 \includegraphics[width=0.95\linewidth]{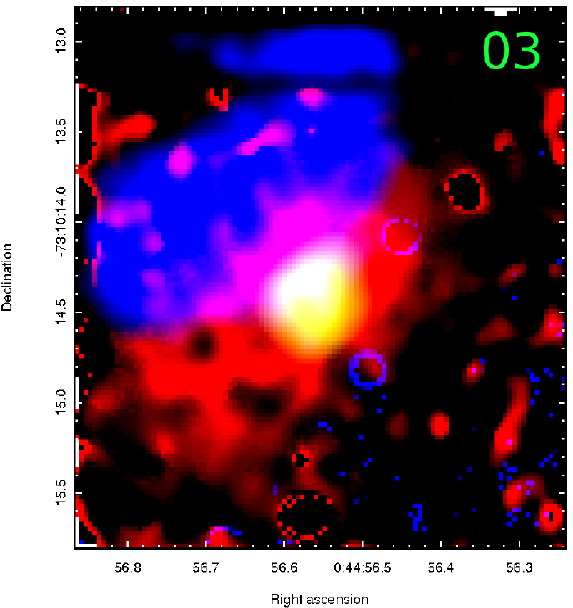}
\caption{Three colour image of source \#03. Red - H$_2$ 1-0S(1), green - \textit{K-}band continuum, blue - Br$\gamma$.}
\end{figure}
The broadening of outflows in massive YSOs leading to a low density ionized region in the centre of the outflow is predicted and modelled by \citet{Kuiper2015} and
\citet{Tanaka2015}, caused by continued accretion onto the central protostar in a YSO environment over time. 
It is therefore possible that this source represents a slightly more evolved state than that of source \#28\,A (Fig. 14).

A comparison of H$_2$ emission line ratios for the continuum source and the nearby off-source emission indicates that source \#03
is dominated by shocked emission towards the continuum source itself with a much lower contribution of shocked excitation further from the continuum source.
This is consistent with a photo-excitation region bordering an H\,{\sc ii} region which is ultimately powered by a broad, uncollimated outflow
in the inner regions of the YSO. The kinematics of source \#03 (see Table 4) indicate that the H$_2$ emission towards the continuum source is blue-shifted with 
respect to the Br$\gamma$ emission and both the extended Br$\gamma$ emission and He\,{\sc i} emission are blue-shifted with respect to the 
continuum source (see Figs. 4 \& 7)  

Finally, source \#03 lies closer to previously detected H$_2$O maser emission \citep{Breen2013} than any other source within our sample. 
The velocity of the maser falls within the range 136.4--143.0 km s$^{-1}$ which is consistent with the velocities we have measured towards the \textit{K-}band 
source.
This emission is detected approximately 3.7$\pm$0.5 arcsec (1.1$\pm$0.1 pc) to the south of the position of the continuum source \#03, outside the SINFONI FOV. 
Therefore the maser emission may not in fact associated with source \#03. 
Still if the Br$\gamma$ emission represents an ionized cavity as the result of powerful outflows from the source then it is conceivable
that associated maser emission may be detected at relatively large distances from the continuum source.

Source \#31 exhibits similar Br$\gamma$ and H$_2$ emission  morphologies to source \#03. However, the properties of these objects as determined by
\citet{Oliveira2013} differ significantly. The \citet{Seale2009} classification of source \#03 is an S type and that of the \citet{Woods2011} scheme
is G1 whilst for source \#31 they are PE and G3.
\,Additionally source \#31 has a bolometric luminosity around 
a tenth that of source \#03 indicative of an object with a significantly lower mass.
Contrary to source \#03, the H$_2$ 1-01S(0) / 1-0S(1) emission line ratio towards source \#31 is consistent with a PDR.
Source \#31, although apparently similar morphologically, possibly represents a much lower mass 
than source \#03.

\subsubsection{Source \#35 (N88\,A)}

\begin{figure}
 \includegraphics[width=0.99\linewidth]{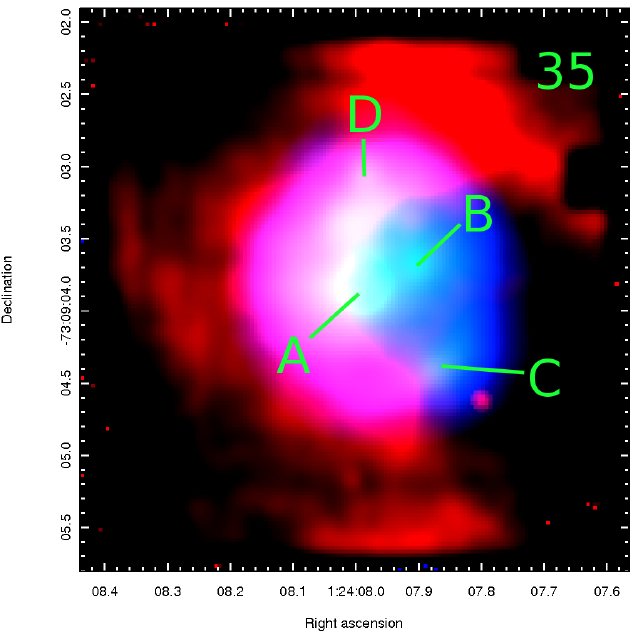}
\caption{Three colour image of source \#35 (N88\,A). Red - H$_2$ 1-0S(1), green - \textit{K-}band continuum, blue - Br$\gamma$.}
\end{figure}

Our observations of N88\,A (source \#35; Fig. 16) are consistent with those by \citet{Testor2010}, with the dominant source of ionization being 
the central bright star (star \#41 in \citealt{Testor2010}). We consistently find that all four bright stars in N88\,A appear brighter in our work than \citet{Testor2010}
in \textit{K}-band continuum magnitude. This is most likely because we do not take into account the high levels of background nebulous emission.
The measured absolute \textit{K-}band magnitude of source \#35\,A is consistent with a ZAMS star with a temperature in excess of 50\,000 K, 
consistent with a star which is more massive than an O3 spectral type object \citep{Hanson1997}.
This is likely to be an overestimate of the mass of the star because the effects of any warm dust excess and nebulous emission are neglected.
The resulting best fitting solution using \textsc{cloudy} is for an exciting source of spectral type O6.

Source \#35\,A exhibits strong Br$\gamma$ emission and He\,{\sc i} emission with some ambient H$_2$ emission originating in the
H$_2$ emission arc (see Fig. 16).
The remaining sources in N88\,A are affected significantly by ambient emission originating from the central bright star \#35\,A.
Peaks in both Br$\gamma$ emission and He\,{\sc i} emission also occur at the positions of sources \#35\,C and 35\,D,
which suggests that there is an additional component originating from these sources.
None of the continuum sources appear to be significant sources of H$_2$ emission, implying that they are not early-stage YSOs.
Source \#35\,B  does not appear to exhibit any intrinsic emission but weak emission cannot be ruled out due to 
its proximity to source \#35\,A.

Towards the H$_2$ emission line arc we find that the measured H$_2$ emission line ratios suggest a combination of collisional excitation and photo-excitation where the 
emission from 1-0 transitions have a significantly higher contribution from photo-excitation compared to the 2-1 transitions.
The collisional component is most likely caused by the impact of the expanding ionized gas 
on the surrounding molecular cloud, forming the H$_2$ emission arc observed in the east of N88\,A; an additional photo-excited component is 
sourced from the massive stars in the centre of the region.

Rather than showing an association with the H\,{\sc i} gas component at 134$\pm$9 km s$^{-1}$ \citep{Testor2010}, both the SINFONI velocity measurements
for H\,{\sc i} and H$_2$ emission and the radio H\,{\sc i} velocities obtained from the data of \citet{Stanimirovic1999} are significantly higher than 
this. The H$_2$ velocities we measure towards the continuum sources in N88\,A range from 162 to 167 km s$^{-1}$ with a typical uncertainty of 8 km s$^{-1}$, 
consistent with the measured 21 cm peak at 161.1$\pm$0.4 km s$^{-1}$.
It therefore appears most likely that N88\,A is associated with the SMC gas component at 167$\pm$8 km s$^{-1}$ \citep{McGee1981}.
 The Br$\gamma$ emission velocities range from 153 to 156 km s$^{-1}$ with a typical uncertainty
of 7 km s$^{-1}$, blue-shifted with respect to the bulk motion of the gas. This is consistent with a compact H\,{\sc ii} region in which the 
ionized gas is expanding outwards due to radiation pressure at a velocity of $\sim$10 km s$^{-1}$, comparable to the compact H\,{\sc ii} regions observed in
LHA 120-N113 in the LMC \citep{Ward2016}.

\subsubsection{Source \#36 (N81)}

We have resolved the central region of N81 into 5 distinct \textit{K}-band continuum sources, a significant improvement on the previous highest 
resolution seeing limited imaging available \citep{Heydari-Malayeri2003} which resolved only two sources.
The most distinctive newly resolved feature of N81 is the bipolar H$_2$ emission centred on (or near to) source \#36\,B (Fig. 17).
This bipolar morphology and the observed velocity gradient indicate a bipolar outflow centred on source \#36\,B, suggestive of 
an early stage massive YSO.
\begin{figure}
 \includegraphics[width=0.99\linewidth]{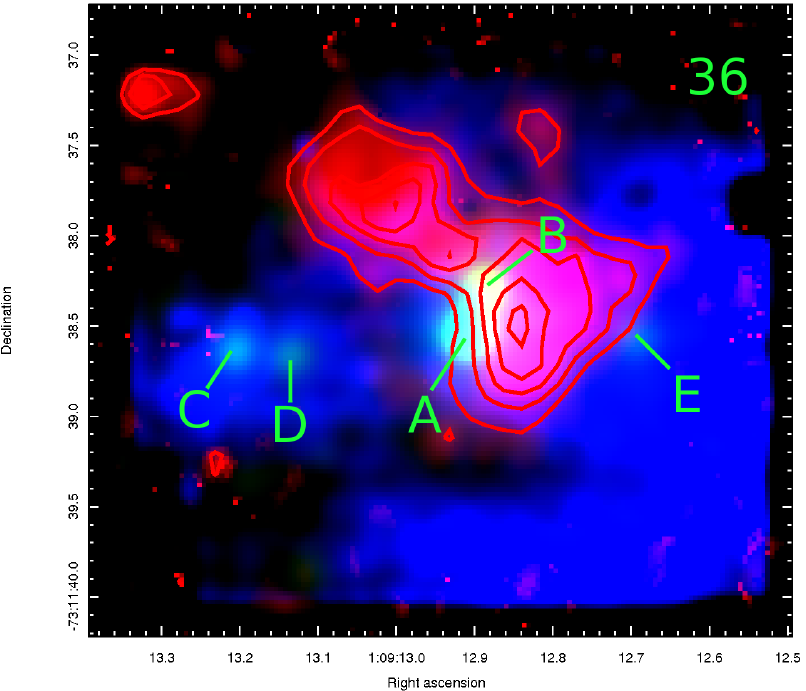}
\caption{Three colour image of source \#36 (N81). Red - H$_2$ 1-0S(1), green - \textit{K}-band continuum, blue - Br$\gamma$.
For clarity the red contours trace the H$_2$ emission.}
\end{figure}

The majority of the Br$\gamma$ emission towards this region appears to be ambient and is likely common to the whole N81 H\,{\sc ii} region. There
may be an additional component from one or both of the fainter (and probably lower mass) sources \#36\,C and 36\,E. 
A similar morphology is observed in the He\,{\sc i} 
emission.
Although it is the brightest continuum source in the region, emission detected towards source \#36\,A appears to be entirely ambient.

The velocities we have determined towards sources in N81 from Br$\gamma$ emission range from 166 to 167 km s$^{-1}$ with uncertainties
of $\sim$9 km s$^{-1}$. These are consistent with both the results of Heydari-Malayeri et al. (1988) determined from H$\beta$
emission and the velocity determined from the radio observations (see Section 4.4) indicating a common bulk motion of the atomic and
ionized gas.
The H$_2$ 2.1218\,$\mu$m emission towards the same sources has been measured to be consistently red-shifted with respect to the Br$\gamma$ emission.

\section{Conclusions}

We have performed \textit{K}-band SINFONI observations for 17 of the 33 massive YSOs in the SMC presented in \citet{Oliveira2013}, as well as two additional
targets in N81 and N88\,A,
revealing a wide range of spectroscopic properties. We have also analysed optical spectra
for 28 of the 33 massive YSO targets of \citet{Oliveira2013} plus sources 35 (N88\,A) and 36 (N81).
\,We summarise our conclusions below.
\begin{enumerate}
 \item 14 of the 17 \textit{Spitzer} YSOs in the SMC observed with SINFONI have been resolved into single 
	\textit{K}-band continuum sources whilst three have been resolved into multiple components in the \textit{K}-band.
Source \#35 (N88\,A) has been resolved into the previously known 4 continuum components.
	N81 (\#36), previously resolved into two components by \citet{Heydari-Malayeri2003}, has now been resolved into five \textit{K}-band continuum sources.
 \item Visual extinctions have been calculated towards all \textit{K}-band sources except source \#04. We find a median visual extinction towards 
	the SMC sources of $A_V =$ 14 mag. The median for the Galactic sources from \citet{Cooper2013} is 44 mag and that for the three 
	YSOs in N113 in the LMC is 20 mag \citep{Ward2016}. 
	This seems to suggest that there is a correlation between extinction towards YSOs and metallicity as
	$Z_{\text{LMC}}\sim$0.5 Z$_{\sun}$ and $Z_{\text{SMC}}\sim$0.2 Z$_{\sun}$.
 \item Whilst consistent with the \citet{Cooper2013} sample, Br$\gamma$ luminosities are high compared with those in the Milky Way,
	suggesting accretion rates which are on average higher than for the Galaxy. 
 \item He\,{\sc i} line luminosities are comparable to the Milky Way sample, indicating that the excess observed in the 
	Br$\gamma$ emission is unlikely to be related to a stronger wind component.
 \item Average velocities have been measured towards the \textit{Spitzer} YSO sources of 
	169$\pm$5 km s$^{-1}$ and 173$\pm$4 km s$^{-1}$ for Br$\gamma$ emission and H$_2$ emission, respectively.
	The Br$\gamma$ emission line velocities are consistent with those of the H\,{\sc i} radio observations, indicating that the average motions
	of the YSOs are bound to the bulk motions of the ISM.
 \item The majority of the \textit{K}-band continuum sources (11/20; excluding observations towards the H\,{\sc ii} regions N81 and N88\,A)
	fall within the unresolved, ultra-compact regime. 
	These sources exhibit a variety of spectral properties and include one probable UCH{\sc ii} region (\#26), a possible Herbig Be star (\#20)
	along with spectroscopically typical massive YSOs and two apparently featureless continuum sources (\#22\,B and 28\,B) which may not actually be YSOs.
 \item Seven sources with extended H$_2$ emission morphologies indicative of outflows have been identified, with velocity gradients
	measured towards four of these. Sources \#22\,A and 36\,B exhibit striking examples of bipolar outflow morphologies.
	Our observations of the well-studied H\,{\sc ii} region N81 are the first to identify the bipolar outflow originating from source \#36\,B. 
\item Source \#28\,A exhibits Br$\gamma$ and H$_2$ emission line morphologies indicative of an ionized cavity in the centre of the outflow. 
	Source \#03 appears to be an extremely broadened (and possibly more evolved) example
	of the same structure. These structures are possibly the result of the broadening of outflows in 
	massive YSOs predicted by the models of \citet{Kuiper2015} and \citet{Tanaka2015}.
 \item CO bandhead emission (commonly used as a tracer of discs) has only been detected towards one source (\#03), 
a detection rate of around one third of that towards 	
massive YSOs in the Milky Way \citep{Cooper2013} for the same range of luminosities. 
This could be due to either the low gas-phase CO abundance of the SMC \citep{Leroy2007}, or conditions within protostellar 
discs which differ significantly (such as higher temperatures or less shielding from dust) from those in the Milky Way, leading to a higher rate of
CO destruction.
	CO absorption is somewhat tentatively observed towards one source (\#02\,B) which could be indicative of a dusty circumstellar environment and possibly 
	suggests a continuum source completely obscured by an edge--on disc.
 \item Optical emission towards all sources (where it is present) appears to originate from the outer edges of the molecular clouds as the
	average extinction measured towards the optical emission is significantly lower than that towards the \textit{K}-band emission for the same targets.
	The optical emission is photo-excited and is therefore unlikely to be produced by the interaction of outflows and winds
	with the ISM.
	This scenario is consistent with relatively large mean free path of high energy photons from the protostar and through the ISM,
	leading to a large number of UV photons leaking from the YSOs. This could occur due to 
 a clumpy and/or torn-up circumstellar medium, consistent with expectations for a low metallicity ISM \citep{Madden2006,Cormier2015,Dimaratos2015}. 
\end{enumerate}

To conclude, we have presented the first study of massive YSOs in the SMC using integral field spectroscopy, studying line emission and
resolving a number of our targets into 
multiple components in the \textit{K}-band for the first time. Through comparison with existing data in the optical, infrared and radio regimes, we have
been able to develop a greater understanding of these objects within an evolutionary context. 
Following from the work on ice chemistry by \citet{Oliveira2013}, we continue to expose differences between YSO properties in our own Galaxy and the low
metallicity SMC, namely those which concern accretion, potential disc properties and the YSO--ISM interplay.

\section*{Acknowledgments}
The authors thank the anonymous referee for his/her useful comments.
JLW acknowledges financial support from the Science and Technology Facilities Council of the UK (STFC) via the PhD studentship programme.
We would like to thank the staff at ESO's Paranal observatory for their support during the observations.
Some of the observations reported in this paper were obtained with the Southern African Large Telescope (SALT).
This paper made use of information from the Red MSX Source survey database at http://rms.leeds.ac.uk/cgi-bin/public/RMS\_DATABASE.cgi 
which was constructed with support from STFC. 
This research has made use of the SIMBAD data base, operated at CDS, Strasbourg, France.

\bibliographystyle{mn2e}
\bibliography{bibliography}

\appendix

\section{Continuum images}

Fig A1 shows the continuum images for all sources observed with SINFONI with the exception of source \#25 (see Section 3.1 for details).
Marked on the images are the positions of all \textit{K-}band continuum sources identified in this work.

\begin{figure*}
\begin{minipage}{175mm}
 \begin{center}
\includegraphics[width=0.4\linewidth]{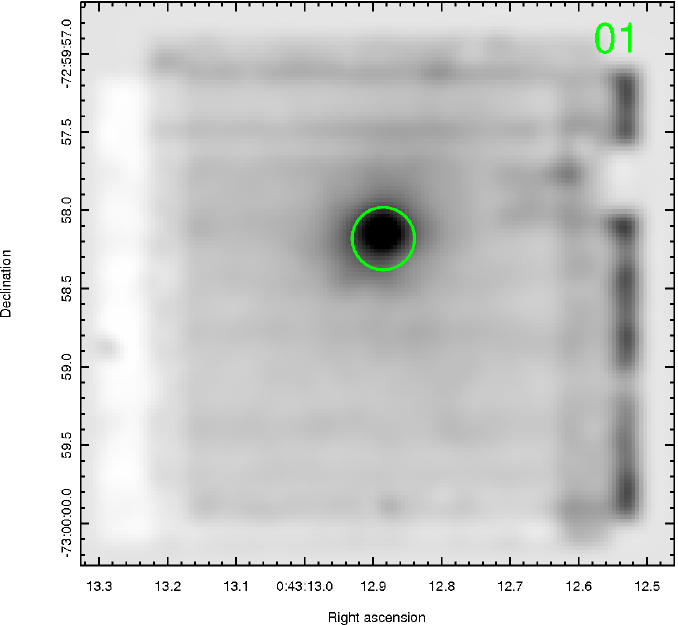}
\includegraphics[width=0.4\linewidth]{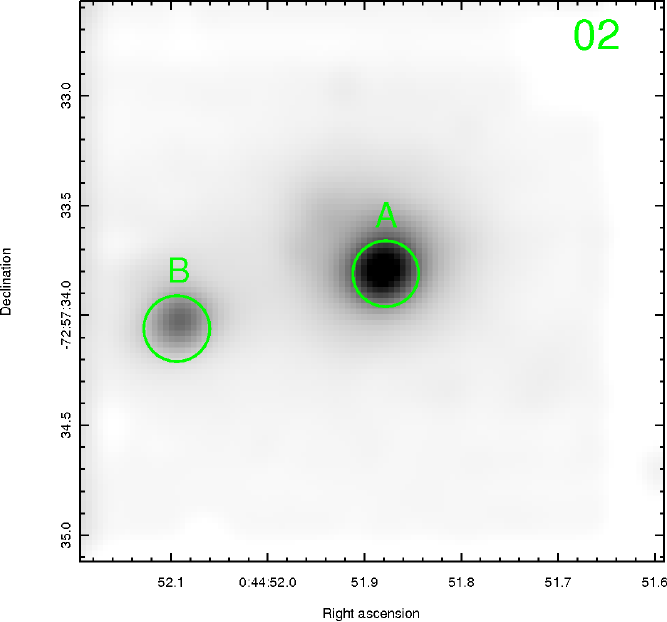}
\includegraphics[width=0.4\linewidth]{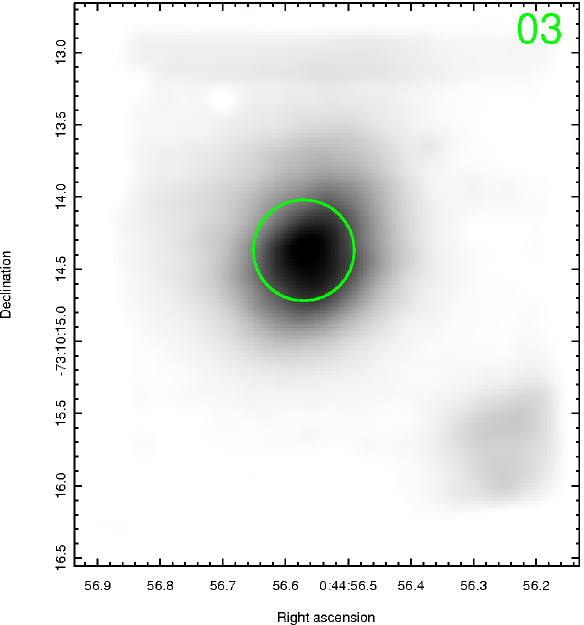}
\includegraphics[width=0.4\linewidth]{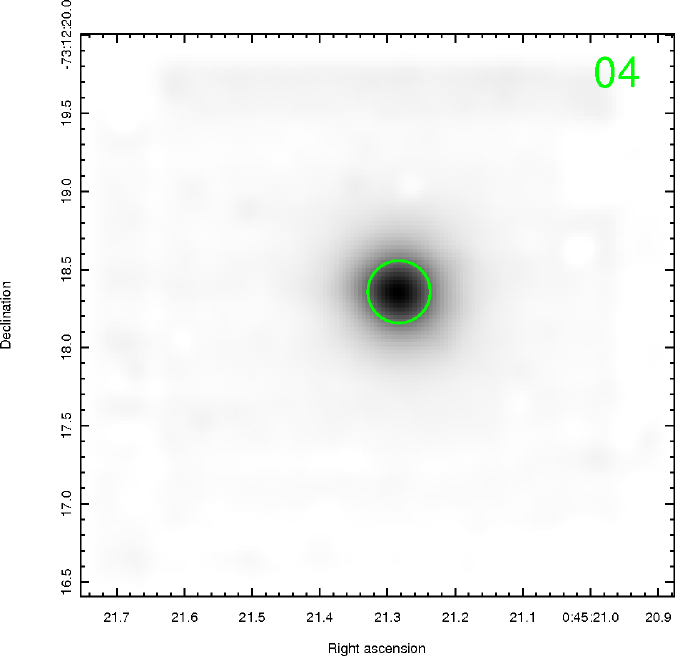}
\includegraphics[width=0.4\linewidth]{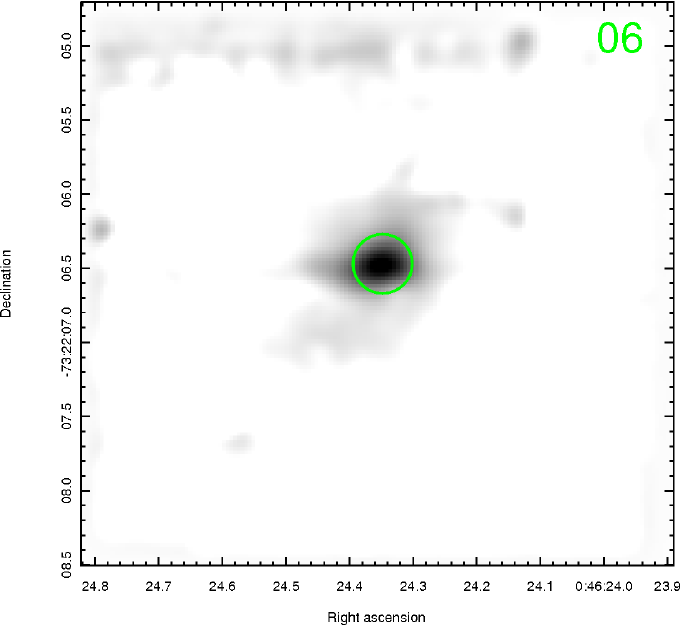}
\includegraphics[width=0.4\linewidth]{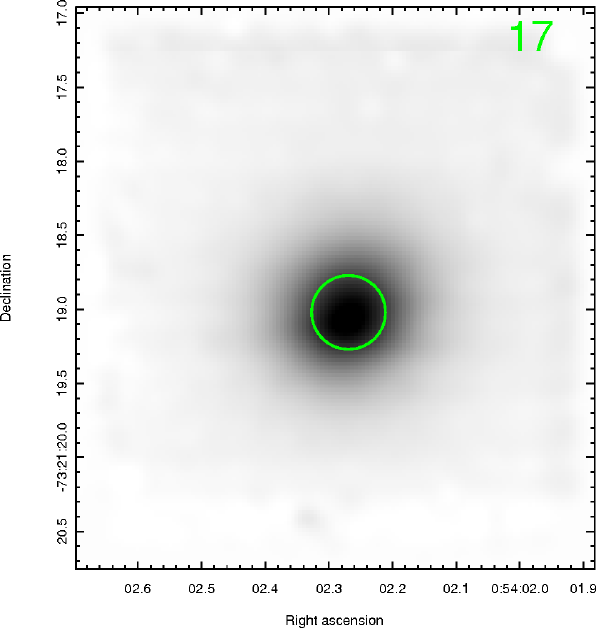}
 \end{center}
\caption{SINFONI \textit{K}-band continuum images for all observed FOVs.}
\end{minipage}
\end{figure*}

\begin{figure*}
\begin{minipage}{175mm}
 \begin{center}
\includegraphics[width=0.4\linewidth]{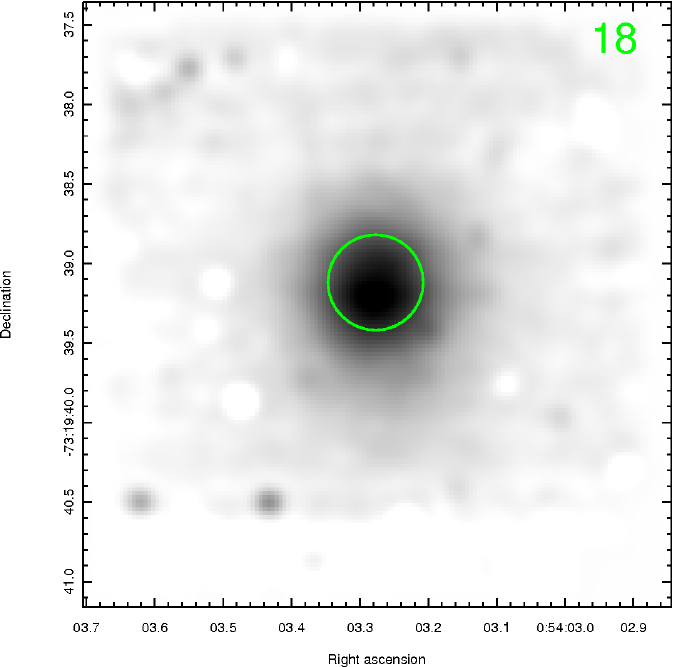}
\includegraphics[width=0.4\linewidth]{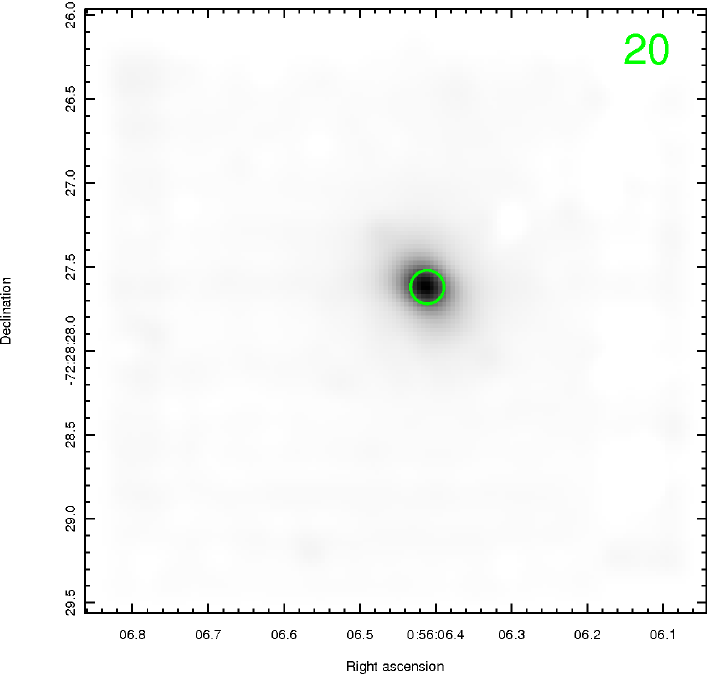}
\includegraphics[width=0.4\linewidth]{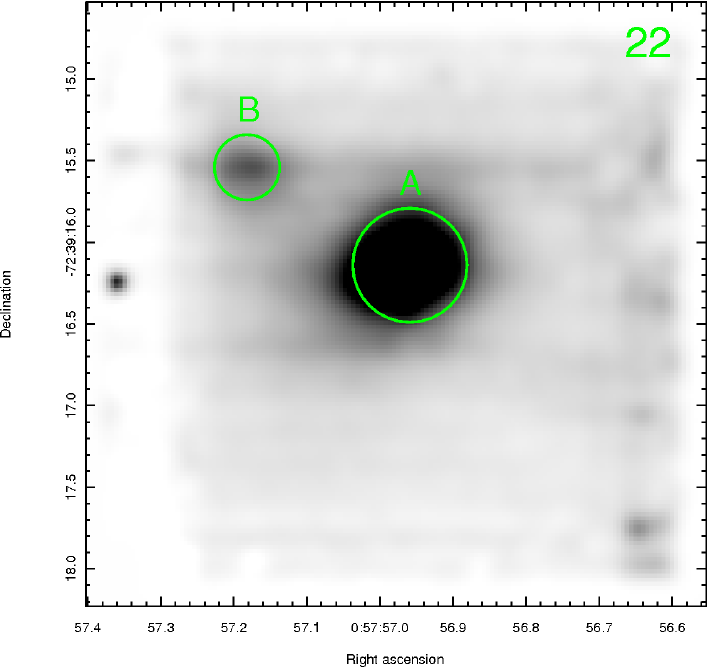}
\includegraphics[width=0.4\linewidth]{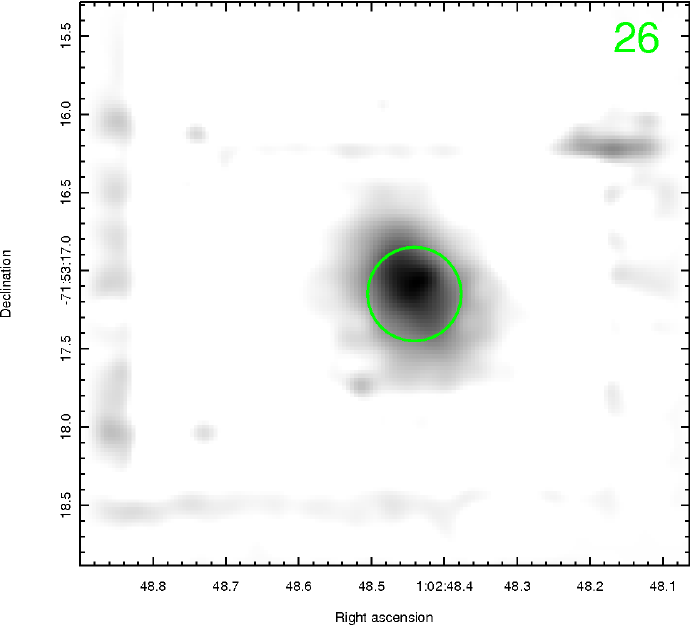}
\includegraphics[width=0.4\linewidth]{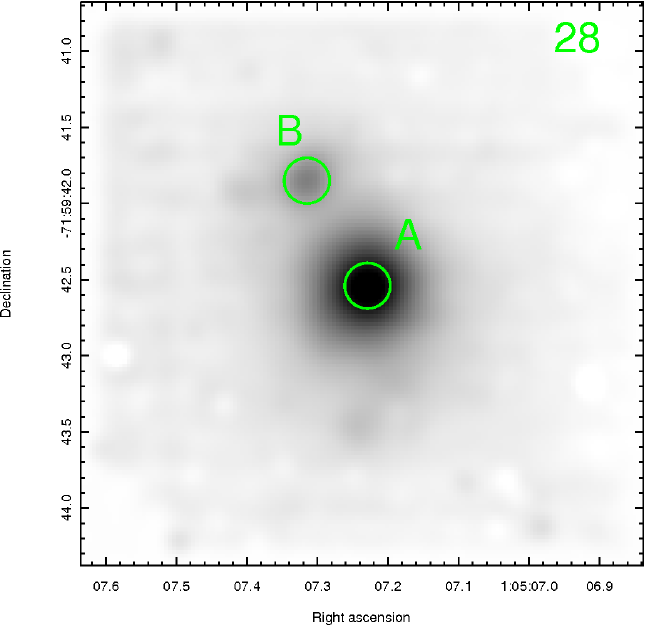}
\includegraphics[width=0.4\linewidth]{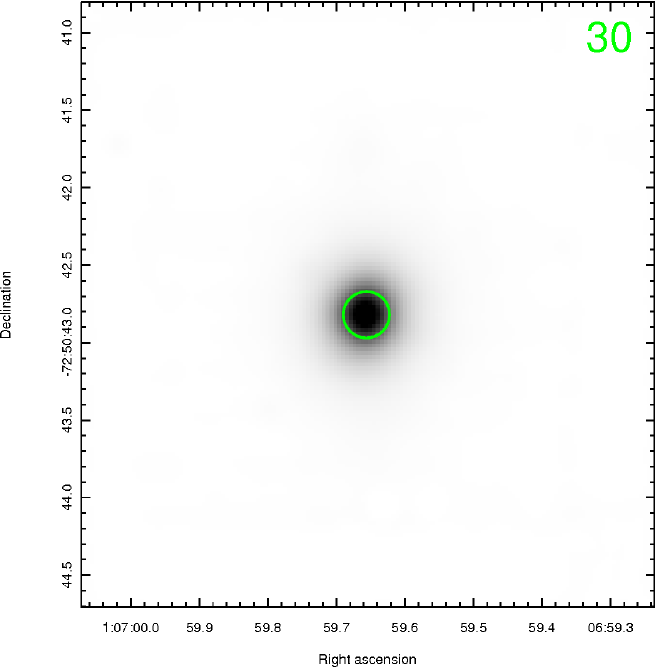}
 \end{center}

\textbf{Fig. A1 cont.} SINFONI \textit{K}-band continuum images for all observed FOVs.

\end{minipage}
\end{figure*}

\begin{figure*}
\begin{minipage}{175mm}
 \begin{center}
\includegraphics[width=0.4\linewidth]{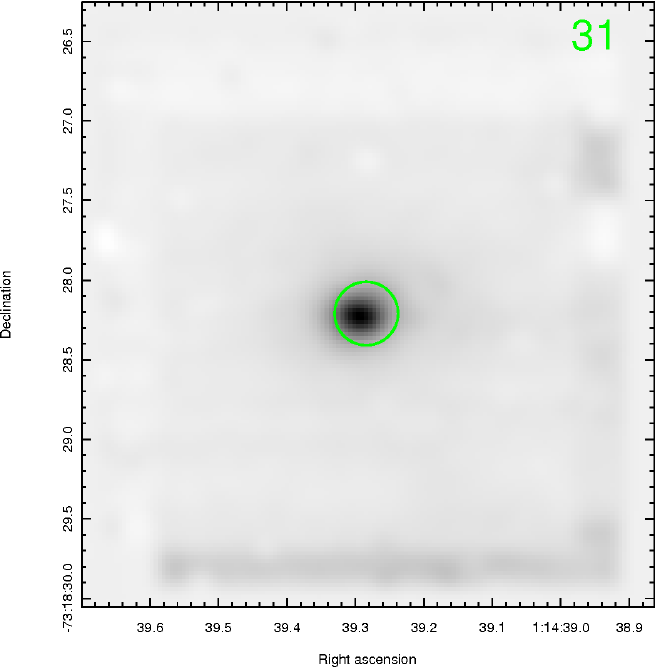}
\includegraphics[width=0.4\linewidth]{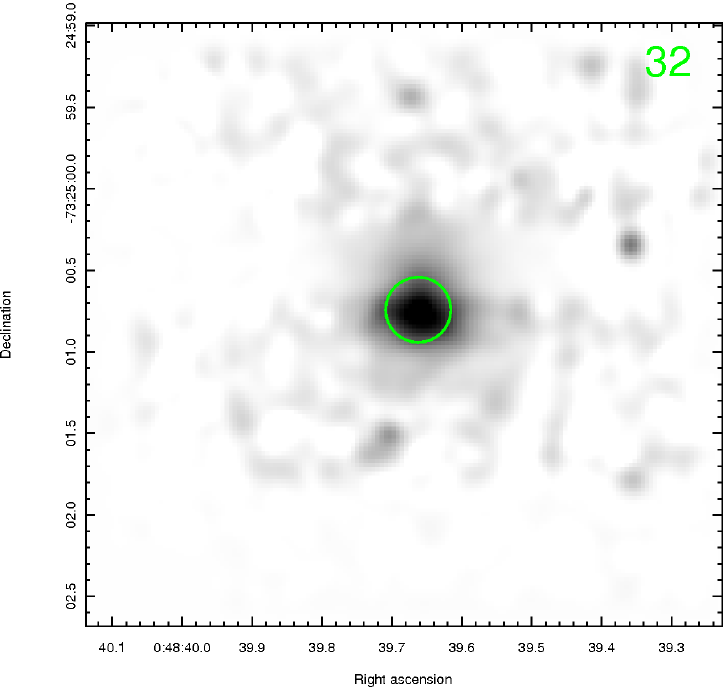}
\includegraphics[width=0.4\linewidth]{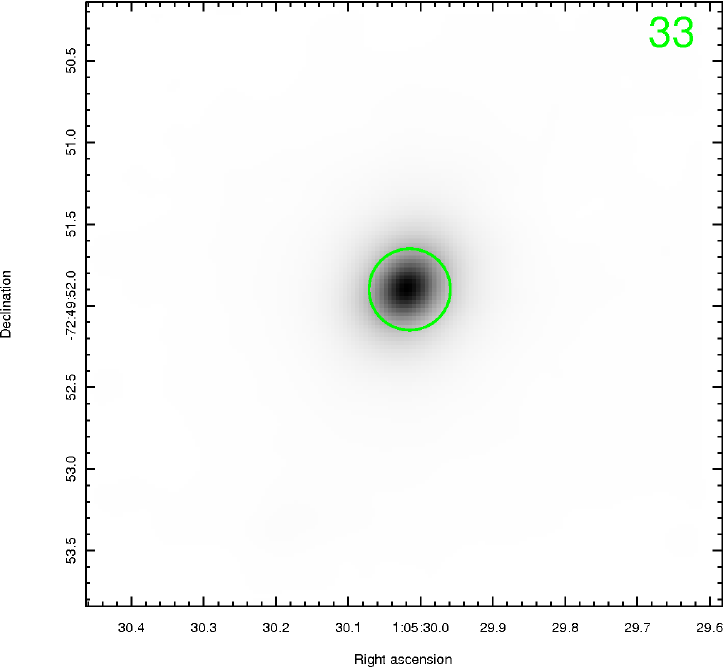}
\includegraphics[width=0.4\linewidth]{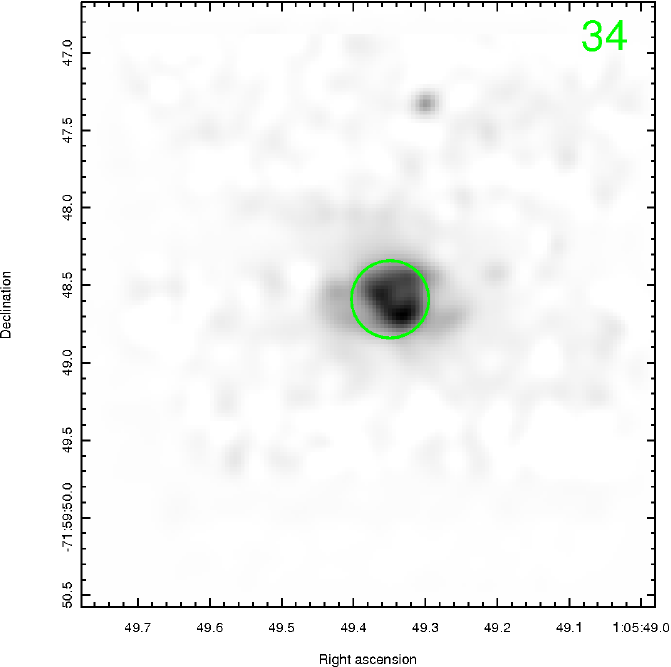}
\includegraphics[width=0.4\linewidth]{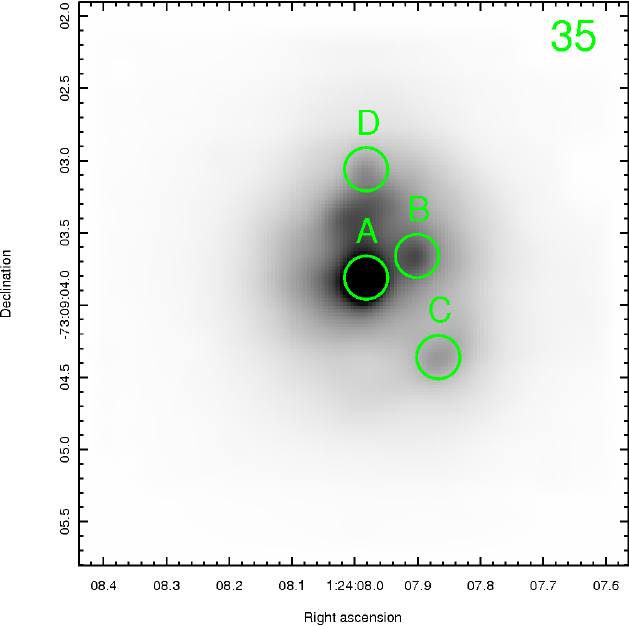}
\includegraphics[width=0.4\linewidth]{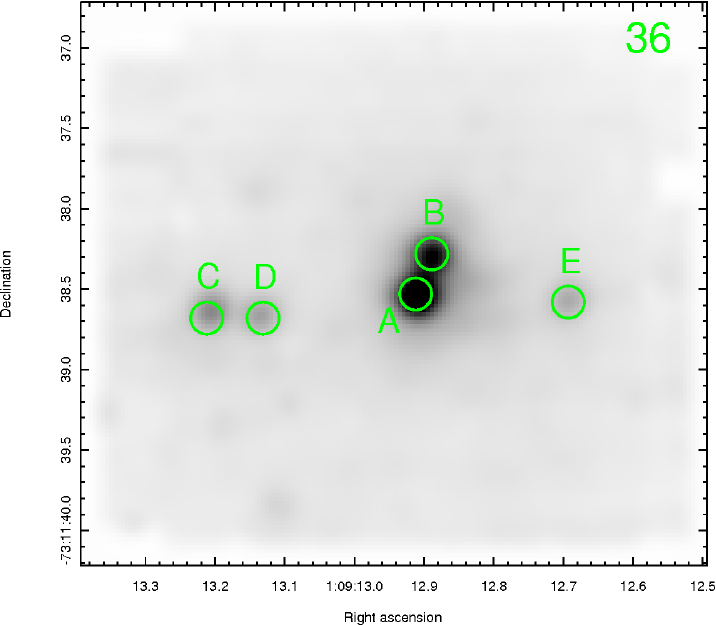}
 \end{center}

\textbf{Fig. A1 cont.} SINFONI \textit{K}-band continuum images for all observed FOVs.
\end{minipage}
\end{figure*}

\section{Extracted \textit{K-}band and optical Spectra}

Figure B1 shows the normalised \textit{K-}band extracted spectra for all of the sources marked in Fig. A1 with the positions of the emission lines of 
interest marked. A number of sky line residuals remain visible in many of the spectra; the positions of these are given in \citet{Ward2016}.
Figures B2, B3 and B4 show the optical spectra obtained with RSS at SALT for N81 and N88\,A. Note that a logarithmic flux scale has been used
for clarity. Emission lines of interest are marked on these spectra.

\begin{figure*}
\begin{minipage}{175mm}
\begin{center}
 \includegraphics[width=1.3\linewidth, angle=90]{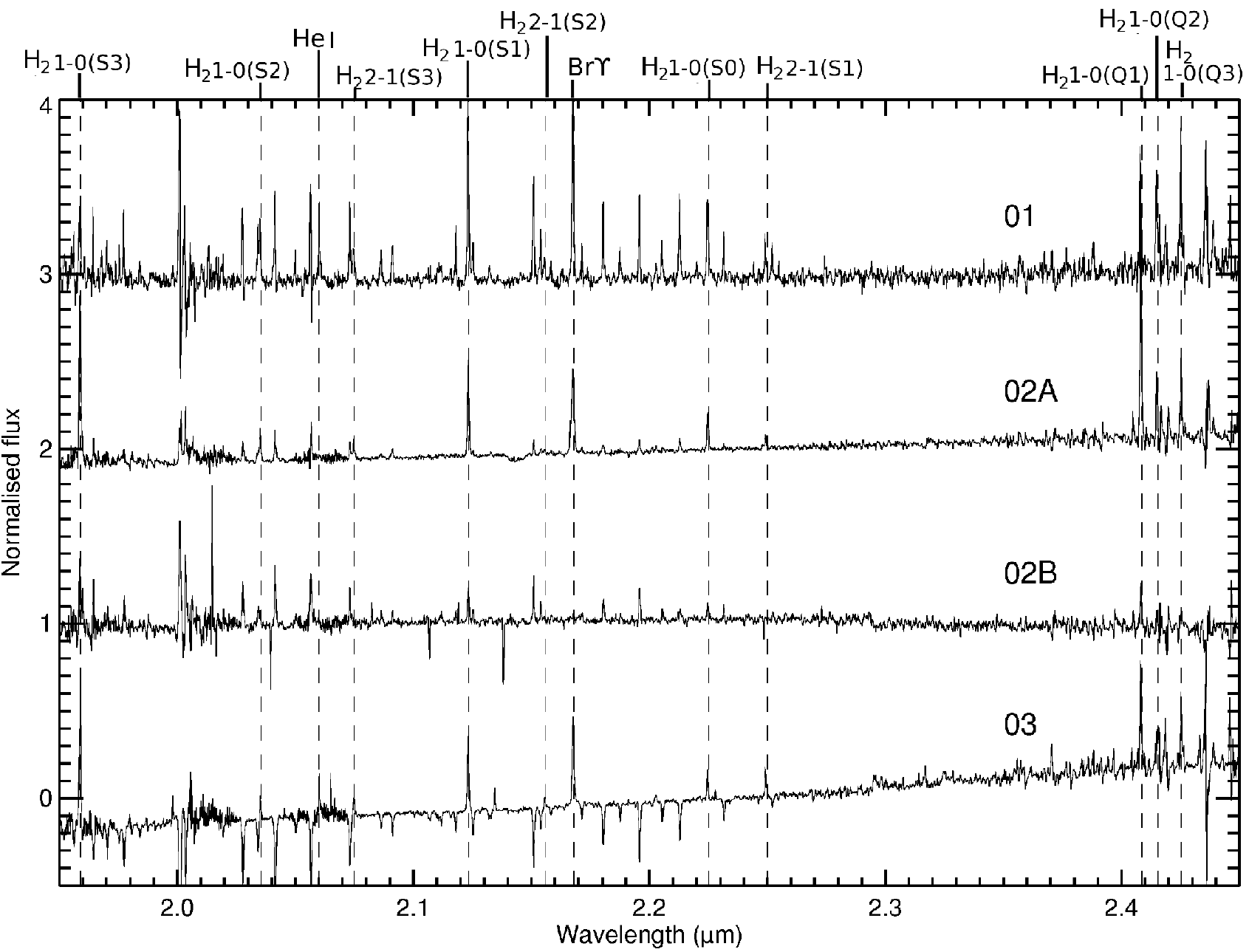}
\end{center}
 \caption{Continuum normalised extracted spectra towards all \textit{K}-band continuum sources detected. 
 Positions of the emission lines studied in this work are marked.}
\end{minipage}
\end{figure*}

\begin{figure*}
\begin{minipage}{175mm}
\begin{center}
 \includegraphics[width=1.3\linewidth, angle=90]{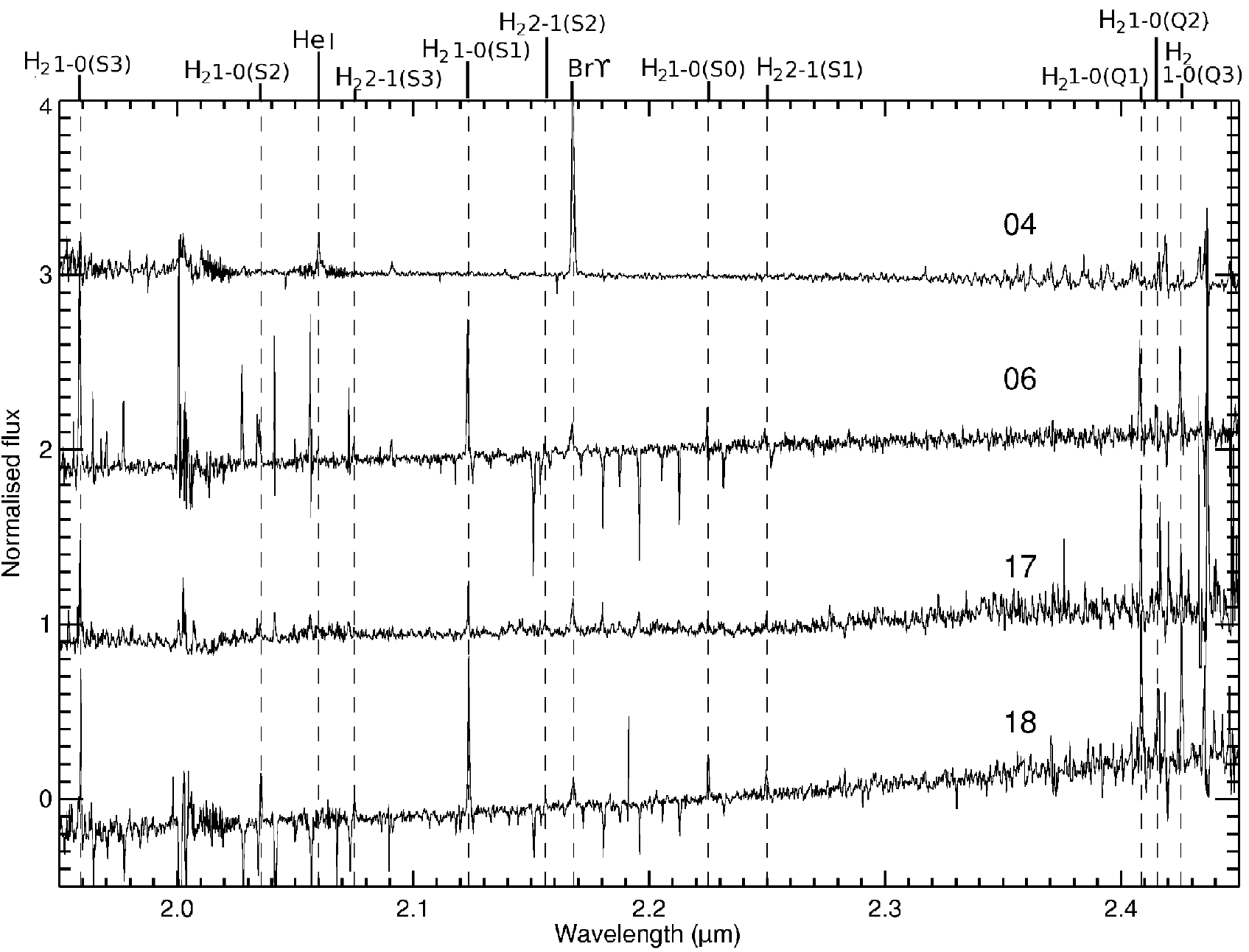}
\end{center}
 \textbf{Fig. B1 cont.} Extracted \textit{K-}band spectra.
\end{minipage}
\end{figure*}

\begin{figure*}
\begin{minipage}{175mm}
\begin{center}
 \includegraphics[width=1.3\linewidth, angle=90]{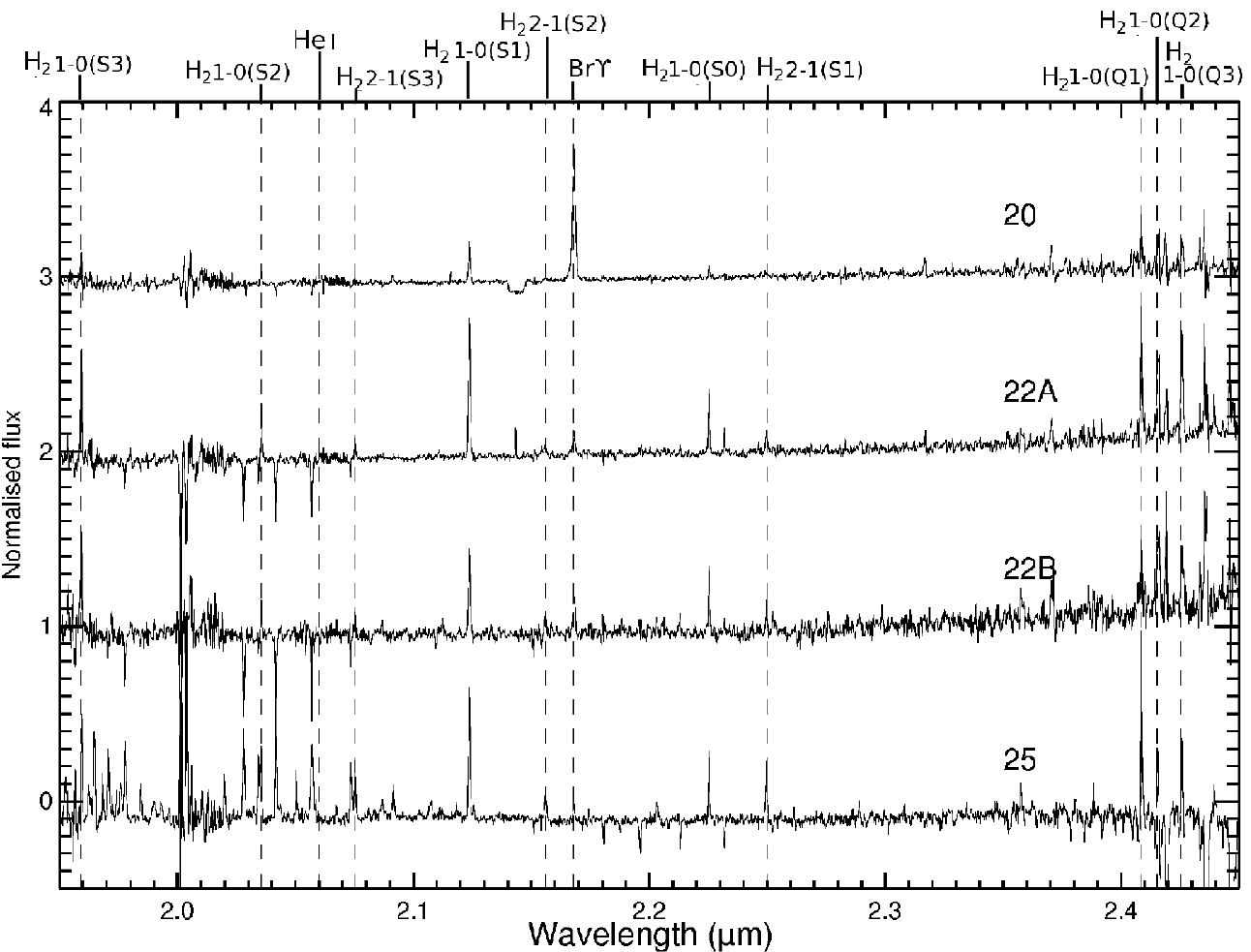}
\end{center}
 \textbf{Fig. B1 cont.} Extracted \textit{K-}band spectra.
\end{minipage}
\end{figure*}

\begin{figure*}
\begin{minipage}{175mm}
\begin{center}
 \includegraphics[width=1.3\linewidth, angle=90]{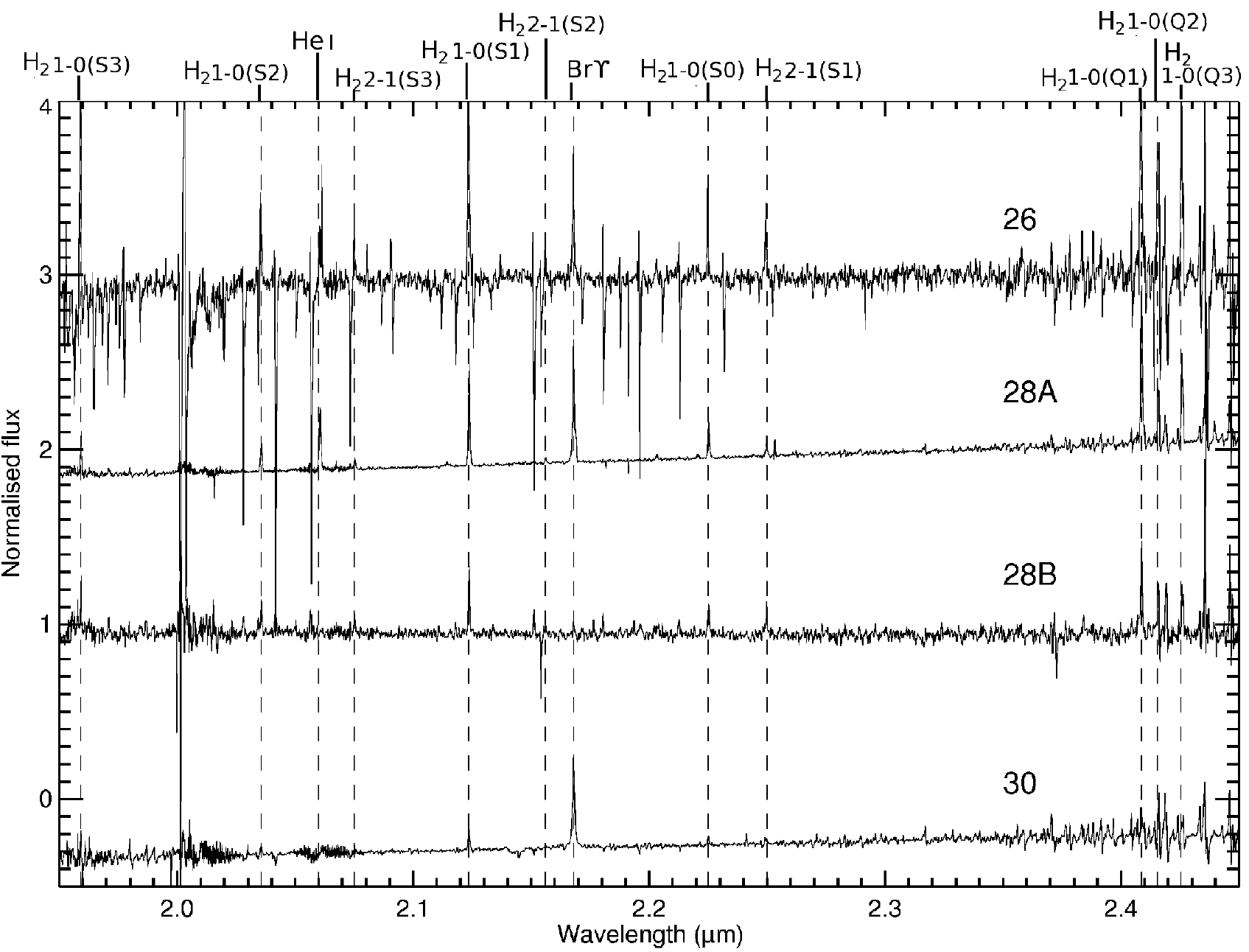}
\end{center}
 \textbf{Fig. B1 cont.} Extracted \textit{K-}band spectra.
\end{minipage}
\end{figure*}

\begin{figure*}
\begin{minipage}{175mm}
\begin{center}
 \includegraphics[width=1.3\linewidth, angle=90]{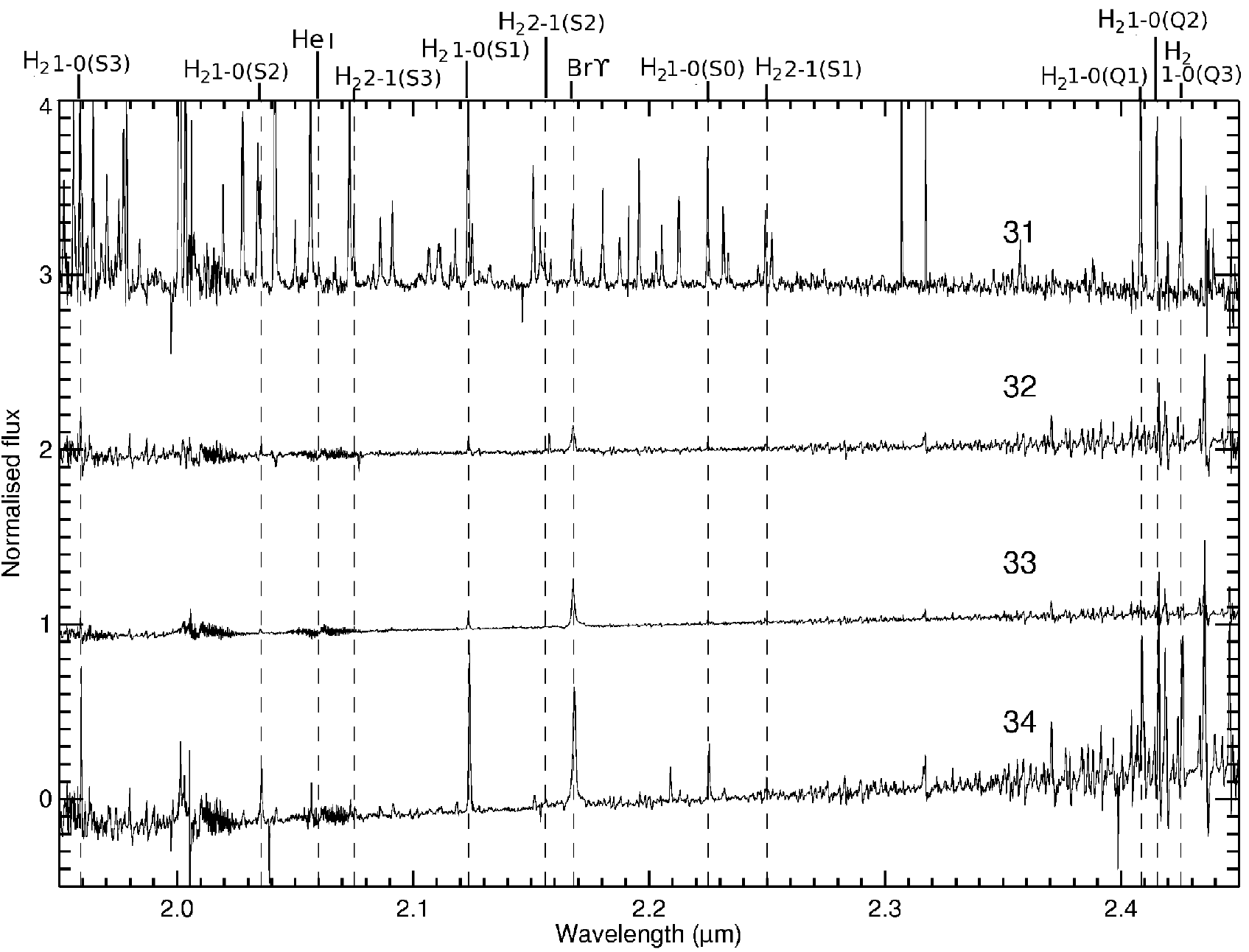}
\end{center}
\textbf{Fig. B1 cont.} Extracted \textit{K-}band spectra.
\end{minipage}
\end{figure*}

\begin{figure*}
\begin{minipage}{175mm}
\begin{center}
 \includegraphics[width=1.3\linewidth, angle=90]{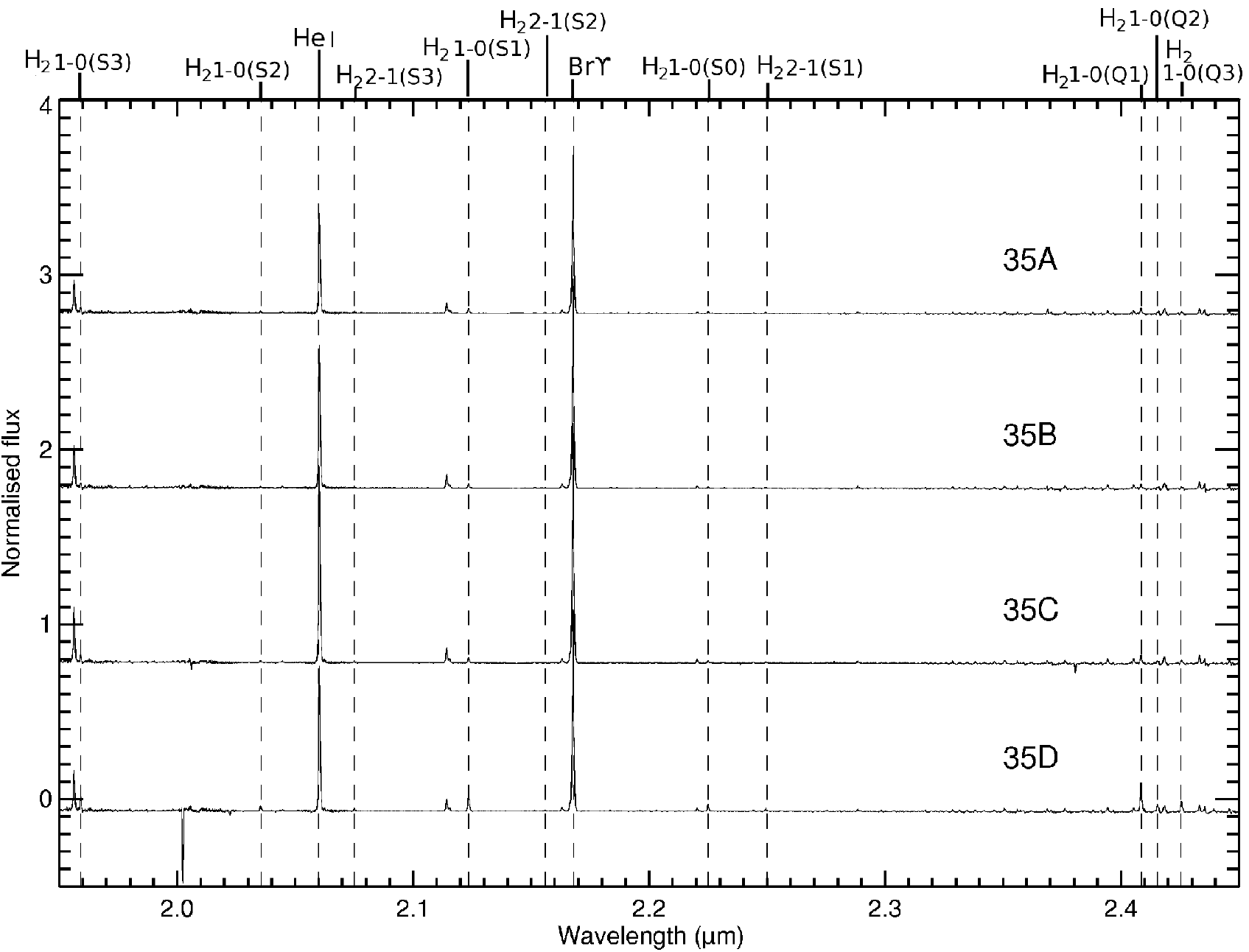}
\end{center}
\textbf{Fig. B1 cont.} Extracted \textit{K-}band spectra.
\end{minipage}
\end{figure*}

\begin{figure*}
\begin{minipage}{175mm}
\begin{center}
 \includegraphics[width=1.3\linewidth, angle=90]{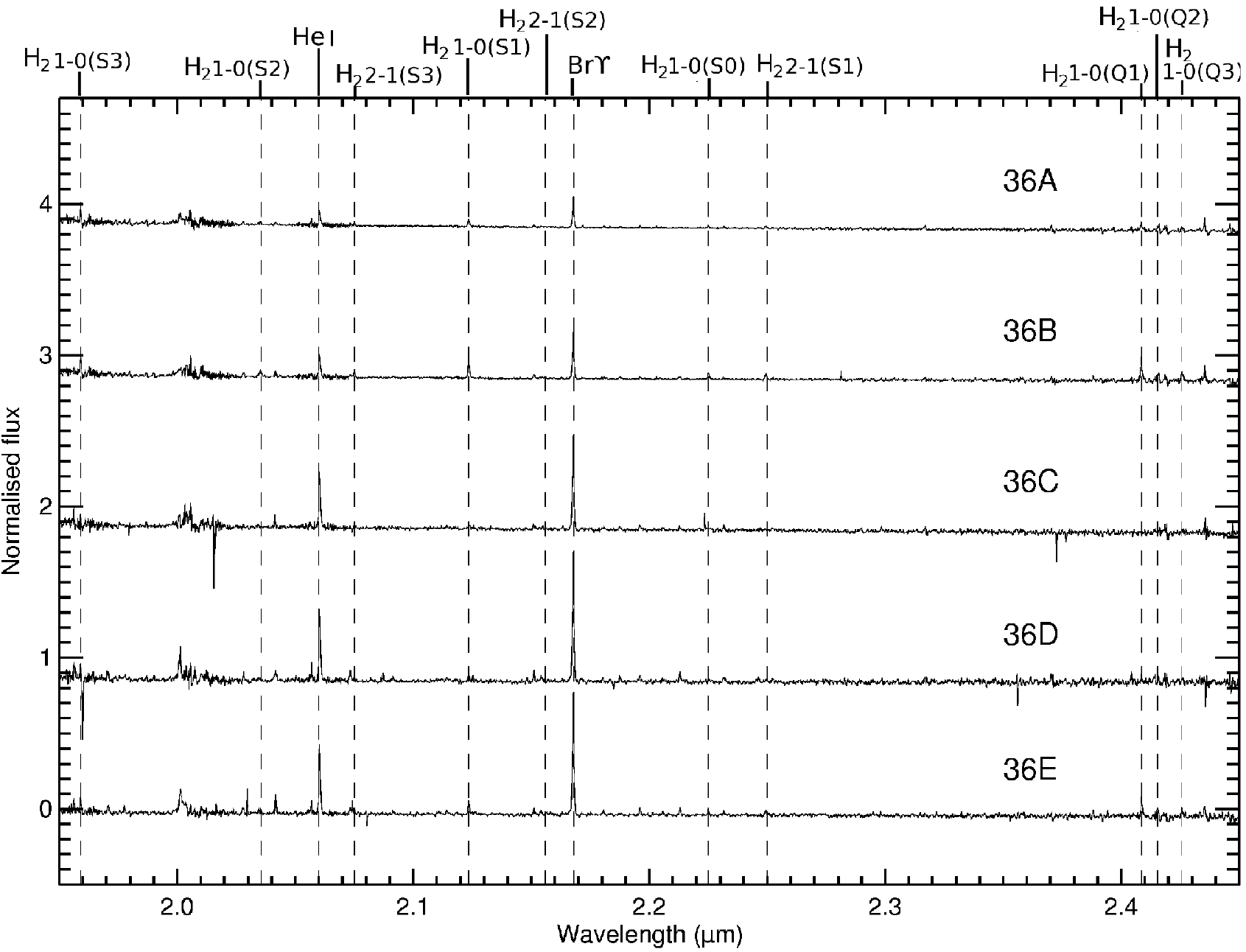}
\end{center}
\textbf{Fig. B1 cont.} Extracted \textit{K-}band spectra.
\end{minipage}
\end{figure*}

\begin{figure*}
\begin{minipage}{175mm}
\begin{center}
 \includegraphics[width=0.9\linewidth]{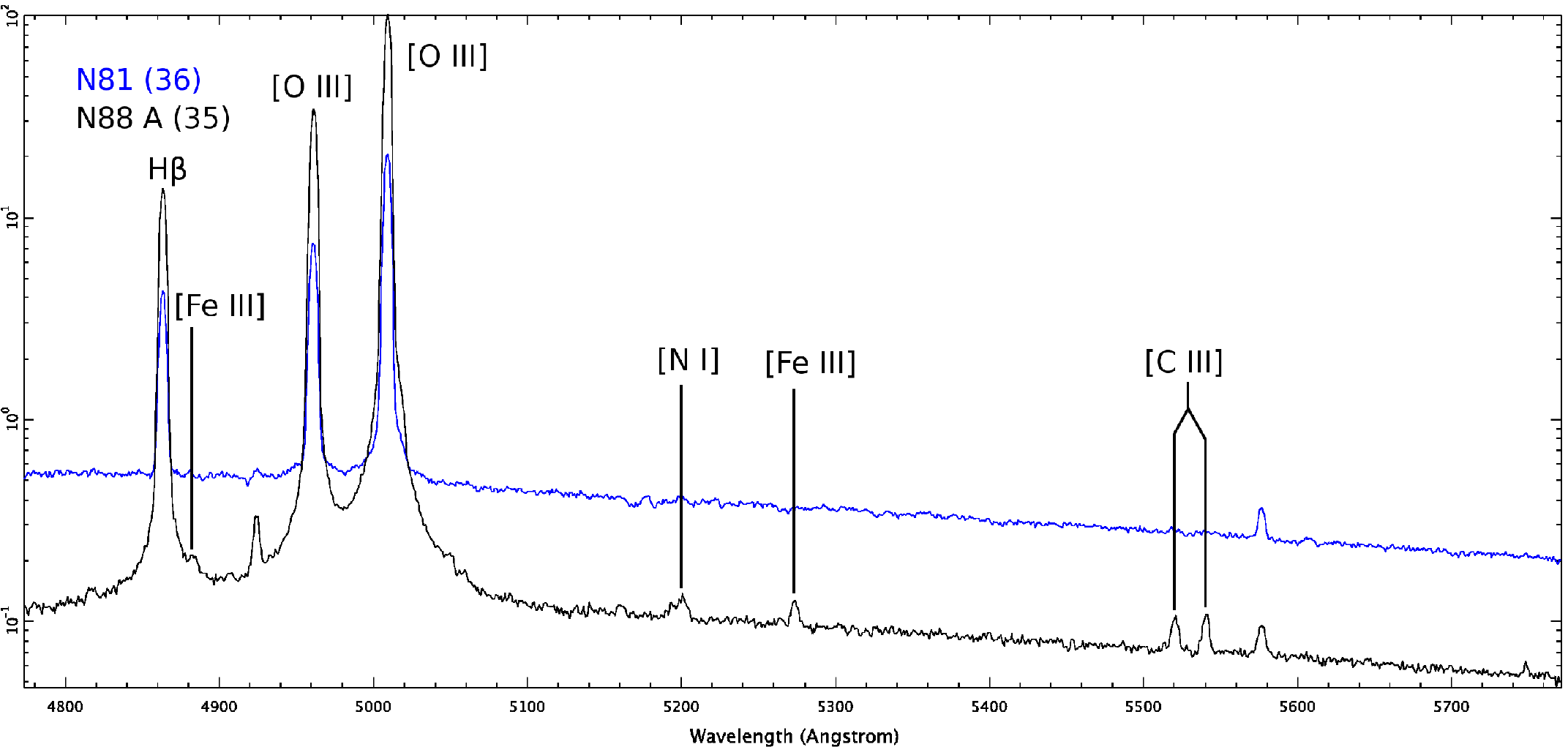}
 \includegraphics[width=0.9\linewidth]{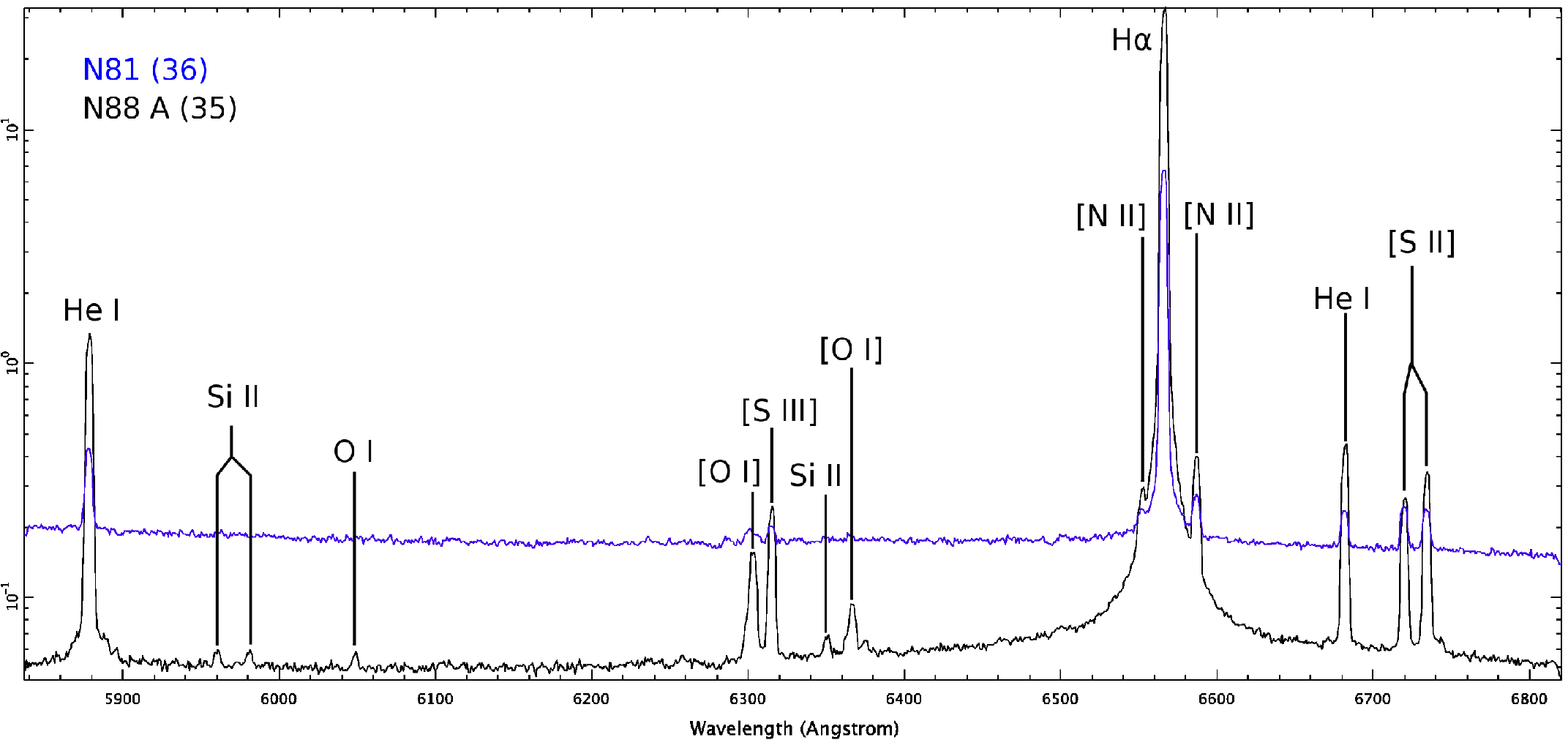}
 \includegraphics[width=0.9\linewidth]{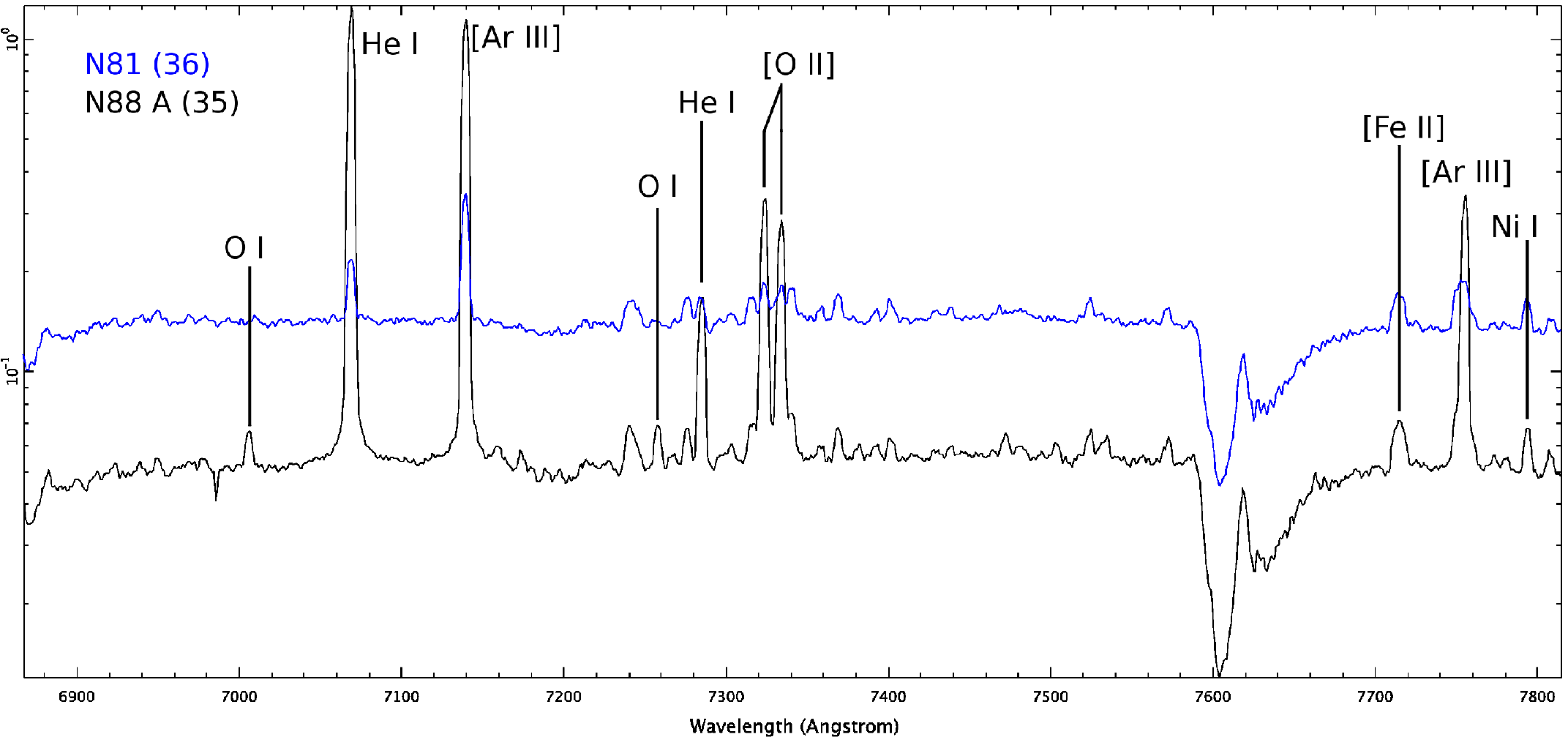}
\end{center}
\caption{Low resolution optical spectra of sources 35 and 36 (N88\,A and N81, respectively)}
\end{minipage}
\end{figure*}

\begin{figure*}
\begin{minipage}{175mm}
\begin{center}
 \includegraphics[width=0.9\linewidth]{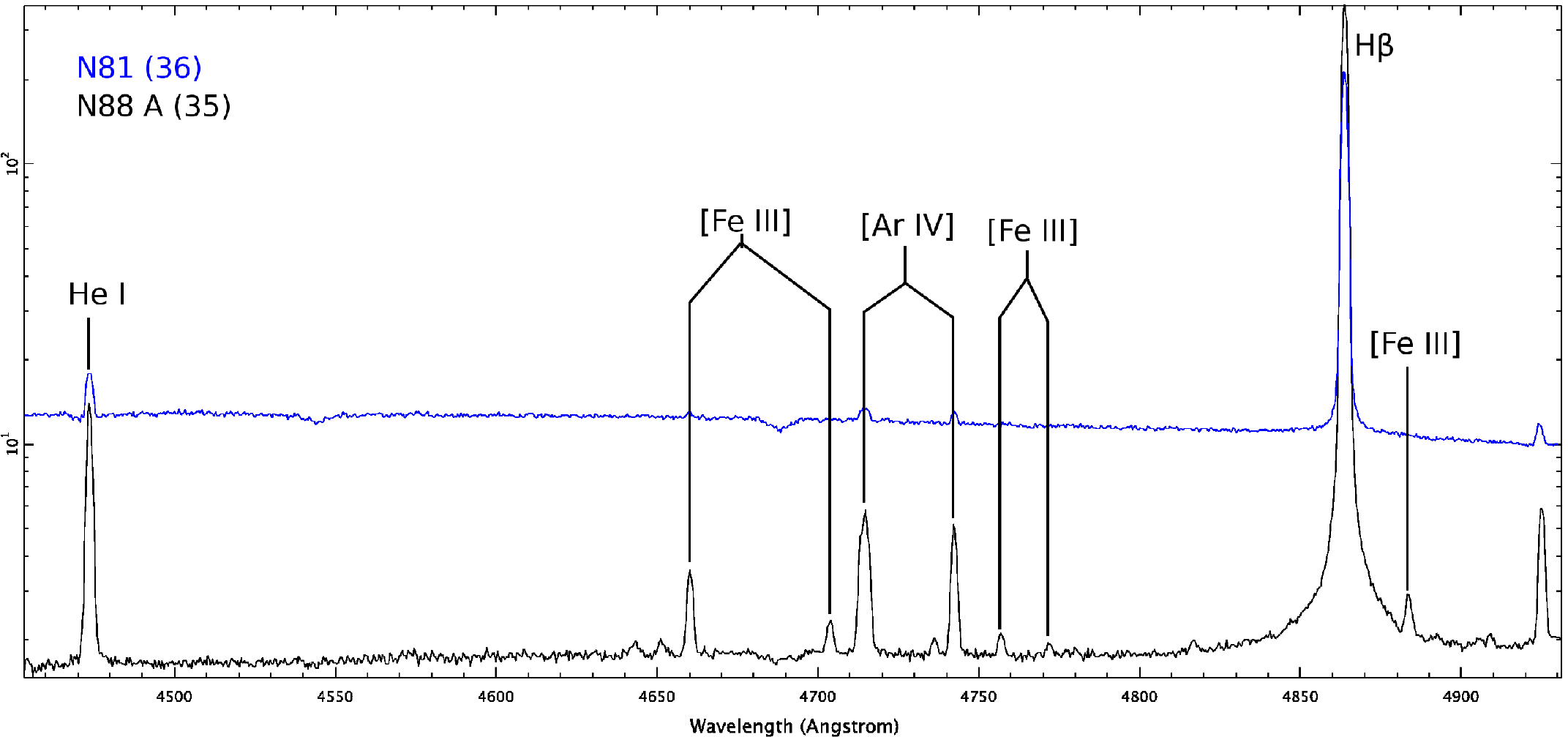}
 \includegraphics[width=0.9\linewidth]{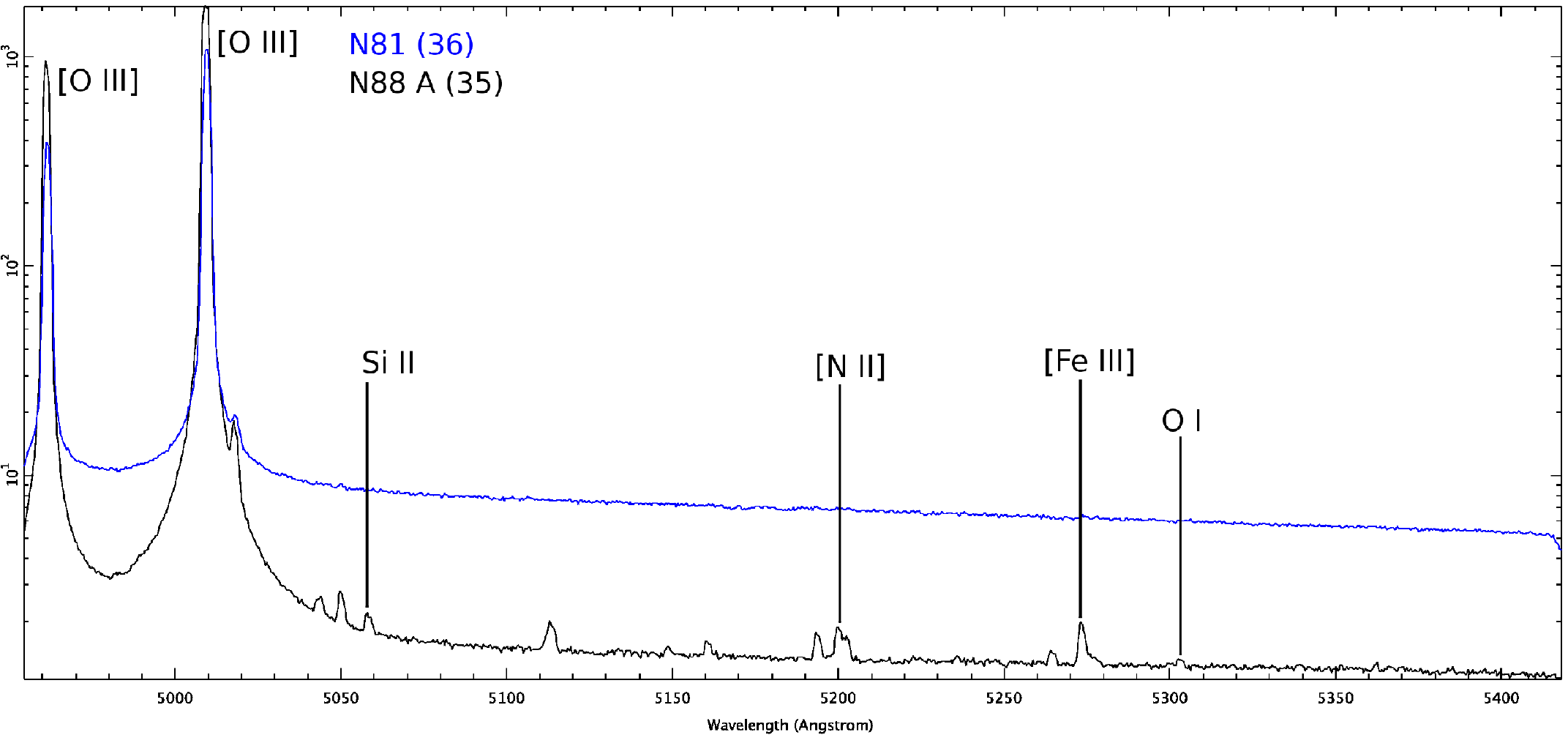}
 \includegraphics[width=0.9\linewidth]{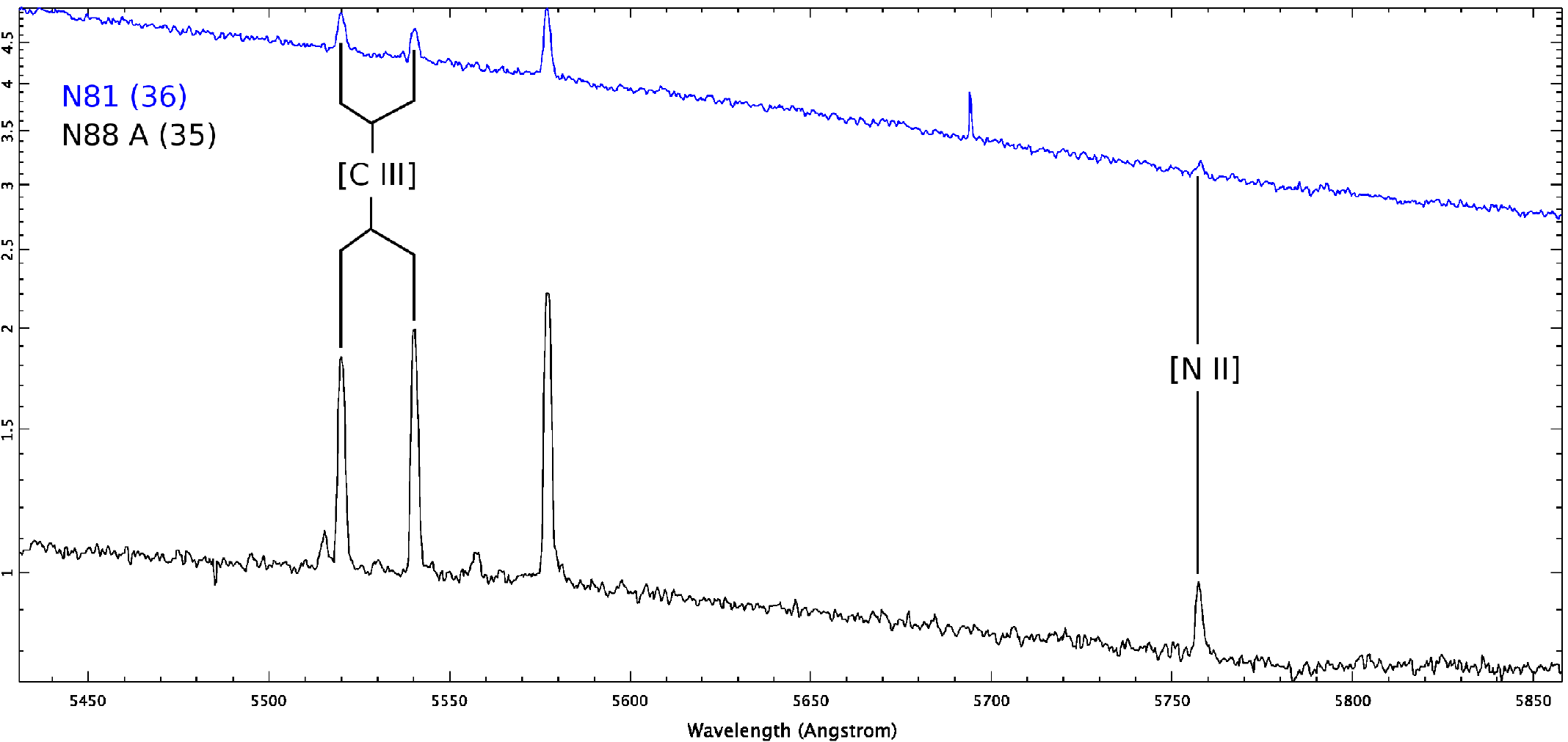}
\end{center}
\caption{Medium resolution blue optical spectra of sources 35 and 36 (N88\,A and N81, respectively)}
\end{minipage}
\end{figure*}

\begin{figure*}
\begin{minipage}{175mm}
\begin{center}
 \includegraphics[width=0.9\linewidth]{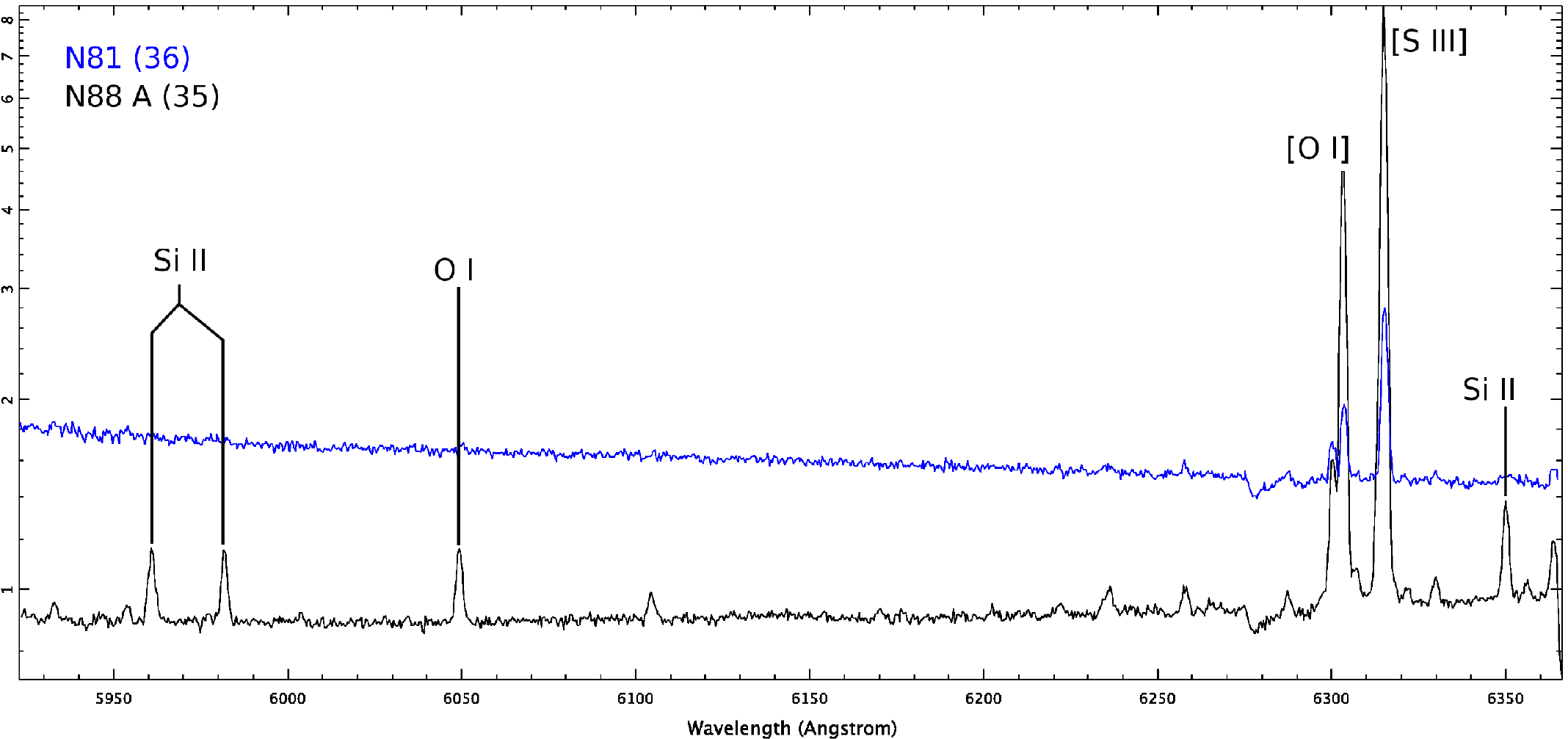}
 \includegraphics[width=0.9\linewidth]{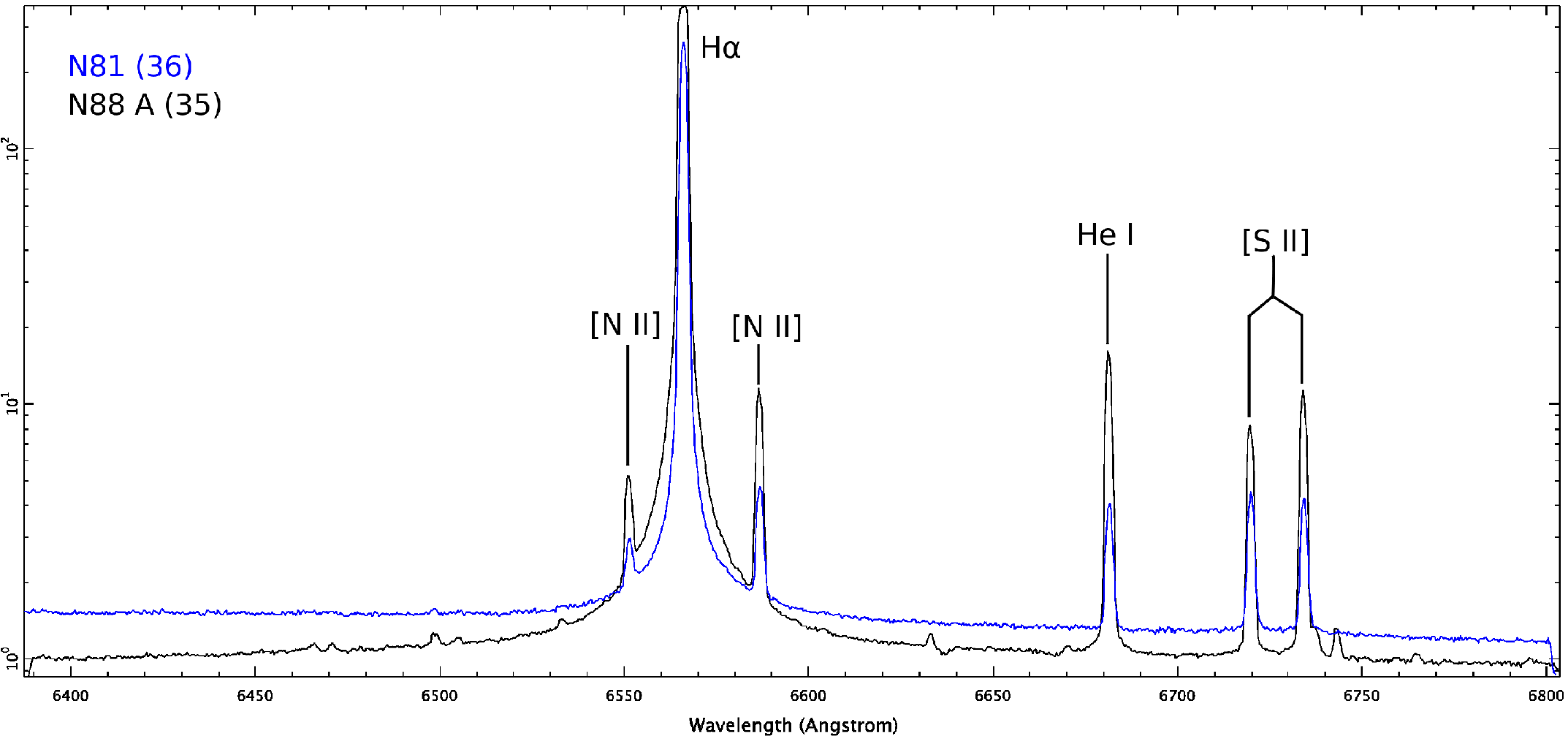}
 \includegraphics[width=0.9\linewidth]{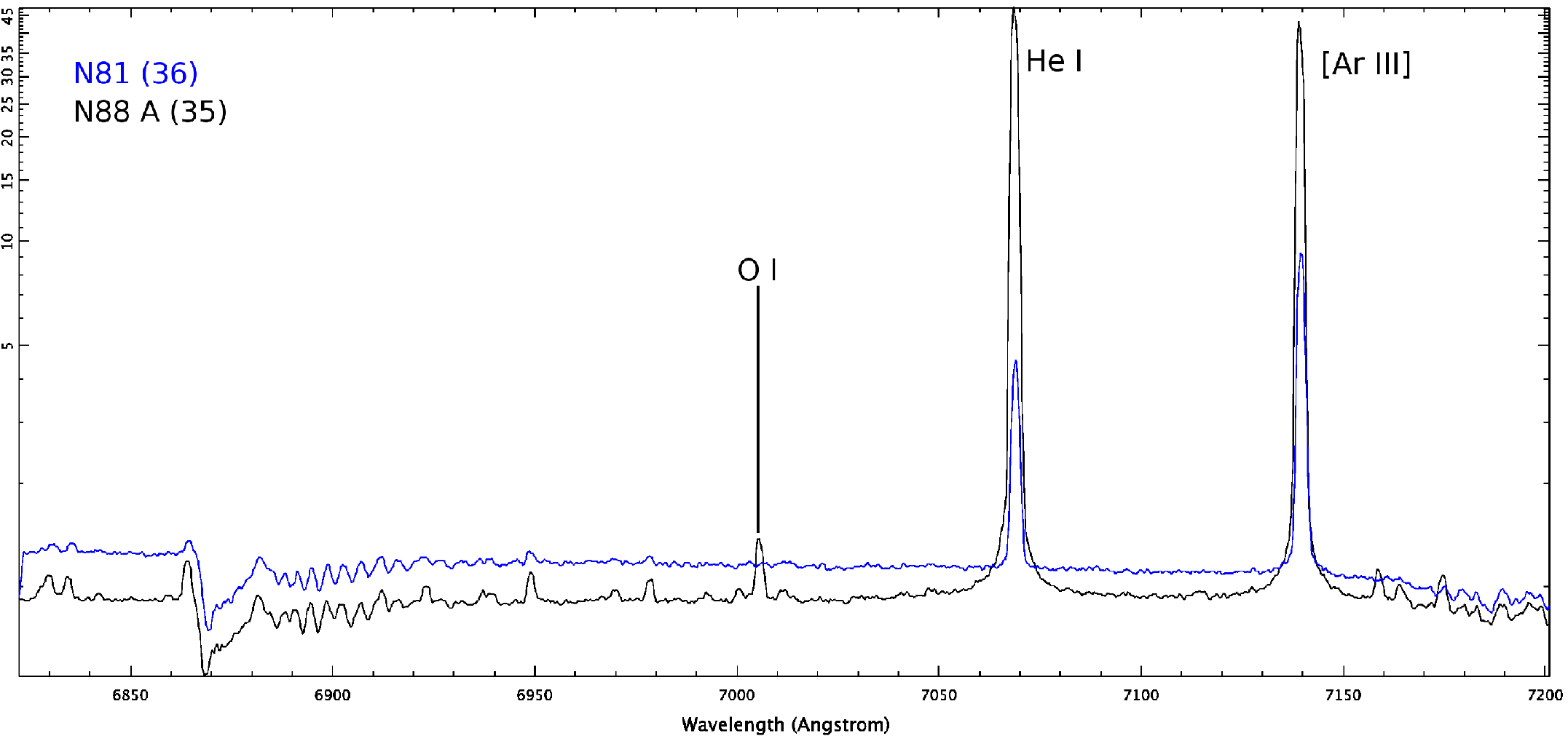}
\end{center}
\caption{Medium resolution red optical spectra of sources 35 and 36 (N88\,A and N81, respectively)}
\end{minipage}
\end{figure*}

\section{Emission line fluxes}

Tables C1 and C2 give the measured emission line fluxes for all spectra. Those for the \textit{K-}band spectra are presented in Table C1
and those for the optical spectra are presented in Table C2. \textquotedblleft p\textquotedblright indicates that an emission line is present but
that a measurement cannot be obtained and \textquotedblleft ?\textquotedblright indicates a significant uncertainty in the identification of the 
emission line.

\begin{table*}
\begin{minipage}{175mm}
\caption{K-band emission line fluxes measured towards all sources. The fluxes have not been corrected for extinction.}
\scriptsize
\begin{tabular}{c c c c c c c}
\hline
Source & He {\sc i} (2.058$\mu$m) & Br$\gamma$ &  H$_2$ 1--0(S0) & H$_2$ 1--0(S1) & H$_2$ 1--0(S2) & H$_2$1--0(S3) \\
	& 10$^{-19}$ W/m$^{2}$	& 10$^{-19}$ W/m$^{2}$	& 10$^{-19}$ W/m$^{2}$	& 10$^{-19}$ W/m$^{2}$	& 10$^{-19}$ W/m$^{2}$	& 10$^{-19}$ W/m$^{2}$ \\
\hline
01	& 	1.6$\pm$0.2	&	5.3$\pm$0.3	&	2.2$\pm$0.2	&	4.4$\pm$0.3	&	1.9$\pm$0.3	&	2.4$\pm$0.4	\\
02\,A	&	\null	&	9.8$\pm$0.6	&	2.3$\pm$0.2	&	6.1$\pm$0.4	&	2.2$\pm$0.4	&	11$\pm$1	\\
02\,B	&	\null	&	\null	&	0.7$\pm$0.2	&	1.18$\pm$0.07	&	0.4$\pm$0.1	&	1.8$\pm$0.4	\\
03	&	5.5$\pm$1.3	&	16.2$\pm$1.2	&	2.8$\pm$0.3	&	11.1$\pm$0.9	&	2.5$\pm$0.8	&	20$\pm$3	\\
04	&	6.3$\pm$1.0	&	36.2$\pm$0.8	&	\null	&	\null	&	\null	&	\null	\\
06	&	\null	&	0.41$\pm$0.06	&	0.14$\pm$0.04	&	0.92$\pm$0.06	&	0.32$\pm$0.08	&	1.4$\pm$0.1	\\
17	&	\null	&	2.0$\pm$0.1	&	\null	&	1.2$\pm$0.2	&	0.4$\pm$0.2	&	2.4$\pm$0.4	\\
18	&	\null	&	0.93$\pm$0.09	&	0.71$\pm$0.05	&	2.4$\pm$0.2	&	0.8$\pm$0.2	&	2.0$\pm$0.4	\\
20	&	\null	&	8.1$\pm$0.2	&	3.0$\pm$0.4	&	1.1$\pm$0.1	&	0.40$\pm$0.08	&	0.8$\pm$0.3 \\
22\,A	&	\null	&	1.3$\pm$0.1	&	1.86$\pm$0.09	&	4.5$\pm$0.2	&	1.7$\pm$0.2	&	2.9$\pm$0.2	\\
22\,B	&	\null	&	0.46$\pm$0.05	&	0.43$\pm$0.04	&	0.60$\pm$0.04	&	0.19$\pm$0.05	&	0.8$\pm$0.1	\\
25	&	\null	&	0.33$\pm$0.06	&	1.05$\pm$0.09	&	2.4$\pm$0.2	&	\null	&	\null	\\
26	&	0.9$\pm$0.2	&	2.3$\pm$0.2	&	1.0$\pm$0.1	&	3.9$\pm$0.4	&	1.1$\pm$0.5	&	3.6$\pm$0.6	\\
28\,A	&	6.4$\pm$0.4	&	13.8$\pm$1.0	&	3.0$\pm$0.3	&	8.6$\pm$0.7	&	2.6$\pm$0.2	&	3.6$\pm$0.4	\\
28\,B	&	\null	&	0.09$\pm$0.03	&	0.25$\pm$0.03	&	0.43$\pm$0.03	&	0.19$\pm$0.07	& 0.3$\pm$0.1	\\
30	&	\null	&	15.3$\pm$0.7	&	0.77$\pm$0.09	&	2.3$\pm$0.3	&	0.8$\pm$0.2	&	2.3$\pm$1.1	\\
31	&	1.0$\pm$0.4	&	5.9$\pm$0.4	&	7.4$\pm$0.6	&	15$\pm$1	&	7$\pm$1	&	16$\pm$4	\\
32	&	\null	&	5.2$\pm$0.4	&	0.44$\pm$0.06	&	1.3$\pm$0.2	&	0.68$\pm$0.05	&	3.6$\pm$0.9	\\
33	&	\null	&	158$\pm$9	&	6.5$\pm$0.9	&	21$\pm$2	&	5$\pm$1	&	p?	\\
34	&	\null	&	20$\pm$1	&	4.2$\pm$0.3	&	13.6$\pm$0.8	&	4.4$\pm$0.5	&	9$\pm$1	\\
35\,A	&	197$\pm$9	&	326$\pm$18	&	2.9$\pm$0.3	&	7.7$\pm$0.8	&	2.3$\pm$0.2	&	7$\pm$3	\\
35\,B	&	91$\pm$5	&	138$\pm$10	&	0.7$\pm$0.2	&	2.2$\pm$0.4	&	0.5$\pm$0.1	&	2.0$\pm$0.7	\\
35\,C	&	100$\pm$6	&	127$\pm$10	&	0.9$\pm$0.2		&	2.3$\pm$0.3	&	1.0$\pm$0.3	&	2.8$\pm$0.8	\\
35\,D	&	68$\pm$4	&	98$\pm$9	&	2.7$\pm$0.3	&	7.2$\pm$0.8	&	2.0$\pm$0.2	&	5.4$\pm$1.0	\\
36\,A	&	3.1$\pm$0.5	&	6.3$\pm$0.3	&	0.4$\pm$0.1	&	1.5$\pm$0.2	&	0.55$\pm$0.06	&	1.7$\pm$0.9	\\
36\,B	&	3.0$\pm$0.3	&	6.1$\pm$0.3	&	0.85$\pm$0.07	&	1.9$\pm$0.2	&	0.27$\pm$0.07	&	1.6$\pm$0.4	\\
36\,C	&	4.1$\pm$0.4	&	6.8$\pm$0.4	&	\null	&	\null	&	\null	&	p	\\
36\,D	&	3.6$\pm$0.3	&	6.2$\pm$0.4	&	\null	&	0.27$\pm$0.04	&	\null	&	0.39$\pm$0.09	\\
36\,E	&	4.4$\pm$0.4	&	8.2$\pm$0.7	&	0.32$\pm$0.06	&	0.72$\pm$0.07	&	\null	&	0.6$\pm$0.2	\\
\hline
Source & H$_2$ 2--1(S1) & H$_2$ 2--1(S2) & H$_2$ 2--1(S3) & 1--0Q(1) & 1--0Q(2) & 1--0Q(3) \\
	& 10$^{-19}$ W/m$^{2}$	& 10$^{-19}$ W/m$^{2}$	& 10$^{-19}$ W/m$^{2}$	& 10$^{-19}$ W/m$^{2}$	& 10$^{-19}$ W/m$^{2}$	& 10$^{-19}$ W/m$^{2}$ \\
\hline
01	& 	0.9$\pm$0.2	&	0.7$\pm$0.1	&	1.5$\pm$0.6	&	2.7$\pm$0.5	&	2.2$\pm$0.4	&	3.3$\pm$0.6	\\
02\,A	&	0.8$\pm$0.1	&	p	&	1.1$\pm$0.4	&	14$\pm$1	&	2.7$\pm$0.4	&	5.2$\pm$0.8	\\
02\,B	&	\null	&	0.2$\pm$0.1	&	\null	&	1.5$\pm$0.3	&	\null	&	0.9$\pm$0.3	\\
03	&	3.1$\pm$0.4	&	1.2$\pm$0.3	&	2.4$\pm$0.6	&	15$\pm$2	& 	\null	&	10$\pm$2 \\
04	&	\null	&	\null	&	\null	&		\null	&	\null	&	\null \\
06	&	0.11$\pm$0.05	&	$<$1.0	&	0.08$\pm$0.03	&	0.76$\pm$0.09	&	0.23$\pm$0.06	&	0.75$\pm$0.08	\\
17	&	p?	&	\null	&	\null	&	3.0$\pm$0.5	&	2.2$\pm$0.6	&	2.2$\pm$0.3	\\
18	&	0.52$\pm$0.09	&	0.18$\pm$0.03	&	0.4$\pm$0.1	&	2.6$\pm$0.5	&	2.2$\pm$0.6	&	2.5$\pm$0.4	\\
20	&	\null	&	\null	&	\null	&	1.7$\pm$0.3	&	?	&	1.2$\pm$0.3	\\
22\,A	&	0.91$\pm$0.09	&	0.6$\pm$0.1	&	0.51$\pm$0.08	&	4.7$\pm$0.6	&	3.8$\pm$0.7	&	3.9$\pm$0.4	\\
22\,B	&	0.21$\pm$0.03	&	0.17$\pm$0.03	&	0.12$\pm$0.04	&	0.6$\pm$0.1	&	1.0$\pm$0.2	&	0.7$\pm$0.2	\\
25	&	1.1$\pm$0.1	&	?	&	?	&	?	&	?	&	1.3$\pm$0.3	\\
26	&	1.1$\pm$0.2	&	0.42$\pm$0.05	&	0.7$\pm$0.1	&	4.7$\pm$0.8	&	3.6$\pm$1.5	&	3.7$\pm$1.0	\\
28\,A	&	1.6$\pm$0.2	&	0.44$\pm$0.05	&	1.0$\pm$0.2	&	14$\pm$2	&	7$\pm$1	&	9$\pm$1	\\
28\,B	&	0.21$\pm$0.05	&	\null	&	0.12$\pm$0.02	&	0.7$\pm$0.2	&	0.5$\pm$0.1	&	0.5$\pm$0.1	\\
30	&	\null	&	\null	&	\null	&	3.0$\pm$0.9	&	4$\pm$2	&	2$\pm$1	\\
31	&	3.6$\pm$0.4	&	2.0$\pm$0.5	&	4.3$\pm$0.8	&	16$\pm$1	&	9.4$\pm$0.8	&	11$\pm$1	\\
32	&	\null	&	\null	&	\null	&	\null	&	\null	&	1.4$\pm$0.5	\\
33	&	p?	&	\null	&	\null	&	14$\pm$4	&	p	&	13$\pm$2 \\
34	&	1.8$\pm$0.4	&	1.4$\pm$0.4	&	\null	&	13$\pm$3	&	14$\pm$3	&	13$\pm$2	\\
35\,A	&	2.6$\pm$0.3	&	0.9$\pm$0.3	&	1.5$\pm$0.5	&	11$\pm$2	&	7$\pm$3	&	6$\pm$2	\\
35\,B	&	0.70$\pm$0.09	&	0.18$\pm$0.08	&	p?	&	2.8$\pm$0.5	&	2.1$\pm$0.8	&	1.4$\pm$0.5	\\
35\,C	&	0.8$\pm$0.1	&	0.22$\pm$0.06	&	0.7$\pm$0.2	&	3.8$\pm$0.5	&	2.0$\pm$0.8	&	1.7$\pm$0.4	\\
35\,D	&	1.2$\pm$0.1	&	0.43$\pm$0.06	&	0.9$\pm$0.2	&	?	&	?	&5.0$\pm$0.7	\\
36\,A	&	0.6$\pm$0.1	&	\null	&	p?	&	1.9$\pm$0.4	&	1.3$\pm$0.6	&	0.9$\pm$0.5	\\
36\,B	&	0.8$\pm$0.1	&	0.27$\pm$0.07	&	0.6$\pm$0.1	&	3.2$\pm$0.5	&	1.5$\pm$0.5	&	1.7$\pm$0.4	\\
36\,C	&	0.29$\pm$0.06	&	p?	&	0.14$\pm$0.03	&	?	&	?	&	0.26$\pm$0.09	\\
36\,D	&	\null	&	\null	&	\null	&	0.5$\pm$0.1	&	?	&	0.4$\pm$0.1	\\
36\,E	&	\null	&	\null	&	\null	&	1.4$\pm$0.2	&	0.6$\pm$0.3	&	0.8$\pm$0.3	\\
\hline
Source  & Pf 20--5 & Pf 21--5 & Pf 22--5 & Pf 23--5 & Pf 24--5 & Pf 25--5\\
&10$^{-19}$ W/m$^{2}$	&10$^{-19}$ W/m$^{2}$	&10$^{-19}$ W/m$^{2}$	&10$^{-19}$ W/m$^{2}$	&10$^{-19}$ W/m$^{2}$ & 10$^{-19}$ W/m$^{2}$	\\
\hline
35\,A	&	9.7$\pm$0.6	&	?	&	5.5$\pm$0.7	&	5.6$\pm$0.9	&	3.5$\pm$1.8	&	6.3$\pm$0.7	\\
35\,B	&	4.1$\pm$0.4	&	5.2$\pm$1.0	&	2.6$\pm$0.3	&	2.6$\pm$0.4	&	1.4$\pm$0.4 &	2.5$\pm$0.3	\\
35\,C	& 	3.8$\pm$0.3	&	4.2$\pm$0.4	&	2.4$\pm$0.4	&	2.3$\pm$0.4	&	1.4$\pm$0.3	&	2.0$\pm$0.4\\
35\,D	& 	3.1$\pm$0.3	&	4.3$\pm$1.1	&	1.7$\pm$0.4	&	1.9$\pm$0.2	&	7.3$\pm$0.7	&	1.7$\pm$0.2	\\
\hline
\end{tabular}
\end{minipage}
\end{table*}

\begin{table*}
\begin{minipage}{175mm}
\caption{Optical emission line fluxes. Fluxes have not been corrected for extinction.}
\scriptsize
\begin{tabular}{c c c c c c c c}
\hline
 Source & H$_{\alpha}$ & H$_{\beta}$ & H$_{\gamma}$ & H$_{\delta}$ & H$_{\eta}$ & He\,{\sc i} 3888 \AA{} & He\,{\sc i} 4464 \AA{} \\
& \multicolumn{7}{c}{$10^{-13}$ erg s$^{-1}$ cm$^{-2}$} \\ 
\hline
01 & 1.77$\pm$0.03 & 0.382$\pm$0.003 & 0.160$\pm$0.005 &0.084$\pm$0.005 & 0.044$\pm$0.004 & &\\
02 & 0.072$\pm$0.003 & \null & \null & \null & \null & \null & \null \\
03 & 7.8$\pm$0.2 & 2.0$\pm$0.1 & 0.73$\pm$0.05 & 0.32$\pm$0.02 & 0.18$\pm$0.03 & 0.16$\pm$0.03 & \\
04 & 1.86$\pm$0.1 & 0.241$\pm$0.001 & 0.0732$\pm$0.0008 & 0.030$\pm$0.003 & & &\\
07 & 0.21$\pm$0.01 & 0.040$\pm$0.006 &  & & & &\\
08 & 0.62$\pm$0.02 & 0.149$\pm$0.008 & 0.057$\pm$0.008 & 0.033$\pm$0.005 & & &\\
09 & 4.3$\pm$0.1 & 1.26$\pm$0.04 & 0.53$\pm$0.04 & 0.24$\pm$0.01 & & &\\
10 & 0.050$\pm$0.009 & & & 0.02$\pm$0.01 & & &\\
11 & 0.06$\pm$0.02 &  & & & & &\\
12 & 0.14$\pm$0.01 & 0.037$\pm$0.009 & & & & &\\
13 & 1.05$\pm$0.08 & 0.157$\pm$0.007 & 0.054$\pm$0.009 & 0.011$\pm$0.006 & & &\\
14 & 0.28$\pm$0.04 &  & & & & &\\
15 & 0.90$\pm$0.05 & 0.237$\pm$0.007 & 0.102$\pm$0.008 & 0.05$\pm$0.01 & 0.029$\pm$0.006 & 0.037$\pm$0.006 &\\
16 & 0.21$\pm$0.02 &  & & & & &\\
17 & 0.038$\pm$0.005 & & & & & &\\
18 & 0.07$\pm$0.02 & & & & & &\\
20 & 1.8$\pm$0.3 & 0.28$\pm$0.07 &  & p? & & &\\
21 & 0.43$\pm$0.02 & 0.060$\pm$0.006 & 0.018$\pm$0.004 &  &  & &\\
22 & 0.54$\pm$0.04 & 0.19$\pm$0.01 & 0.07$\pm$0.01 & 0.024$\pm$0.004 & & &\\
23 & 0.055$\pm$0.006 &  & & & & &\\
25 & 1.18$\pm$0.02 & 0.35$\pm$0.02 & 0.12$\pm$0.02 & & & &\\
26 & 6.1$\pm$0.3 & 2.6$\pm$0.2 & 0.82$\pm$0.04 & 0.41$\pm$0.07 & 0.27$\pm$0.03 & 0.33$\pm$0.06 & \\
27 & 0.07$\pm$0.01 &  & & & & &\\
28 & 0.90$\pm$0.05 & 0.24$\pm$0.02 & 0.081$\pm$0.006 & 0.03$\pm$0.01 & & &\\
30 & 0.36$\pm$0.02 & 0.054$\pm$0.005 & 0.013$\pm$ 0.005 & & & &\\
32 & 0.21$\pm$0.04 &  & & & & &\\
33 & 0.20$\pm$0.07 & & & & & &\\
34 & 0.082$\pm$0.007 & & & p? & & &\\
35 & 97$\pm$6 & 44$\pm$4 & & & & &\\
36 & 19$\pm$1 & 11.9$\pm$0.3 & & & & &\\
\hline
Source & He\,{\sc i} 5876 \AA{} & He\,{\sc i} 6679 \AA{} & He\,{\sc i} 7066 \AA{} & [N\,{\sc ii}] 6548 \AA{} & [N\,{\sc ii}] 6584 \AA{} & [S\,{\sc ii}] 6717 \AA{} & [S\,{\sc ii}] 6731 \AA{} \\
& \multicolumn{7}{c}{$10^{-13}$ erg s$^{-1}$ cm$^{-2}$} \\ 
\hline
01 & \null & \null & \null & 0.049$\pm$0.003 & 0.160$\pm$0.004 & 0.12$\pm$0.07 & 0.119$\pm$0.003\\
02 & \null & \null & \null & \null & \null & \null & \null \\
03 & 1.14$\pm$0.02 & 0.033$\pm$0.004 & \null & 0.19$\pm$0.02 & 0.62$\pm$0.01 & 0.4$\pm$0.1 & 0.38$\pm$0.03 \\
04 & \null & \null & \null &\null & \null & 0.04$\pm$0.01 & 0.028$\pm$0.002 \\
07 & \null & \null & \null & \null & 0.023$\pm$0.009 & 0.021$\pm$0.004 & 0.020$\pm$0.001\\
08 & 0.02$\pm$0.01 & 0.009$\pm$0.003 & 0.005$\pm$0.002 & 0.009$\pm$0.004 & 0.036$\pm$0.005 & 0.03$\pm$0.02 & 0.025$\pm$0.002\\
09 & \null & \null & \null & 0.13$\pm$0.02 & 0.43$\pm$0.03 & 0.35$\pm$0.09 & 0.30$\pm$0.02\\
12 & \null & \null & \null & \null & 0.018$\pm$0.003 & 0.024$\pm$0.003 & 0.016$\pm$0.003\\
13 & \null & \null & \null & 0.014$\pm$0.005 & 0.065$\pm$0.008 & 0.04$\pm$0.01 & 0.044$\pm$0.006\\
15 & \null & 0.005$\pm$0.003 & \null & 0.014$\pm$0.007 & 0.062$\pm$0.006 & 0.063$\pm$0.009 & 0.05$\pm$0.01\\
16 & \null & \null & \null & \null & \null & 0.009$\pm$0.003 & \null\\
21 & \null & \null & \null & \null & 0.033$\pm$0.006 & 0.012$\pm$0.003 & 0.017$\pm$0.002\\
22 & \null & \null & \null & \null & \null & 0.04$\pm$0.01 & 0.033$\pm$0.009\\
23 & \null & \null & \null & \null & \null & \null & 0.002$\pm$0.001 \\
25 & \null & \null & \null & 0.036$\pm$0.006 & 0.097$\pm$0.007 & 0.14$\pm$0.01 & 0.105$\pm$0.008\\
26 & 0.31$\pm$0.01 & 0.073$\pm$0.009 & 0.063$\pm$0.007 & \null & 0.159$\pm$0.008 & 0.20$\pm$0.01 & 0.16$\pm$0.02\\
27 & \null & \null & \null & 0.006$\pm$0.003 & 0.007$\pm$0.003 & 0.019$\pm$0.002 & 0.015$\pm$0.003 \\
28 & 0.015$\pm$0.005 & 0.0027$\pm$0.0009 & \null & 0.025$\pm$0.006 & 0.060$\pm$0.007 & 0.07$\pm$0.01 & 0.07$\pm$0.01\\
34 & \null & \null & \null & \null & \null & \null & 0.002$\pm$0.001 \\
35 & 3.8$\pm$0.4 & \null & 3.5$\pm$0.4 & 0.42$\pm$0.03 & 1.11$\pm$0.06 & 0.68$\pm$0.04 & 0.94$\pm$0.06\\
36 & \null & \null & 0.23$\pm$0.01 & 0.32$\pm$0.05 & 0.35$\pm$0.03 & 0.27$\pm$0.03 & 0.27$\pm$0.03 \\
\hline
 Source &  [O\,{\sc ii}] 7319/7319 \AA{} & [O\,{\sc ii}] 7330/7331 \AA{} & [O\,{\sc iii}] 4959 \AA{} & [O\,{\sc iii}] 5007 \AA{} & [Ne\,{\sc iii}] 3869 \AA{} & [Ar\,{\sc iii}] 7136 \AA{} & [Ar\,{\sc iii}] 7751 \AA{} \\
& \multicolumn{7}{c}{$10^{-13}$ erg s$^{-1}$ cm$^{-2}$} \\ 
\hline
01 & 0.031$\pm$0.006 & \null & \null & \null & \null & \null & \null \\
03 & 0.16$\pm$0.06 & \null & 0.45$\pm$0.01 & 1.28$\pm$0.06 & \null & 0.095$\pm$0.001 & 0.028$\pm$0.005 \\
08 & 0.021$\pm$0.004 & \null & 0.054$\pm$0.008 & 0.19$\pm$0.01 & \null & 0.010$\pm$0.003 & \null \\
09 & 0.07$\pm$0.02 & \null & \null & \null & \null & \null & \null \\
12 & \null & \null & \null & 0.027$\pm$0.005 & \null & \null & \null \\
13 & 0.030$\pm$0.003 & \null & 0.014$\pm$0.006 & 0.018$\pm$0.005 & 0.03$\pm$0.02 & \null & \null\\
15 & & & & & & &\\
17 & & & & & & &\\
21 & & & & & & &\\
22 & & & & & & &\\
25 & & & & & & &\\
26  & & & & & & &\\
27 & & & & & & &\\
28 & & & & & & &\\
34 & & & & & & &\\
35 & 0.95$\pm$0.05 & 0.79$\pm$0.03 & 110$\pm$10 & 330$\pm$20 & & 3.3$\pm$0.4 & 9.3$\pm$0.1\\
36 & 0.14$\pm$0.01 & 0.21$\pm$0.06 & 22.4$\pm$0.5 & 65$\pm$1 &  & 0.61$\pm$0.03 & 0.32$\pm$0.02\\
\hline
\end{tabular}
\end{minipage}
\end{table*}

\label{lastpage}
\end{document}